\documentclass{article}
\usepackage[letterpaper,top=2cm,bottom=2cm,left=3cm,right=3cm,marginparwidth=1.75cm]{geometry}
\usepackage[labelformat=simple]{subcaption}
\usepackage[colorlinks=true, linkcolor=blue, citecolor=blue]{hyperref}
\usepackage{amssymb,amsmath,amsfonts,amsthm,dsfont}
\usepackage{bbm}
\usepackage{physics}
\usepackage{graphicx}
\usepackage{authblk}
\usepackage[superscript]{cite}
\usepackage{setspace}
\DeclareMathAlphabet{\mathcal}{OMS}{cmsy}{m}{n}

\DeclareGraphicsExtensions{.pdf, .png}
\begin{document}
\title{Variational Quantum Approximate Support Vector Machine with Inference Transfer}

\author[1]{Siheon Park}
\author[2]{Daniel K. Park}
\author[1,3,*]{June-Koo Kevin Rhee}
\affil[1]{KAIST, School of Electrical Engineering, Daejeon, 34141, South Korea}
\affil[2]{Yonsei University, Dept. of Applied Statistics \& Dept. of Statistics and Data Science, Seoul, 03722, South Korea}
\affil[3]{Qunova Computing, Inc., Daejeon, 34051, South Korea}

\affil[*]{rhee.jk@kaist.edu}
\date{}

\maketitle


\begin{abstract}
A kernel-based quantum classifier is the most practical and influential quantum machine learning technique for the hyper-linear classification of complex data. We propose a Variational Quantum Approximate Support Vector Machine (VQASVM) algorithm that demonstrates empirical sub-quadratic run-time complexity with quantum operations feasible even in NISQ computers. We experimented our algorithm with toy example dataset on cloud-based NISQ machines as a proof of concept. We also numerically investigated its performance on the standard Iris flower and MNIST datasets to confirm the practicality and scalability.
\end{abstract}

\flushbottom
%
%
\thispagestyle{empty}

\section*{Introduction}

    Quantum computing opens up new exciting prospects of quantum advantages in machine learning in terms of sample and computation complexity.\cite{qnflprl22, qram2, qml, schuld2018supervised, qsvm} One of the foundations of these quantum advantages is the ability to form and manipulate data efficiently in a large quantum feature space, especially with kernel functions used in classification and other classes of machine learning.\cite{ schuld2017implementing,schuld2019quantum,ibm-qsvm,lloyd2020quantum, blank2020quantum,park2020theory,schuld2021supervised,liu2021rigorous,blank2022compact}

    The support vector machine (henceforth SVM)\cite{svm} is one of the most comprehensive models that help conceptualize the basis of supervised machine learning. SVM classifies data by finding the optimal hyperplane associated with the widest margin between the two classes in a feature space. SVM can also perform highly nonlinear classifications using what is known as the kernel trick.\cite{boser1992training, vapnik1997support, guyon1992capacity} The convexity of SVM guarantees global optimization.

    One of the first quantum algorithms exhibiting an exponential speed-up capability is the least-square quantum support vector machine (LS-QSVM).\cite{qsvm}
    However, the quantum advantage of LS-QSVM strongly depends on costly quantum subroutines such as density matrix exponentiation\cite{qpca} and quantum matrix inversion\cite{hhl, aaronson2015read} as well as components such as quantum random access memory (QRAM) \cite{qram1, qram2}.
    Because the corresponding procedures require quantum computers to be fault-tolerant, LS-QSVM is unlikely to be realized in noisy intermediate-scale quantum (NISQ) devices\cite{nisq}.
    On the other hand, there are a few quantum kernel-based machine-learning algorithms for near-term quantum applications. Well-known examples are quantum kernel estimators (QKE)\cite{ibm-qsvm}, variational quantum classifiers (VQC),\cite{ibm-qsvm} and Hadamard or SWAP test classifiers (HTC, STC).\cite{park2020theory, blank2020quantum}
    These algorithms are applicable to NISQ, as there are no costly operations needed.
    However, the training time complexity is even worse than in the classical SVM case. For example, the number of measurements required to generate only the kernel matrix evaluation of QKE scales with the number of training samples to the power of four.\cite{ibm-qsvm} 

    Here, we propose a novel quantum kernel-based classifier that is feasible with NISQ devices and that can exhibit a quantum advantage in terms of accuracy and training time complexity as exerted in Ref.~\citenum{ibm-qsvm}. Specifically, we have discovered distinctive designs of quantum circuits that can evaluate the objective and decision functions of SVM. 
    The number of measurements for these circuits with a bounded error is independent from the number of training samples. The depth of these circuits scales also linearly with the size of the training dataset. Meanwhile, the exponentially fewer parameters of parameterized quantum circuits (PQCs)\cite{pqc} encodes the Lagrange multipliers of SVM. Therefore, the training time of our model with a variational quantum algorithm (VQA)\cite{vqa} scales as sub-quadratic, which is asymptotically lower than that of the classical SVM case.\cite{qsvm,chang2011libsvm, Coppersmith1990MatrixProgressions} Our model also shows an advantage in classification due to its compatibility with any typical quantum feature map.
    

\section*{Results}\label{sec:result}

\subsection*{Support Vector Machine (SVM)}\label{sec:review-svm}

    Data classification infers the most likely class of an unseen data point $\hat{\mathbf{x}} \in \mathbbm{C}^N$ given a training dataset $\mathcal{S} = \left\{ \left(\mathbf{x}_i, y_i\right) \right\}_{i=0}^{M-1}$ $\subset$ $\mathcal{X}\times\mathcal{Y}$. Here, $\mathcal{X}\subset \mathbbm{C}^N$ and $\mathcal{Y}=\{0,1,\ldots,L-1\}$. Although the data is real-valued in practical machine learning tasks, we allow complex-valued data without a loss of generality. We focus on binary classification. (i.e., $\mathcal{Y}=\{-1, 1\}$), because multi-class classification can be conducted with a multiple binary SVM via a one-versus-all or a one-versus-on scheme\cite{giuntini2021quantum}.  We assume that $\mathcal{S}$ is linearly separable in the higher dimensional Hilbert space $\mathcal{H}$ given some feature map $\phi:\mathcal{X}\mapsto\mathcal{H}$. Then, there should exist two parallel supporting hyperplanes $\langle\mathbf{w}, \phi(\cdot)\rangle+b=y\in\mathcal{Y}$ that divide training data. The goal is to find hyperplanes for which the margin between them is maximized. To maximize the margin even further, the linearly separable condition can be relaxed so that some of the training data can penetrate into the ``soft" margin. Because the margin is given as $2/\norm{\mathbf{w}}$ by simple geometry, the mathematical formulation of SVM\cite{liu2021rigorous} is given as 
    \begin{equation}\label{eq:primal_svm}
        p^\star=\min_{\mathbf{w}, b, \boldsymbol{\xi}}\frac{1}{2}\norm{\mathbf{w}}^2+\frac{C}{2}\sum_{i=0}^{M-1}{\xi_i^2}~:~ y_i\left(\langle\mathbf{w}, \phi(\mathbf{x}_i)\rangle+b\right)\ge1-\xi_i,
    \end{equation}
    where the slack variable $\xi$ is introduced to represent a violation of the data in the linearly separable condition. The dual formulation of SVM is expressed as\cite{boyd2004convex}
    \begin{equation}\label{eq:dual_svm}
        d^\star=\max_{\boldsymbol{\beta}\succeq0}{\sum_{i=0}^{M-1}{\beta_i}-\frac{1}{2}\sum_{i,j=0}^{M-1}{\beta_i\beta_jy_iy_jk(\mathbf{x}_i, \mathbf{x}_j)}-\frac{1}{2C}\sum_{i=0}^{M-1}{\beta_i^2}}~:~\sum_{i=0}^{M-1}{\beta_iy_i}=0,
    \end{equation}
    where the positive semi-definite (PSD) kernel is $k(\mathbf{x}_1, \mathbf{x}_2) = \langle{\phi(\mathbf{x}_1)},{\phi(\mathbf{x}_2)}\rangle$ for $\mathbf{x}_{1,2} \in \mathcal{X}$.  The $\beta_i$ values are non-negative Karush-Kuhn-Tucker multipliers. This formulation employs an implicit feature map uniquely determined by the kernel. The global solution $\boldsymbol{\beta}^\star$ is obtained in polynomial time due to convexity.\cite{boyd2004convex} After optimization, the optimum bias is recovered as $b^{\star}=y_q(1-C^{-1}\beta^\star_q)-\sum_{i=0}^{M-1}{\beta^{\star}_i y_i k(\mathbf{x}_q, \mathbf{x}_i)}$ for any $\beta^\star_q>0$. Such training data $\mathbf{x}_q$ with non-zero weight $\beta_q$ are known as the support vectors. We estimate the labels of unseen data with a binary classifier: 
    \begin{equation}\label{eq:class_form}
        \hat{y} = \mathrm{sgn}\left\{\sum_{i=0}^{M-1}{\beta^{\star}_i y_i k({\mathbf{x}_i},{\hat{\mathbf{x}})}} +b^{\star}\right\}.
    \end{equation}
    In a first-hand principle analysis, the complexity of solving Eq.~(\ref{eq:dual_svm}) is $\order{M^2(N+M)\log(1/\delta)}$ with accuracy of $\delta$. A kernel function with complexity of $\order{N}$ is queried $M(M-1)/2$ times to construct the kernel matrix, and quadratic programming takes $\order{M^3\log(1/\delta)}$ to find $\boldsymbol{\beta}^\star$ for a non-sparse kernel matrix\cite{qsvm}. Although the complexity of SVM decreases when employing modern programming methods\cite{Coppersmith1990MatrixProgressions, chang2011libsvm}, it is still higher than or equal to $\order{M^2N}$ due to kernel matrix generation and quadratic programming. Thus, a quantum algorithm that evaluates all terms in Eq.~(\ref{eq:dual_svm}) for $\order{MN}$ time and achieves a minimum of fewer than $\order{M}$ evaluations would have lower complexity than classical algorithms. We apply two forms of transformations to Eq.~(\ref{eq:dual_svm}) to realize an efficient quantum algorithm.

\subsection*{Change of Variable and Bias Regularization}\label{sec:bias}
    Constrained programming, such as that in the SVM case, is often transformed into unconstrained programming by adding penalty terms of constraints to the objective function. Although there are well-known methods such as an interior point method,\cite{boyd2004convex} we prefer the strategies of a `change of variables' and `bias regularization' to maintain the quadratic form of SVM. Although motivated to eliminate constraints, the results appear likely to lead to an efficient quantum SVM algorithm.
    
    First, we change optimization variable $\boldsymbol{\beta}$ to $(\boldsymbol{\alpha}, B)$, where $B:=\sum_{i=0}^{M-1}{\beta_i}$ and $\boldsymbol{\alpha}:=\boldsymbol{\beta}/B$ to eliminate inequality constraints. The $l_1$-normalized variable $\boldsymbol{\alpha}$ is an $M$-dimensional probability vector given that $0\le\alpha_i\le1, \forall{i}\in\left\{0,\cdots,M-1\right\}$ and $\sum_{i=0}^{M-1}{\alpha_i}=1$. Let us define $W_k(\boldsymbol{\alpha};\mathcal{S}):=\sum_{i,j=0}^{M-1}{} \alpha_i \alpha_j y_i y_j k\left(\mathbf{x}_i, \mathbf{x}_j\right) + C^{-1}\sum_{i=0}^{M-1}{}\alpha^2_i$. We substitute the variables into Eq.~(\ref{eq:dual_svm}):
    \begin{equation}\label{eq:sub_dual_svm}
        \max_{\boldsymbol{\alpha}\in {PV}_M}\max_{B\ge0}\left\{B-\frac{1}{2}B^2W_k(\boldsymbol{\alpha};\mathcal{S})\right\}:\sum_{i=0}^{M-1}{\alpha_i y_i}=0,
    \end{equation}
    where ${PV}_M$ is a set of $M$-dimensional probability vectors. Because $W_k(\boldsymbol{\alpha};\mathcal{S})\ge0$ for an arbitrary $\boldsymbol{\alpha}$ due to the property of the positive semi-definite kernel, $B^{\star} = 1/W_k(\boldsymbol{\alpha};\mathcal{S})$ is a partial solution that maximizes Eq.~(\ref{eq:sub_dual_svm}) on $B$. Substituting $B^\star$ with Eq.~(\ref{eq:sub_dual_svm}), we have
    \begin{equation}
        \max_{\boldsymbol{\alpha}\in {PV}_M}{\frac{1}{2W_k(\boldsymbol{\alpha};\mathcal{S})}}:\sum_{i=0}^{M-1}{\alpha_i y_i}=0.
    \end{equation}
    Finally, because maximizing $1/2W_k(\boldsymbol{\alpha};\mathcal{S})$ is identical to minimizing $W_k(\boldsymbol{\alpha};\mathcal{S})$, we have a simpler formula that is equivalent to Eq.~(\ref{eq:dual_svm}):
    \begin{equation}\label{eq:pv_dual_svm}
        \Tilde{d}^\star = \min_{\boldsymbol{\alpha}\in {PV}_M}{\sum_{i,j=0}^{M-1}{\alpha_i\alpha_j y_i y_j k(\mathbf{x}_i, \mathbf{x}_j)}+\frac{1}{C}\sum_{i=0}^{M-1}{\alpha^2_i}}:\sum_{i=0}^{M-1}{\alpha_i y_i}=0,
    \end{equation}
    The above Eq.~(\ref{eq:pv_dual_svm}) implies that instead of optimizing $M$ numbers of bounded free parameters $\boldsymbol{\beta}$ or $\boldsymbol{\alpha}$, we can optimize the $\log(M)$-qubit quantum state $\ket{\psi_{\boldsymbol{\alpha}}}$ and define $\alpha_i:=\abs{\braket{i}{\psi_{\boldsymbol{\alpha}}}}^2$. Therefore, if there exists an efficient quantum algorithm that evaluates the objective function of Eq.~(\ref{eq:pv_dual_svm}) given $\ket{\psi_{\boldsymbol{\alpha}}}$, the complexity of SVM would be improved. In fact, in \hyperref[sec:circuits]{the later section}, we propose quantum circuits with linearly scaling complexity for that purpose.
    
    The equality constraint is relaxed after adding the $l_2$-regularization term of the bias to Eq.~(\ref{eq:primal_svm}). Motivated by the loss function and regularization perspectives of SVM\cite{mml-book}, this technique was introduced\cite{mangasarian1999successive, frie1998kernel} and developed\cite{hsu2002simple,hsu2002comparison} previously. The primal and dual forms of SVM become
    \begin{equation}\label{eq:reduced_primal_svm}
        p^\star=\min_{\mathbf{w}, b, \boldsymbol{\xi}}\frac{1}{2}\norm{\mathbf{w}}^2+\frac{\lambda}{2}b^2+\frac{C}{2}\sum_{i=0}^{M-1}{\xi_i^2}~:~ y_i\left(\langle\mathbf{w}, \phi(\mathbf{x}_i)\rangle+b\right)\ge1-\xi_i,
    \end{equation}
    \begin{equation}\label{eq:reduced_dual_svm}
       d^\star=\max_{\boldsymbol{\beta}\succeq0}{\sum_{i=0}^{M-1}{\beta_i}-\frac{1}{2}\sum_{i,j=0}^{M-1}{\beta_i\beta_jy_iy_j\left[k(\mathbf{x}_i, \mathbf{x}_j)+\frac{1}{\lambda}\right]}-\frac{1}{2C}\sum_{i=0}^{M-1}{\beta_i^2}}.
    \end{equation}
    Note that $k(\cdot, \cdot)+\lambda^{-1}$ is a positive definite. As shown earlier, changing the variables causes Eq.~(\ref{eq:reduced_dual_svm}) to become another equivalent optimization problem:
        \begin{equation}\label{eq:VQASVM_optim}
            \Tilde{d}^\star = \min_{\boldsymbol{\alpha}\in {PV}_M}{\sum_{i,j=0}^{M-1}{\alpha_i\alpha_jy_iy_j\left[k(\mathbf{x}_i, \mathbf{x}_j)+\frac{1}{\lambda}\right]}+\frac{1}{C}\sum_{i=0}^{M-1}{\alpha_i^2}}.
        \end{equation}
    As the optimal bias is given as $b^{\star}=\lambda^{-1}\sum_{i=0}^{M-1}{\alpha^{\star}_i y_i}$ according to the Karush-Kuhn-Tucker condition, the classification formula inherited from Eq.~(\ref{eq:class_form}) is expressed as
        \begin{equation}\label{eq:VQASVM_class}
            \hat{y} = \mathrm{sgn}\left\{\sum_{i=0}^{M-1}{\alpha^{\star}_i y_i {\left[k(\mathbf{x}_i, \mathbf{x}_j)+\frac{1}{\lambda}\right]}}\right\}.
        \end{equation}
    Eqs. (\ref{eq:reduced_dual_svm}) and (\ref{eq:VQASVM_optim}) can be viewed as Eqs.~(\ref{eq:dual_svm}) and (\ref{eq:pv_dual_svm}) with a quadratic penalizing term on the equality constraint such that they become equivalent in terms of the limit of $\lambda\to0$. Thus, Eqs. (\ref{eq:reduced_primal_svm}), (\ref{eq:reduced_dual_svm}), and (\ref{eq:VQASVM_optim}) are more relaxed SVM optimization problems with an additional hyperparameter $\lambda$.
    
\begin{figure}[t]
    \centering
    \includegraphics[width=\textwidth]{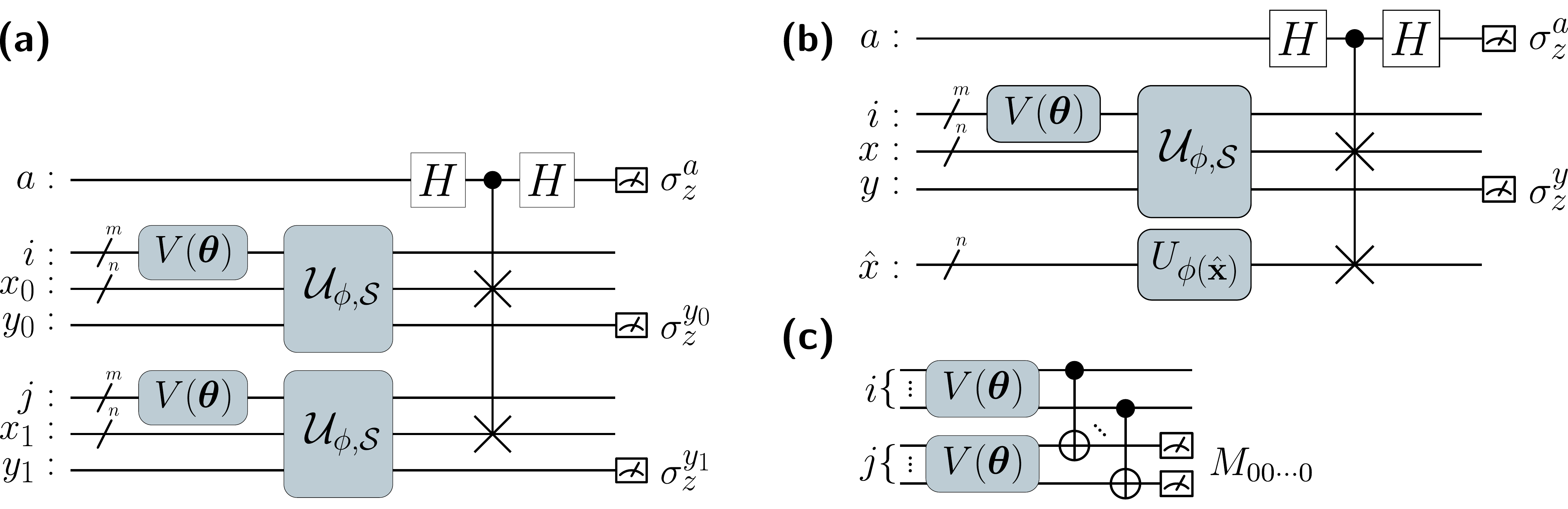}
    \caption{Circuit Architecture of VQASVM. Loss, decision, and regularization circuits are shown in the order of panel \textbf{\textsf{(a)}}, \textbf{\textsf{(b)}}, and \textbf{\textsf{(c)}} All qubits of index registers $i$ and $j$ are initialized to $\ket{+}=\left(\ket{0}+\ket{1}\right)/\sqrt{2}$, and the rest to $\ket{0}$. Ansatz ${V(\boldsymbol{\theta})}$ is a PQC of $m=\log(M)$ qubits that encodes probability vector $\boldsymbol{\alpha}$. $\mathcal{U}_{\phi, \mathcal{S}}$ embeds a training data set $\mathcal{S}$ with a quantum feature map $U_{\phi(\hat{\mathbf{x}})}$, which embeds classical data $\hat{\mathbf{x}}$ to a quantum state $\ket{\phi(\hat{\mathbf{x}})}$. $n$ denotes the number of qubits for the quantum feature map, which is usually $N$, but can be reduced to $\log(N)$ if an amplitude encoding feature map is used. }
    \label{fig:circuits}
\end{figure}  

\subsection*{Variational Quantum Approximate Support Vector Machine}\label{sec:circuits}
    One way to generate the aforementioned quantum state $\ket{\psi_{\boldsymbol{\alpha}}}$ is to use amplitude encoding: $\ket{\psi_{\boldsymbol{\alpha}}}=\sum_{i=0}^{M-1}{\sqrt{\alpha_i}\ket{i}}$. However, doing so would be inefficient because the unitary gate of amplitude encoding has a complex structure that scales as $\mathcal{O}(\mathrm{poly}(M))$.\cite{mottonen2004transformation} Another way to generate $\ket{\psi_{\boldsymbol{\alpha}}}$ is to use a parameterized quantum circuit (PQC), known as an ansatz in this case. Because there is no prior knowledge in the distribution of $\alpha_i^\star$s, the initial state should be $\ket{++\cdots+}=\frac{1}{\sqrt{M}}\sum_{i=0}^{M-1}\ket{i}$. The ansatz $V(\boldsymbol{\theta})$ can transform the initial state into other states depending on gate parameter vector $\boldsymbol{\theta}$: $\ket{\psi_{\boldsymbol{\alpha}}}=V(\boldsymbol{\theta})\ket{++\cdots+}$. In other words, optimization parameters encoded by $\boldsymbol{\theta}$ with the ansatz are represented as $\alpha_i(\boldsymbol{\theta})=\abs{\bra{i}{V(\boldsymbol{\theta})}\ket{++\cdots+}}^2$. Given the lack of prior information, the most efficient ansatz design can be a hardware-efficient ansatz (HEA), which consists of alternating local rotation layers and entanglement layers.\cite{sim2019expressibility, kandala2017hardware} The number of qubits and the depth of this ansatz are $\mathcal{O}(\mathrm{polylog}(M))$.

    We discovered the quantum circuit designs that compute Eqs. (\ref{eq:VQASVM_optim}) and (\ref{eq:VQASVM_class}) within $\order{M}$ time. Conventionally, the quantum kernel function is defined as the Hilbert Schmidt inner product: $k(\cdot,\cdot)=\abs{\braket{\phi(\cdot)}{\phi(\cdot)}}^2$.\cite{ibm-qsvm,schuld2019quantum,schuld2018supervised,schuld2021supervised,park2020theory,blank2020quantum} First, we divide the objective function in Eq.~(\ref{eq:VQASVM_optim}) into the loss and regularizing functions of $\boldsymbol{\theta}$ using the above ansatz encoding:
    \begin{equation}\label{eq:loss_function}
        \mathcal{L}_{\phi, \lambda}\left(\boldsymbol{\theta};\mathcal{S}\right) = \sum_{i, j=0}^{M-1}{\alpha_i(\boldsymbol{\theta})\alpha_j(\boldsymbol{\theta}) y_iy_j \left[\abs{\braket{\phi(\mathbf{x}_i)}{\phi(\mathbf{x}_j)}}^2 + \frac{1}{\lambda}\right]},~\mathcal{R}(\boldsymbol{\theta}) = \sum_{i=0}^{M-1}{\alpha_i(\boldsymbol{\theta})^2}.
    \end{equation}
    Specifically, the objective function is equal to $\mathcal{L}_{\phi, \lambda}+{C^{-1}}\mathcal{R}$. Similarly, the decision function in Eq.~(\ref{eq:VQASVM_class}) becomes
    \begin{equation}\label{eq:decision_function}
        f_{\phi, \lambda}\left(\mathbf{x};\boldsymbol{\theta},\mathcal{S}\right) = \sum_{i=0}^{M-1}\alpha_i(\boldsymbol{\theta})y_i{\left[\abs{\braket{\phi(\mathbf{x}_i)}{\phi(\mathbf{x})}}^2 + \frac{1}{\lambda}\right]}.
    \end{equation}
    Inspired by STC\cite{blank2020quantum, park2020theory}, the quantum circuits in Fig.~\ref{fig:circuits} efficiently evaluate $\mathcal{L}_{\phi, \lambda}, \mathcal{R}$ and $f_{\phi, \lambda}$. The quantum gate $\mathcal{U}_{\phi, \mathcal{S}}$ embeds the entire training dataset with the corresponding quantum feature map $U_{\phi(\mathbf{x})}\ket{00\cdots0}=\ket{\phi(\mathbf{x})}$, so that  $\mathcal{U}_{\phi, \mathcal{S}}\ket{i}\otimes\ket{00\cdots0}\otimes\ket{0}=\ket{i}\otimes\ket{\phi(\mathbf{x}_i)}\otimes\ket{y_i}$. Therefore, the quantum state after state preparation is $\ket{\Psi}=\sum_{i=0}^{M-1}{}\sqrt{\alpha_i}\ket{i}\otimes\ket{\phi(\mathbf{x}_i)}\otimes\ket{y_i}$.
    We apply a SWAP test and a joint $\sigma_z$ measurement in the loss and decision circuits to evaluate $\mathcal{L}_{\phi, \lambda}$ and $f_{\phi, \lambda}$:
    \begin{equation}\label{eq:thrm1}
        \mathcal{L}_{\phi, \lambda}\left(\boldsymbol{\theta};\mathcal{S}\right)=\expval{\sigma_z^a \sigma_z^{y_1} \sigma_z^{y_2}}_{\boldsymbol{\theta}}+\frac{1}{\lambda}\expval{\sigma_z^{y_1} \sigma_z^{y_2}}_{\boldsymbol{\theta}},~
        f_{\phi, \lambda}\left(\mathbf{x};\boldsymbol{\theta},\mathcal{S}\right)=\expval{\sigma_z^a \sigma_z^y}_{\mathbf{x};\boldsymbol{\theta}}+\frac{1}{\lambda}\expval{\sigma_z^y}_{\mathbf{x};\boldsymbol{\theta}}, \mathcal{R}(\boldsymbol{\theta})=\langle M_{00\cdots0} \rangle_{\boldsymbol{\theta}}.
    \end{equation}
    Here, $\sigma_z$ is a Pauli Z operator and $M_{00\cdots0}$ is a projection measurement operator of state $\ket{0}^{\otimes{\log(M)}}$. See \hyperref[method:proof]{Method} for specified derivations. 
    The asymptotic complexities of the loss and decision circuits are linear with regard to the amount of training data. $\mathcal{U}_{\phi, \mathcal{S}}$ can be prepared with $\order{MN}$ operations.\cite{schuld2018supervised} See \hyperref[method:ucsq]{Method} for the specific realization used in this article. Because $V(\boldsymbol{\theta})$ has $\order{\mathrm{polylog}(M)}$ depth and the SWAP test requires only $\order{N}$ operations, the overall run-time complexity of evaluating $\mathcal{L}_{\phi, \lambda}(\boldsymbol{\theta};\mathcal{S})$ and $f_{\phi, \lambda}(\mathbf{x}; \boldsymbol{\theta}, \mathcal{S})$ with bounded error $\epsilon$ is $\order{\epsilon^{-2}MN}$ (See the Supplementary Fig. S5 online for numerical verification of the scaling). Similarly, the complexity of estimating $\mathcal{R}(\boldsymbol{\theta})$ with accuracy $\epsilon$ is $\order{\epsilon^{-2}\mathrm{polylog}(M)}$ due to two parallel ${V(\boldsymbol{\theta})}$s and $\order{\log(M)}$ CNOT gates.

    
\begin{figure}[t]
    \centering
    \includegraphics[width=\textwidth]{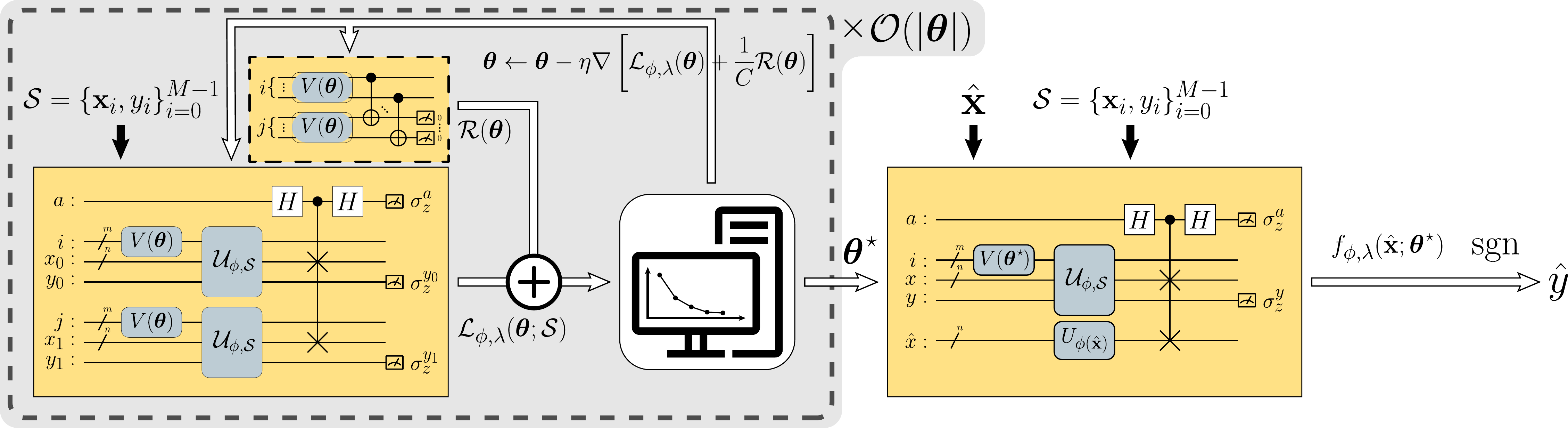}
    \caption{Variational Quantum Approximated Support Vector Machine. The white round boxes represent classical calculations whereas the yellow round boxes represent quantum operations. The white arrows represent the flow of classical data whereas the black arrows represent the embedding of classical data. The grey areas indicate the corresponding training phase of each iteration. The regularization circuit in the black dashed box can be omitted for a hard-margin case where $C\to\infty$.}
    \label{fig:algorithms}
\end{figure}    

    We propose a variational quantum approximate support vector machine (VQASVM) algorithm that solves the SVM optimization problem with VQA\cite{vqa} and transfers the optimized parameters to classify new data effectively. Fig.~\ref{fig:algorithms} summarizes the process of VQASVM. We estimate $\boldsymbol{\theta}^\star$, which minimizes the objective function; this is then used for classifying unseen data:
    \begin{equation}\label{eq:VQASVM_proc}
        \boldsymbol{\theta}^{\star} = \mathop{\mathrm{arg\,min}}_{\boldsymbol{\theta}}{\mathcal{L}_{\phi, \lambda}\left(\boldsymbol{\theta};\mathcal{S}\right)}+\frac{1}{C}\mathcal{R}(\boldsymbol{\theta}),~\hat{y}=\mathrm{sgn}\left\{f_{\phi, \lambda}\left(\hat{\mathbf{x}};\boldsymbol{\theta}^{\star}, \mathcal{S}\right)\right\}.
    \end{equation}
    Following the general scheme of VQA, the gradient descent (GD) algorithm can be applied; classical processors update the parameters of $\theta_i$, whereas the quantum processors evaluate the functions for computing gradients. Because the objective function of VQASVM can be expressed as the expectation value of a Hamiltonian, i.e., 
    \begin{equation}\label{eq:cost_fn}
        \mathcal{L}(\boldsymbol{\theta};\mathcal{S})+\frac{1}{C}\mathcal{R}(\boldsymbol{\theta})=\bra{\boldsymbol{+}}V(\boldsymbol{\theta})^\dagger\otimes V(\boldsymbol{\theta})^\dagger H V(\boldsymbol{\theta})\otimes V(\boldsymbol{\theta}) \ket{\boldsymbol{+}},
    \end{equation}
    where $H=\sum_{ij=0}^{M-1}\left[y_iy_jk(\mathbf{x}_i,\mathbf{x}_j)+\frac{1}{\lambda}y_iy_j+\frac{1}{C}\delta_{ij}\right]\ket{i}\bra{i}\otimes\ket{j}\bra{j}$,
    the exact gradient can be obtained by the modified parameter-shift rule\cite{param-shift1,param-shift2}.
    GD converges to a local minimum after $\order{\log(1/\delta)}$ iterations with the difference $\delta$\cite{boyd2004convex} given that estimation error of the objective function is smaller than $\delta$. Therefore, the total run-time complexity of VQASVM is $\order{\epsilon^{-2}\log({1/\epsilon}) M N \mathrm{polylog}(M)}$ with error of $\epsilon$ as the number of parameters is $\order{\mathrm{polylog}(M)}$.

    \begin{figure}[t]
        \centering
        \includegraphics[width=\textwidth]{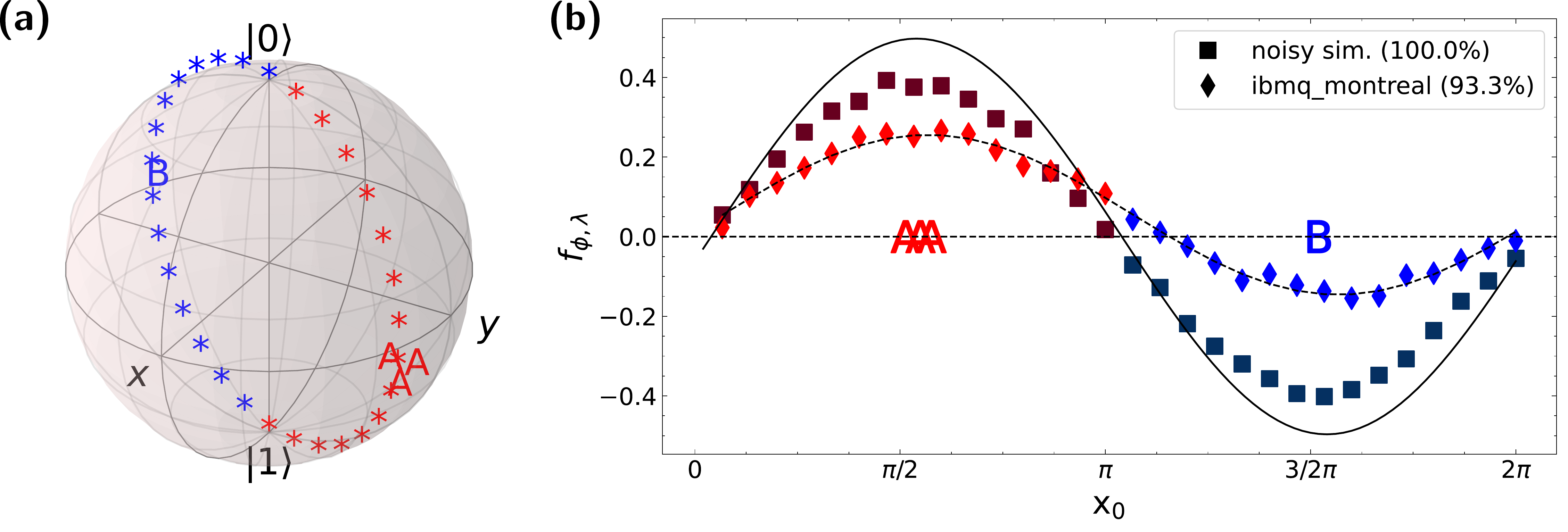}
        \caption{Experiments on a \textit{ibmq\_montreal} cloud NISQ processor. \textbf{\textsf{(a)}} The toy training (letters) and test (asterisk) data are shown here in a Bloch sphere. The color indicates the true class label of the data; i.e., red=class A, and blue=class B. The letters A represent the training data of class A, and letter B represents the training datum of class B. \textbf{\textsf{(b)}} classification results performed on \textit{ibmq\_montreal} QPU (diamonds) and a simulation with noise (squares) compared to theoretical values(solid line). $f_{\phi, \lambda}$ is the decision function value of each test datum. The letters A and B represent the training data, located at their longitudinal coordinates on the Bloch sphere ($\mathrm{x}_0$). Curved dashed lines are the sine-fitting of the \textit{ibmq\_montreal} results. Values inside the round brackets in the legend are the classification accuracy rates.}\label{fig:montreal}
    \end{figure}
    
\subsection*{Experiments on IBM Quantum Processors}\label{sec:experiment}
    We demonstrate the classification of a toy dataset using the VQASVM algorithm on NISQ computers as a proof-of-concept. Our example dataset is mapped to a Bloch sphere, as shown in Fig. \ref{fig:montreal}a. Due to decoherence, we set the data dimension to $N=2$ and number of training data instances to $M=4$. First, we randomly choose the greatest circle on the Bloch sphere that passes $\ket{0}$ and $\ket{1}$. Then, we randomly choose two opposite points on the circle to be the center of two classes, A and B. Subsequently, four training data instances are generated close to each class center in order to avoid overlaps between the test data and each other. This results in a good training dataset with the maximum margin such that soft-margin consideration is not needed. In addition, thirty test data instances are generated evenly along the great circle and are labelled as 1 or -1 according to the inner products with the class centers. In this case, we can set hyperparameter $C\to\infty$ and the process hence requires no regularization circuit evaluation. The test dataset is non-trivial to classify given that the test data are located mostly in the margin area; convex hulls of both training datasets do not include most of the test data.
    
    We choose a quantum feature map that embeds data $(\mathrm{x}_0, \mathrm{x}_1)$ into a Bloch sphere instead of $N=2$ qubits: $U_{\phi(\mathrm{x}_0, \mathrm{x}_1)}=R_z(\mathrm{x}_1)R_y(\mathrm{x}_0)$. Features $\mathrm{x}_0$ and $\mathrm{x}_1$ are the latitude and the longitude of the Bloch sphere. We use two qubits ($q_0$ and $q_1$) \textit{RealAmplitude}\cite{Qiskit} PQC as the ansatz: $V(\boldsymbol{\theta})=R^{q_0}_y(\theta_2)\otimes R^{q_1}_y(\theta_3)$ ${CNOT}_{q_0\to q_1}$ $R^{q_0}_y(\theta_0)\otimes R^{q_1}_y(\theta_1)$. In this experiment, we use \textit{ibmq\_montreal}, which is one of the IBM Quantum Falcon processors. The Methods section presents the specific techniques for optimizing quantum circuits against decoherence. The simultaneous perturbation stochastic approximation (SPSA) algorithm is selected to optimize $V(\boldsymbol{\theta})$ due to its rapid convergence and good robustness to noise.\cite{spall1997one, spall2000adaptive}
    The measurements of each circuits are repeated $R=8192$ times to estimate expectation values, which was the maximum possible option for \textit{ibmq\_montreal}. Due to long queue time of a cloud-based QPU, we reduce the QPU usage by applying warm-start and early-stopping techniques explained in the Method section.
    
    Fig.~\ref{fig:montreal} shows the classification result. Theoretical decision function values were calculated by solving Eq.~(\ref{eq:VQASVM_optim}) with convex optimization. A noisy simulation is a classical simulation that emulates an actual QPU based on a noise parameter set estimated from a noise measurement. Although the scale of the decision values is reduced, the impact on the classification accuracy is negligible given that only the signs of decision values matter. This logic has been applies to most NISQ-applicable quantum binary classifiers. The accuracy would improve with additional error mitigation and offset calibration processes on the quantum device. Other VQASVM demonstrations with different datasets can be found as Supplementary Fig. S2 online.

    \begin{figure}[t]
        \centering
        \includegraphics[width=\textwidth]{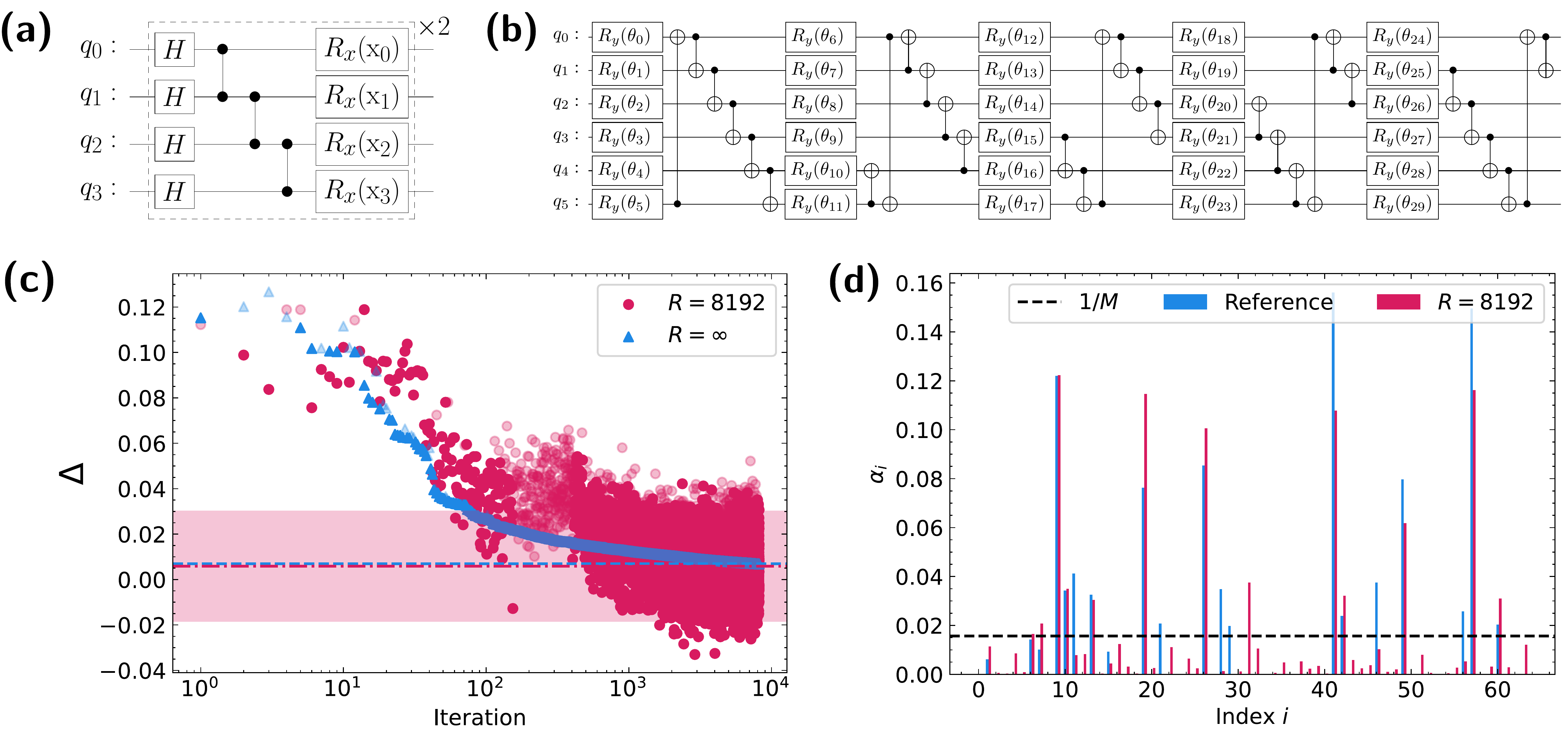}
        \caption{Numerical analysis on the iris dataset ($\lambda=C=10^4$). \textbf{\textsf{(a)}} Custom quantum feature map for the iris dataset. \textbf{\textsf{(b)}} PQC design for $V(\boldsymbol{\theta})$ with $5\times\log(M)=30$ gate parameters. \textbf{\textsf{(c)}} The shaded circles and triangles depict the training convergence outcomes of the residual losses. At the final iteration, ($t=2^{13}$) the residual loss $\Delta$ for $R=8192$ repeated measurements (red dot-dashed line) is almost equal to the case of $R=\infty$ (blue dashed line), where the error when estimating the expectation value is 0. The red shaded area represents the 95\% credible intervals of the last 16 residual losses for the $R=8192$ case.
        \textbf{\textsf{(d)}} The spectrum of optimized weights, $\alpha_i$s, of $\mathbf{x}_i$ for the theoretical and $R=8192$ cases are compared. The dashed black line indicates the level of the uniform weight $\alpha_i=1/M=1/64$. `Reference' in the legend refers to the theoretical values of the $\alpha_i$s obtained by convex optimization.}
        \label{fig:numanal_iris}
    \end{figure}
    
\subsection*{Numerical Simulation}\label{sec:iris_numanal}
    In a practical situation, the measurement on quantum circuits is repeated $R$ times to estimate the expectation value within $\epsilon=\order{1/\sqrt{R}}$ error, which could interfere with the convergence of VQA. However, the numerical analysis with the Iris dataset\cite{fisher1936iris} confirmed that VQASVM converges even with the noise in objective function estimation exists. The following paragraphs describe the details of the numerical simulation, such as data preprocessing and the choice of the quantum kernel and the ansatz.
    
    We assigned the labels +1 to Iris setosa and  -1 to Iris versicolour and Iris virginica for binary classification. The features of the data were scaled so that the range became $[-\pi, \pi]$. We sampled $M=64$ training data instances from the total dataset and treated the rest as the test data. The training kernel matrix constructed with our custom quantum feature map in Fig.~\ref{fig:numanal_iris}a is learnable; i.e., the singular values of the kernel matrix decay exponentially. After testing the PQC designs introduced in Ref.~\cite{sim2019expressibility}, we chose the PQC exhibited in Fig.~\ref{fig:numanal_iris}b as the ansatz for this simulation (see Supplementary Figs. S6-S9 online). The number of PQC parameters is 30, which is less than $M$. In this simulation, the SPSA optimizer was used for training due to its fast convergence and robustness to noise\cite{spall1997one, spall2000adaptive}. 
    
    The objective and decision functions were evaluated in two scenarios. The first case samples a finite number of measurement results $R=8192$ to estimate the expectation values such that the error of the estimation is non-zero. The second case directly calculates the expectation values with zero estimation error, which corresponds to sampling infinitely many measurement results; i.e., $R=\infty$. We defined the residual loss of training as $\Delta=\mathcal{L}_{\phi, \lambda}(\boldsymbol{\theta}^t;\mathcal{S})+C^{-1}\mathcal{R}(\boldsymbol{\theta}^t)-\Tilde{d}^\star$ at iteration $t$ to compare the convergence. Here, $\Tilde{d}^\star$ is the theoretical minimum of Eq.~(\ref{eq:VQASVM_optim}) as obtained by convex optimization. 
    
    Although containing some uncertainty, Fig.~\ref{fig:numanal_iris}c shows that SPSA converges to a local minimum despite the estimation noise. Both the $R=8192$ and $R=\infty$ cases show the critical convergence rule of SPSA; $\Delta\sim\order{\abs{\boldsymbol{\theta}}/t}$ for a sufficiently large number of iterations $t$. More vivid visualization can be found as Supplementary Fig. S10 online. In addition, the spectrum of the optimized Lagrange multipliers $\alpha_i$s mostly coincides with the theory, especially for the significant support vectors. (Fig.~\ref{fig:numanal_iris}d) Therefore, we concluded that training VQASVM within a finite number of measurements is achievable. The classification accuracy was 95.34\% for $R=8192$ and 94.19\% for $R=\infty$.
    

    The empirical speed-up of VQASVM is possible because the number of optimization parameters is exponentially reduced by the PQC encoding. However, in the limit of large number of qubits (i.e., $\log(M)$), it is unclear that such PQC can be trained with VQA to well-approximate the optimal solution. Therefore, we performed numerical analysis to empirically verify that VQASVM with $\order{\mathrm{polylog}(M)}$ parameters achieves bounded classification error even for large $M$. For this simulation, the MNIST dataset\cite{deng2012mnist} was used instead of the Iris dataset because there are not enough Iris data points to clarify the asymptotic behavior of VQASVM. In this setting, the number of maximum possible MNIST training data instances for VQASVM is $M=2^{13}$.
    
    A binary image data of `0's and `1's with a $28\times28$ image size were selected for binary classification, the features of which were then reduced to 10 by means of a principle component analysis (PCA). The well-known quantum feature map introduced in Ref.\cite{ibm-qsvm} was chosen for the simulation: $U_{\phi(\mathbf{x})}=W_{\phi(\mathbf{x})}H^{\otimes n}W_{\phi(\mathbf{x})}H^{\otimes n}$, where $W_{\phi(\mathbf{x})}=\exp{i\sum_{G\subset[n],\abs{G}=2}g_G(\mathbf{x})\Pi_{i\in G}\sigma_z^i}$ and $g_{\{i\}}(\mathbf{x})=x_i, g_{\{i,j\}}(\mathbf{x})=(\pi-x_i)(\pi-x_j)\delta_{i+1,j}$. The visualization of the feature map can be found as Supplementary Fig. S4 online. The ansatz architecture used for this simulation was the PQC template shown in Fig.\ref{fig:numanal_iris}b with 19 layers; i.e., the first part of the PQC in Fig.\ref{fig:numanal_iris}b is repeated 19 times. Thus, the number of optimization parameters is $19\times\log(M)$.
    
    The numerical simulation shows that although residual training loss $\Delta$ linearly increases with the number of PQC qubits, the rate is extremely low, i.e., $\Delta\sim0.00024\times\log(M)$. Moreover, we could not observe any critical difference in classification accuracy against the reference, i.e., theoretical accuracy obtained by convex optimization. Therefore, at least for $M\leq 2^{13}$, we conclude that run-time complexity of $\order{M\mathrm{polylog(M)}}$ with bounded classification accuracy is empirically achievable for VQASVM due to $\order{\mathrm{polylog}(M)}$ number of parameters (see Supplementary Fig. S11 online for a visual summarization of the result).

\section*{Discussion}\label{sec:discussion}
    In this work, we propose a novel quantum-classical hybrid supervised machine learning algorithm that achieves run-time complexity of $\order{M\mathrm{polylog}(M)}$, whereas the complexity of the modern classical algorithm is $\order{M^2}$.
    The main idea of our VQASVM algorithm is to encode optimization parameters that represent the normalized weight for each training data instance in a quantum state using exponentially fewer parameters.
    We numerically confirmed the convergence and feasibility of VQASVM even in the presence of expectation value estimation error using SPSA. We also observed the sub-quadratic asymptotic run-time complexity of VQASVM numerically. Finally, VQASVM was tested on cloud-based NISQ processors with a toy example dataset to highlight its practical application potential. 
    

    Based on the numerical results, we presume that our variational algorithm can bypass the issues evoked by the expressibility\cite{sim2019expressibility} and trainability relationship of PQCs; i.e., the highly expressive PQC is not likely to be trained with VQA due to the vanishing gradient variance,\cite{holmes2021connecting} a phenomenon known as the barren plateau.\cite{mcclean2018barren} 
    This problem has become a critical barrier for most VQAs utilizing a PQC to generate solution states. However, given that the SVM is learnable (i.e., the singular values of the kernel matrix decay exponentially\cite{qsvm}), only a few Lagrange multipliers corresponding to the significant support vectors are critically large; $\alpha_i\gg 1/M$.\cite{mml-book, boyd2004convex} For example, Figure \ref{fig:numanal_iris}d illustrates the statement. Thus, the PQCs encoding optimal multipliers should not necessarily be highly expressive. We speculate that there exists an ansatz generating these sparse probability distributions. Moreover, the optimal solution would have exponential degeneracy because only measurement probability matters instead of the amplitude of the state itself. Therefore, we could not observe a critical decrease in classification accuracy for these reasons even though the barren plateau exists, i.e., $\Delta=\order{\log(M)}$. More analytic discussion on the trainability of VQASVM and the ansatz generating sparse probability distribution should be investigated in the future.
    
    
    The VQASVM method manifests distinctive features compared to LS-QSVM, which solves the linear algebraic optimization problem of least-square SVM\cite{ls-svm} with the quantum algorithm for linear systems of equations (HHL)\cite{hhl}. Given that the fault-tolerant universal quantum computers and efficient quantum data loading with QRAM\cite{qram1,qram2} are possible, the run-time complexity of LS-QSVM is exponentially low: $\order{\kappa_{\mathrm{eff}}^3\epsilon^{-3}\log(MN)}$ with error $\epsilon$ and effective condition number $\kappa_{\mathrm{eff}}$. However, near-term implementation of LS-QSVM is infeasible due to lengthy quantum subroutines, which VQASVM has managed to avoid. Also, training LS-QSVM has to be repeated for each query of unseen data because the solution state collapses after the measurements at the end; transferring the solution state to classify multiple test data violates the no-cloning theorem. VQASVM can overcome these drawbacks. VQASVM is composed of much shorter operations; VQASVM circuits are much shallower than HHL circuits with the same moderate system size when decomposed in the same universal gate set. The classification phase of VQASVM can be separated from the training phase and performed simultaneously; training results are classically saved and transferred to a decision circuit in other quantum processing units (QPUs).
    
    We continue the discussion on the advantage of our method compared to other quantum kernel-based algorithms, such as a variational quantum classifier (VQC) and quantum kernel estimator (QKE), which are expected to be realized in the near-term NISQ devices\cite{ibm-qsvm}.
    VQC estimates the label of data $\mathbf{x}$ as $\hat{y}=\mathrm{sgn}\{\bar{f}(\mathbf{x};\boldsymbol{\theta})+b\}$, where $\bar{f}(\mathbf{x};\boldsymbol{\theta})$ is the empirical average of a binary function $f(z)$ such that $z$ is the $N$-bit computational basis measurement result of quantum circuit $W(\boldsymbol{\theta})\ket{\phi(\mathbf{x})}$. 
    The parameters $\boldsymbol{\theta}$ and $b$ are trained with variational methods that minimize the empirical risk: $R_{\mathrm{emp}}(\boldsymbol{\theta}, b)=\sum_{(\mathbf{x}, y)\in\mathcal{S}}\Pr\left[y\neq\mathrm{sgn}\{\bar{f}(\mathbf{x};\boldsymbol{\theta})+b\}\right]/{\abs{\mathcal{S}}}$. This requires $\order{MN\times\abs{\boldsymbol{\theta}}}$ quantum circuit measurements per iteration.\cite{schuld2018supervised} Subsequently, the complexity of VQC would match VQASVM from the heuristic point of view. However, VQC does not take advantage of the strong duality; the optimal state that $W(\boldsymbol{\theta^\star})^\dagger$ should generate has no characteristic, whereas the optimal distribution of $\alpha^\star_i$ that ansatz of VQASVM should generate is very sparse, i.e., most $\alpha_i$s are close to zero. Therefore, optimizing VQC in terms of the sufficiently large number of qubits would be vulnerable to issues such as local minima and barren plateaus. On the other hand,  QKE estimates the kernel matrix elements $\hat{K}_{ij}$ from the empirical probability of measuring a $N$-bit zero sequence on quantum circuit ${U_{\phi({\mathbf{x}_i})}}^\dagger{U_{\phi({\mathbf{x}_i})}}\ket{\mathbf{0}}$, given the kernel matrix ${K}_{ij}=\abs{\braket{\phi(\mathbf{x}_i)}{\phi(\mathbf{x}_j)}}^2$. The estimated kernel matrix is feed into the classical kernel-based algorithms, such as SVM. That is why we used QKE as the reference for the numerical analysis.
    The kernel matrix can be estimated in $\order{\epsilon^{-2}M^4}$ measurements with the bounded error of $||K-\hat{K}||\leq\epsilon$. Thus, QKE has much higher complexity than both VQASVM and classical SVM.\cite{ibm-qsvm}
    In addition, unlike QKE, the generalized error converges to zero as $M\to\infty$ due to the exponentially fewer parameters of VQASVM, strengthening the reliability of the training.\cite{caro2022generalization} The numerical linearity relation with the decision function error $\mathcal{E}_f$ and $\Delta$ supports the claim (see Supplementary Fig. S11 online). 

    VQASVM can be enhanced further with kernel optimization. Like other quantum-kernel-based methods, the choice of the quantum feature map is crucial for VQASVM. Unlike previous methods (e.g., quantum kernel alignment\cite{qka2021}), VQASVM can optimize a quantum feature map online during the training process. Given that $U_{\phi(\cdot)}$ is tuned with other parameters $\boldsymbol{\varphi}$, optimal parameters should be the saddle point: $(\boldsymbol{\theta}^\star, \boldsymbol{\varphi}^\star) = \mathop{\mathrm{arg\,min}}_{\boldsymbol{\theta}}\mathop{\mathrm{max}}_{\boldsymbol{\varphi}}\mathcal{L}_{\phi[\boldsymbol{\varphi}], \lambda}(\boldsymbol{\theta})+C^{-1}\mathcal{R}(\boldsymbol{\theta})$. 
    In addition, tailored quantum kernels (e.g., $k(\cdot,\cdot)=\abs{\braket{\phi(\cdot)}{\phi(\cdot)}}^{2r}$) can be adapted with the simple modification\cite{blank2020quantum} on the quantum circuits for VQASVM to improve classification accuracy. However, because the quantum advantage in classification accuracy derived from the power of quantum kernels is not the scope of this paper, we leave the remaining discussion for the future.
    Another method to improve VQASVM is boosting. Since VQASVM is not a convex problem, the performance may depend on the initial point and not be immune to the overfitting problem, like other kernel-based algorithms. A boosting method can be applied to improve classification accuracy by cascading low-performance VQASVMs. Because each VQASVM model only requires $\order{\log(M)+N}$ space, ensemble methods such as boosting are suitable for VQASVM.
    
\section*{Methods}

\subsection*{Proof of Eq. (\ref{eq:thrm1})}\label{method:proof}
    First, we note that a SWAP test operation $(H_a\cdot\mathrm{cSWAP}_{a\to b, c}\cdot H_a)$ in Fig.~\ref{fig:circuits} measures the Hilbert-Schmidt inner product between two pure states by estimating $\langle\sigma_z^a\rangle=\abs{\braket{\phi}{\psi}}^2$, where $a$ is the control qubit and $\ket{\phi}$ and $\ket{\psi_{\boldsymbol{\alpha}}}$ are states on target qubits $b$ and $c$, respectively. Quantum registers $i$ and $j$  in Fig.~\ref{fig:circuits} are traced out because measurements are performed on only $a$ and $y$ qubits. 
    The reduced density matrix on $x$ and $y$ quantum registers before the controlled-SWAP operation is $\rho_{x, y} = \sum_{i=0}^{M-1}\alpha_i \ket{\phi(\mathbf{x}_i)}\bra{\phi(\mathbf{x}_i)}\otimes\ket{y_i}\bra{y_i}$, which is the statistical sum of quantum states $\ket{\phi(\mathbf{x}_i)}\otimes\ket{y_i}$ with probability $\alpha_i$. 
    Let us first consider the decision circuit (Fig.~\ref{fig:circuits}b
    ). Given that the states $\ket{\phi(\mathbf{x}_i)}_{x}\otimes\ket{y_i}_{y}$ and $\ket{\phi(\hat{\mathbf{x}})}_{\hat{x}}$ are prepared, $\expval{\mathcal{A}_{a\to x,\hat{x}}\sigma_z^y}=\expval{\mathcal{A}_{a\to x,\hat{x}}}\expval{\sigma_z^y}=y_i\expval{\mathcal{A}_{a\to x, \hat{x}}}$ due to separability. Here, $\mathcal{A}_{a\to x,\hat{x}}$ can be $(H_a\cdot\mathrm{cSWAP}_{a\to x,\hat{x}}\cdot H_a)^\dagger \sigma_z^a (H_a\cdot\mathrm{cSWAP}_{a\to x,\hat{x}}\cdot H_a)$ or $\frac{1}{\lambda}\sigma^0_a$. Similarly, for the loss circuit (Fig.~\ref{fig:circuits}a 
    ), we have states $\ket{\phi(\mathbf{x}_i)}_{x_0}\otimes\ket{y_i}_{y_0}$ and $\ket{\phi(\mathbf{x}_j)}_{x_1}\otimes\ket{y_j}_{y_1}$ with probability $\alpha_i\alpha_j$ such that $\expval{\mathcal{A}_{a\to x_0,x_1}\sigma_z^{y_0}\sigma_z^{y_1}}=\expval{\mathcal{A}_{a\to x_0,x_1}}\expval{\sigma_z^{y_0}\sigma_z^{y_1}}=y_iy_j\expval{\mathcal{A}_{a\to x_0, x_1}}$. Therefore, from the definition of our quantum kernel, each term in Eq. (\ref{eq:thrm1}) matches the loss and decision functions in Eqs.~(\ref{eq:loss_function}) and (\ref{eq:decision_function}). More direct proof is provided in Ref.~\citenum{blank2020quantum, park2020theory} and in the Supplementary Information section B (online). 
    
\subsection*{Realization of Quantum Circuits}\label{method:ucsq}
    In this article, $\mathcal{U}_{\phi, \mathcal{S}}$ is realized using uniformly controlled one-qubit gates,  which require at most $M-1$ CNOT gates, $M$ one-qubit gates, and a single diagonal $(\log(M) + 1)$-qubit gate\cite{qiskit_uc, mottonen2004transformation}. 
    We compiled the quantum feature map with a basis gate set composed of Pauli rotations and CNOT. $\mathcal{U}_{\phi, \mathcal{S}}$ can be efficiently implemented by replacing all Pauli rotations with uniformly controlled Pauli rotations. The training data label embedding of $\mathcal{U}_{\phi, \mathcal{S}}$ can also be easily implemented using a uniformly controlled Pauli X rotation (i.e., setting the rotation angle to $\pi$ if the label is positive and $0$ otherwise). Although this procedure allows one to incorporate existing quantum feature maps, the complexity can increase to $\order{MN^2}$ if the quantum feature map contains all-to-all connecting parameterized two-qubit gates. Nonetheless, such a value of $\mathcal{U}_{\phi, \mathcal{S}}$ has linear complexity proportional to the number of training data instances.
    
\subsection*{Application to IBM Quantum Processors}\label{method:nisq}
    Because IBM quantum processors are based on superconducting qubits, all-to-all connections are not possible. Additional SWAP operations among qubits for distant interactions would shorten the effective decoherence time and increase the noise. We carefully selected the physical qubits of \textit{ibmq\_montreal} in order to reduce the number of SWAP operations. For $M=4$ and single qubit embedding, $m=2$ and $n=1$. Thus, multi-qubit interaction is required for the following connections: $(a, x_0, x_1)$,  $([i_0, i_1], x_0)$, $([j_0, j_1], x_1)$, $([i_0, i_1], y_0)$, and $([j_0, j_1], y_1)$. 
    We initially selected nine qubits connected in a linear topology such that the overall estimated single and two-qubit gate errors are lowest among all other possible options. The noise parameters and topology of \textit{ibmq\_montreal} are provided by IBM Quantum.
    For instance, the physical qubits indexed as 1, 2, 3, 4, 5, 8, 11, 14, and 16 in \textit{ibmq\_montreal} were selected in this article (see Supplementary Fig. S1 online) We then assign a virtual qubit in the order of $y_0, i_0, i_1, x_0, a, x_1, j_0, j_1, y_1$ so that the aforementioned required connections can be made between qubits next to each other. 
    In conclusion, mapping from virtual qubits to physical qubits proceeds as $\{a\mapsto5, i_0\mapsto2, i_1\mapsto1, x_i\mapsto3, y_i\mapsto4, j_0\mapsto11, j_1\mapsto14, x_j\mapsto8, y_j\mapsto16\}$ in this experiment. We report that with this arrangement, the circuit depths of loss and decision circuits are correspondingly 60 and 59 for a balanced dataset and 64 and 63 for the an unbalanced dataset in the basis gate set of \textit{ibmq\_montreal}: $\left\{R_z, \sqrt{X}, X, CNOT\right\}$.
    
\subsection*{Additional Techniques on SPSA}\label{method:spsa}
    The conventional SPSA algorithm has been adjusted for faster and better convergence. 
    First, the \textit{blocking} technique was introduced. Assuming that the variance $\sigma^2$ of objective function $\mathcal{L}_{\phi, \lambda}+C^{-1}\mathcal{R}$ is uniform on parameter $\boldsymbol{\theta}$, the next iteration $t+1$ is rejected if $\left[\mathcal{L}_{\phi, \lambda}+C^{-1}\mathcal{R}\right](\boldsymbol{\theta}^{t+1})\geq\left[\mathcal{L}_{\phi, \lambda}+C^{-1}\mathcal{R}\right](\boldsymbol{\theta}^{t})+2\sigma$. SPSA would converge more rapidly with blocking by preventing its objective function from becoming too large with some probability (see Supplementary Fig. S10). 
    Second, \textit{Early-stopping} is applied. Iterations are terminated if certain conditions are satisfied. Specifically, we stop SPSA optimization if the average of last 16 recorded training loss values is greater than or equal to the last 32 recorded values. Early stopping reduces the training time drastically, especially when running on a QPU. 
    Last, we averaged the last 16 recorded parameters to yield the result $\boldsymbol{\theta}^{\star}=\frac{1}{16}\sum_{i=0}^{15}\boldsymbol{\theta}^{t-i}$. Combinations of these techniques were selected for better optimization. We adopted all these techniques for the experiments and simulations as the default condition.
    
\subsection*{Warm-start optimization}\label{method:warm-start}
    We report cases in which the optimization of IBM Q Quantum Processors yields vanishing kernel amplitudes due to the constantly varying error map problem. The total run time should be minimized to avoid this problem. Because accessing a QPU takes a relatively long queue time, we apply a `warm-start' technique, which reduces number of QPU uses. First, we initialize and proceed a few iterations (32) with a noisy simulation on a CPU and then evaluate the functions on a QPU for the remaining iterations. Note that an SPSA optimizer requires heavy initialization computation, such as when the initial variance is calculated. With this warm-start method, we are able to obtain better results on some trials.

\bibliographystyle{naturemag}
\bibliography{main}

\begin{thebibliography}{10}
\expandafter\ifx\csname url\endcsname\relax
  \def\url#1{\texttt{#1}}\fi
\expandafter\ifx\csname urlprefix\endcsname\relax\def\urlprefix{URL }\fi
\providecommand{\bibinfo}[2]{#2}
\providecommand{\eprint}[2][]{\url{#2}}

\bibitem{qnflprl22}
\bibinfo{author}{Sharma, K.} \emph{et~al.}
\newblock \bibinfo{title}{{Reformulation of the No-Free-Lunch Theorem for
  Entangled Datasets}}.
\newblock \emph{\bibinfo{journal}{Physical Review Letters}}
  \textbf{\bibinfo{volume}{128}}, \bibinfo{pages}{070501}
  (\bibinfo{year}{2022}).
\newblock
  \urlprefix\url{https://link.aps.org/doi/10.1103/PhysRevLett.128.070501}.

\bibitem{qram2}
\bibinfo{author}{Lloyd, S.}, \bibinfo{author}{Mohseni, M.} \&
  \bibinfo{author}{Rebentrost, P.}
\newblock \bibinfo{title}{{Quantum algorithms for supervised and unsupervised
  machine learning}}.
\newblock \emph{\bibinfo{journal}{arXiv preprint}}  (\bibinfo{year}{2013}).
\newblock \urlprefix\url{http://arxiv.org/abs/1307.0411}.

\bibitem{qml}
\bibinfo{author}{Biamonte, J.} \emph{et~al.}
\newblock \bibinfo{title}{{Quantum machine learning}}.
\newblock \emph{\bibinfo{journal}{Nature}} \textbf{\bibinfo{volume}{549}},
  \bibinfo{pages}{195--202} (\bibinfo{year}{2017}).
\newblock \urlprefix\url{http://www.nature.com/articles/nature23474}.

\bibitem{schuld2018supervised}
\bibinfo{author}{Schuld, M.} \& \bibinfo{author}{Petruccione, F.}
\newblock \emph{\bibinfo{title}{Supervised learning with quantum computers}},
  vol.~\bibinfo{volume}{17} (\bibinfo{publisher}{Springer},
  \bibinfo{year}{2018}).

\bibitem{qsvm}
\bibinfo{author}{Rebentrost, P.}, \bibinfo{author}{Mohseni, M.} \&
  \bibinfo{author}{Lloyd, S.}
\newblock \bibinfo{title}{{Quantum Support Vector Machine for Big Data
  Classification}}.
\newblock \emph{\bibinfo{journal}{Physical Review Letters}}
  \textbf{\bibinfo{volume}{113}}, \bibinfo{pages}{130503}
  (\bibinfo{year}{2014}).
\newblock
  \urlprefix\url{https://link.aps.org/doi/10.1103/PhysRevLett.113.130503}.

\bibitem{schuld2017implementing}
\bibinfo{author}{Schuld, M.}, \bibinfo{author}{Fingerhuth, M.} \&
  \bibinfo{author}{Petruccione, F.}
\newblock \bibinfo{title}{{Implementing a distance-based classifier with a
  quantum interference circuit}}.
\newblock \emph{\bibinfo{journal}{EPL (Europhysics Letters)}}
  \textbf{\bibinfo{volume}{119}}, \bibinfo{pages}{60002}
  (\bibinfo{year}{2017}).
\newblock
  \urlprefix\url{https://iopscience.iop.org/article/10.1209/0295-5075/119/60002}.

\bibitem{schuld2019quantum}
\bibinfo{author}{Schuld, M.} \& \bibinfo{author}{Killoran, N.}
\newblock \bibinfo{title}{{Quantum Machine Learning in Feature Hilbert
  Spaces}}.
\newblock \emph{\bibinfo{journal}{Physical Review Letters}}
  \textbf{\bibinfo{volume}{122}}, \bibinfo{pages}{040504}
  (\bibinfo{year}{2019}).
\newblock
  \urlprefix\url{https://link.aps.org/doi/10.1103/PhysRevLett.122.040504}.

\bibitem{ibm-qsvm}
\bibinfo{author}{Havl{\'{i}}{\v{c}}ek, V.} \emph{et~al.}
\newblock \bibinfo{title}{{Supervised learning with quantum-enhanced feature
  spaces}}.
\newblock \emph{\bibinfo{journal}{Nature}} \textbf{\bibinfo{volume}{567}},
  \bibinfo{pages}{209--212} (\bibinfo{year}{2019}).
\newblock \urlprefix\url{http://www.nature.com/articles/s41586-019-0980-2}.

\bibitem{lloyd2020quantum}
\bibinfo{author}{Lloyd, S.}, \bibinfo{author}{Schuld, M.},
  \bibinfo{author}{Ijaz, A.}, \bibinfo{author}{Izaac, J.} \&
  \bibinfo{author}{Killoran, N.}
\newblock \bibinfo{title}{{Quantum embeddings for machine learning}}.
\newblock \emph{\bibinfo{journal}{arXiv preprint}}  (\bibinfo{year}{2020}).
\newblock \urlprefix\url{http://arxiv.org/abs/2001.03622}.

\bibitem{blank2020quantum}
\bibinfo{author}{Blank, C.}, \bibinfo{author}{Park, D.~K.},
  \bibinfo{author}{Rhee, J.-K.~K.} \& \bibinfo{author}{Petruccione, F.}
\newblock \bibinfo{title}{{Quantum classifier with tailored quantum kernel}}.
\newblock \emph{\bibinfo{journal}{npj Quantum Information}}
  \textbf{\bibinfo{volume}{6}}, \bibinfo{pages}{41} (\bibinfo{year}{2020}).
\newblock \urlprefix\url{http://www.nature.com/articles/s41534-020-0272-6}.

\bibitem{park2020theory}
\bibinfo{author}{Park, D.~K.}, \bibinfo{author}{Blank, C.} \&
  \bibinfo{author}{Petruccione, F.}
\newblock \bibinfo{title}{{The theory of the quantum kernel-based binary
  classifier}}.
\newblock \emph{\bibinfo{journal}{Physics Letters A}}
  \textbf{\bibinfo{volume}{384}}, \bibinfo{pages}{126422}
  (\bibinfo{year}{2020}).
\newblock
  \urlprefix\url{https://linkinghub.elsevier.com/retrieve/pii/S0375960120302541}.

\bibitem{schuld2021supervised}
\bibinfo{author}{Schuld, M.}
\newblock \bibinfo{title}{{Supervised quantum machine learning models are
  kernel methods}}.
\newblock \emph{\bibinfo{journal}{arXiv preprint}}  (\bibinfo{year}{2021}).
\newblock \urlprefix\url{http://arxiv.org/abs/2101.11020}.

\bibitem{liu2021rigorous}
\bibinfo{author}{Liu, Y.}, \bibinfo{author}{Arunachalam, S.} \&
  \bibinfo{author}{Temme, K.}
\newblock \bibinfo{title}{{A rigorous and robust quantum speed-up in supervised
  machine learning}}.
\newblock \emph{\bibinfo{journal}{Nature Physics}}
  \textbf{\bibinfo{volume}{17}}, \bibinfo{pages}{1013--1017}
  (\bibinfo{year}{2021}).
\newblock \urlprefix\url{https://www.nature.com/articles/s41567-021-01287-z}.

\bibitem{blank2022compact}
\bibinfo{author}{Blank, C.}, \bibinfo{author}{da~Silva, A.~J.},
  \bibinfo{author}{de~Albuquerque, L.~P.}, \bibinfo{author}{Petruccione, F.} \&
  \bibinfo{author}{Park, D.~K.}
\newblock \bibinfo{title}{Compact quantum kernel-based binary classifier}.
\newblock \emph{\bibinfo{journal}{Quantum Science and Technology}}
  \textbf{\bibinfo{volume}{7}}, \bibinfo{pages}{045007} (\bibinfo{year}{2022}).

\bibitem{svm}
\bibinfo{author}{Cortes, C.} \& \bibinfo{author}{Vapnik, V.}
\newblock \bibinfo{title}{{Support-vector networks}}.
\newblock \emph{\bibinfo{journal}{Machine Learning}}
  \textbf{\bibinfo{volume}{20}}, \bibinfo{pages}{273--297}
  (\bibinfo{year}{1995}).
\newblock \urlprefix\url{http://link.springer.com/10.1007/BF00994018}.

\bibitem{boser1992training}
\bibinfo{author}{Boser, B.~E.}, \bibinfo{author}{Guyon, I.~M.} \&
  \bibinfo{author}{Vapnik, V.~N.}
\newblock \bibinfo{title}{{A training algorithm for optimal margin
  classifiers}}.
\newblock In \emph{\bibinfo{booktitle}{Proceedings of the fifth annual workshop
  on Computational learning theory}}, \bibinfo{pages}{144--152}
  (\bibinfo{year}{1992}).

\bibitem{vapnik1997support}
\bibinfo{author}{Vapnik, V.}, \bibinfo{author}{Golowich, S.~E.} \&
  \bibinfo{author}{Smola, A.}
\newblock \bibinfo{title}{{Support vector method for function approximation,
  regression estimation, and signal processing}}.
\newblock In \emph{\bibinfo{booktitle}{Advances in Neural Information
  Processing Systems}}, \bibinfo{pages}{281--287} (\bibinfo{year}{1997}).

\bibitem{guyon1992capacity}
\bibinfo{author}{Guyon, I.}, \bibinfo{author}{Vapnik, V.},
  \bibinfo{author}{Boser, B.}, \bibinfo{author}{Bottou, L.} \&
  \bibinfo{author}{Solla, S.}
\newblock \bibinfo{title}{{Capacity control in linear classifiers for pattern
  recognition}}.
\newblock In \emph{\bibinfo{booktitle}{Proceedings., 11th IAPR International
  Conference on Pattern Recognition. Vol.II. Conference B: Pattern Recognition
  Methodology and Systems}}, \bibinfo{pages}{385--388}
  (\bibinfo{publisher}{IEEE Comput. Soc. Press}, \bibinfo{year}{1992}).
\newblock \urlprefix\url{http://ieeexplore.ieee.org/document/201798/}.

\bibitem{qpca}
\bibinfo{author}{Lloyd, S.}, \bibinfo{author}{Mohseni, M.} \&
  \bibinfo{author}{Rebentrost, P.}
\newblock \bibinfo{title}{{Quantum principal component analysis}}.
\newblock \emph{\bibinfo{journal}{Nature Physics}}
  \textbf{\bibinfo{volume}{10}}, \bibinfo{pages}{631--633}
  (\bibinfo{year}{2014}).
\newblock \urlprefix\url{http://www.nature.com/articles/nphys3029}.

\bibitem{hhl}
\bibinfo{author}{Harrow, A.~W.}, \bibinfo{author}{Hassidim, A.} \&
  \bibinfo{author}{Lloyd, S.}
\newblock \bibinfo{title}{{Quantum Algorithm for Linear Systems of Equations}}.
\newblock \emph{\bibinfo{journal}{Physical Review Letters}}
  \textbf{\bibinfo{volume}{103}}, \bibinfo{pages}{150502}
  (\bibinfo{year}{2009}).
\newblock
  \urlprefix\url{https://link.aps.org/doi/10.1103/PhysRevLett.103.150502}.

\bibitem{aaronson2015read}
\bibinfo{author}{Aaronson, S.}
\newblock \bibinfo{title}{{Read the fine print}}.
\newblock \emph{\bibinfo{journal}{Nature Physics}}
  \textbf{\bibinfo{volume}{11}}, \bibinfo{pages}{291--293}
  (\bibinfo{year}{2015}).
\newblock \urlprefix\url{http://www.nature.com/articles/nphys3272}.

\bibitem{qram1}
\bibinfo{author}{Giovannetti, V.}, \bibinfo{author}{Lloyd, S.} \&
  \bibinfo{author}{Maccone, L.}
\newblock \bibinfo{title}{{Quantum Random Access Memory}}.
\newblock \emph{\bibinfo{journal}{Physical Review Letters}}
  \textbf{\bibinfo{volume}{100}}, \bibinfo{pages}{160501}
  (\bibinfo{year}{2008}).
\newblock
  \urlprefix\url{https://link.aps.org/doi/10.1103/PhysRevLett.100.160501}.

\bibitem{nisq}
\bibinfo{author}{Preskill, J.}
\newblock \bibinfo{title}{{Quantum Computing in the NISQ era and beyond}}.
\newblock \emph{\bibinfo{journal}{Quantum}} \textbf{\bibinfo{volume}{2}},
  \bibinfo{pages}{79} (\bibinfo{year}{2018}).
\newblock \urlprefix\url{https://quantum-journal.org/papers/q-2018-08-06-79/}.

\bibitem{pqc}
\bibinfo{author}{Benedetti, M.}, \bibinfo{author}{Lloyd, E.},
  \bibinfo{author}{Sack, S.} \& \bibinfo{author}{Fiorentini, M.}
\newblock \bibinfo{title}{Parameterized quantum circuits as machine learning
  models}.
\newblock \emph{\bibinfo{journal}{Quantum Science and Technology}}
  \textbf{\bibinfo{volume}{4}}, \bibinfo{pages}{043001} (\bibinfo{year}{2019}).

\bibitem{vqa}
\bibinfo{author}{Cerezo, M.} \emph{et~al.}
\newblock \bibinfo{title}{{Variational quantum algorithms}}.
\newblock \emph{\bibinfo{journal}{Nature Reviews Physics}}
  \textbf{\bibinfo{volume}{3}}, \bibinfo{pages}{625--644}
  (\bibinfo{year}{2021}).
\newblock \urlprefix\url{https://www.nature.com/articles/s42254-021-00348-9}.

\bibitem{chang2011libsvm}
\bibinfo{author}{Chang, C.-C.} \& \bibinfo{author}{Lin, C.-J.}
\newblock \bibinfo{title}{{LIBSVM}}.
\newblock \emph{\bibinfo{journal}{ACM Transactions on Intelligent Systems and
  Technology}} \textbf{\bibinfo{volume}{2}}, \bibinfo{pages}{1--27}
  (\bibinfo{year}{2011}).
\newblock \urlprefix\url{https://dl.acm.org/doi/10.1145/1961189.1961199}.

\bibitem{Coppersmith1990MatrixProgressions}
\bibinfo{author}{Coppersmith, D.} \& \bibinfo{author}{Winograd, S.}
\newblock \bibinfo{title}{{Matrix multiplication via arithmetic progressions}}.
\newblock \emph{\bibinfo{journal}{Journal of Symbolic Computation}}
  \textbf{\bibinfo{volume}{9}}, \bibinfo{pages}{251--280}
  (\bibinfo{year}{1990}).
\newblock
  \urlprefix\url{https://linkinghub.elsevier.com/retrieve/pii/S0747717108800132}.

\bibitem{giuntini2021quantum}
\bibinfo{author}{Giuntini, R.} \emph{et~al.}
\newblock \bibinfo{title}{{Quantum State Discrimination for Supervised
  Classification}}.
\newblock \emph{\bibinfo{journal}{arXiv preprint}}  (\bibinfo{year}{2021}).
\newblock \urlprefix\url{http://arxiv.org/abs/2104.00971}.

\bibitem{boyd2004convex}
\bibinfo{author}{Boyd, S.~P.} \& \bibinfo{author}{Vandenberghe, L.}
\newblock \emph{\bibinfo{title}{{Convex optimization}}}
  (\bibinfo{publisher}{Cambridge university press}, \bibinfo{year}{2004}).

\bibitem{mml-book}
\bibinfo{author}{Deisenroth, M.~P.}, \bibinfo{author}{Faisal, A.~A.} \&
  \bibinfo{author}{Ong, C.~S.}
\newblock \emph{\bibinfo{title}{{Mathematics for Machine Learning}}}
  (\bibinfo{publisher}{Cambridge University Press}, \bibinfo{year}{2020}).
\newblock
  \urlprefix\url{https://www.cambridge.org/highereducation/books/mathematics-for-machine-learning/5EE57FD1CFB23E6EB11E130309C7EF98#contents}.

\bibitem{mangasarian1999successive}
\bibinfo{author}{Mangasarian, O.} \& \bibinfo{author}{Musicant, D.}
\newblock \bibinfo{title}{{Successive overrelaxation for support vector
  machines}}.
\newblock \emph{\bibinfo{journal}{IEEE Transactions on Neural Networks}}
  \textbf{\bibinfo{volume}{10}}, \bibinfo{pages}{1032--1037}
  (\bibinfo{year}{1999}).
\newblock \urlprefix\url{http://ieeexplore.ieee.org/document/788643/}.

\bibitem{frie1998kernel}
\bibinfo{author}{Frie, T.-T.}, \bibinfo{author}{Cristianini, N.} \&
  \bibinfo{author}{Campbell, C.}
\newblock \bibinfo{title}{The kernel-adatron algorithm: a fast and simple
  learning procedure for support vector machines}.
\newblock In \emph{\bibinfo{booktitle}{Machine learning: proceedings of the
  fifteenth international conference (ICML'98)}}, \bibinfo{pages}{188--196}
  (\bibinfo{year}{1998}).

\bibitem{hsu2002simple}
\bibinfo{author}{Hsu, C.-W.} \& \bibinfo{author}{Lin, C.-J.}
\newblock \bibinfo{title}{{A Simple Decomposition Method for Support Vector
  Machines}}.
\newblock \emph{\bibinfo{journal}{Machine Learning}}
  \textbf{\bibinfo{volume}{46}}, \bibinfo{pages}{291--314}
  (\bibinfo{year}{2002}).
\newblock \urlprefix\url{https://doi.org/10.1023/A:1012427100071}.

\bibitem{hsu2002comparison}
\bibinfo{author}{{Chih-Wei Hsu}} \& \bibinfo{author}{{Chih-Jen Lin}}.
\newblock \bibinfo{title}{{A comparison of methods for multiclass support
  vector machines}}.
\newblock \emph{\bibinfo{journal}{IEEE Transactions on Neural Networks}}
  \textbf{\bibinfo{volume}{13}}, \bibinfo{pages}{415--425}
  (\bibinfo{year}{2002}).
\newblock \urlprefix\url{http://ieeexplore.ieee.org/document/991427/}.

\bibitem{mottonen2004transformation}
\bibinfo{author}{Mottonen, M.}, \bibinfo{author}{Vartiainen, J.~J.},
  \bibinfo{author}{Bergholm, V.} \& \bibinfo{author}{Salomaa, M.~M.}
\newblock \bibinfo{title}{{Transformation of quantum states using uniformly
  controlled rotations}}.
\newblock \emph{\bibinfo{journal}{arXiv preprint}}  (\bibinfo{year}{2004}).
\newblock \urlprefix\url{http://arxiv.org/abs/quant-ph/0407010}.

\bibitem{sim2019expressibility}
\bibinfo{author}{Sim, S.}, \bibinfo{author}{Johnson, P.~D.} \&
  \bibinfo{author}{Aspuru‐Guzik, A.}
\newblock \bibinfo{title}{{Expressibility and Entangling Capability of
  Parameterized Quantum Circuits for Hybrid Quantum‐Classical Algorithms}}.
\newblock \emph{\bibinfo{journal}{Advanced Quantum Technologies}}
  \textbf{\bibinfo{volume}{2}}, \bibinfo{pages}{1900070}
  (\bibinfo{year}{2019}).
\newblock
  \urlprefix\url{https://onlinelibrary.wiley.com/doi/10.1002/qute.201900070}.

\bibitem{kandala2017hardware}
\bibinfo{author}{Kandala, A.} \emph{et~al.}
\newblock \bibinfo{title}{{Hardware-efficient variational quantum eigensolver
  for small molecules and quantum magnets}}.
\newblock \emph{\bibinfo{journal}{Nature}} \textbf{\bibinfo{volume}{549}},
  \bibinfo{pages}{242--246} (\bibinfo{year}{2017}).
\newblock \urlprefix\url{https://doi.org/10.1038/nature23879}.

\bibitem{param-shift1}
\bibinfo{author}{Mitarai, K.}, \bibinfo{author}{Negoro, M.},
  \bibinfo{author}{Kitagawa, M.} \& \bibinfo{author}{Fujii, K.}
\newblock \bibinfo{title}{{Quantum circuit learning}}.
\newblock \emph{\bibinfo{journal}{Physical Review A}}
  \textbf{\bibinfo{volume}{98}}, \bibinfo{pages}{032309}
  (\bibinfo{year}{2018}).
\newblock \urlprefix\url{https://link.aps.org/doi/10.1103/PhysRevA.98.032309}.

\bibitem{param-shift2}
\bibinfo{author}{Schuld, M.}, \bibinfo{author}{Bergholm, V.},
  \bibinfo{author}{Gogolin, C.}, \bibinfo{author}{Izaac, J.} \&
  \bibinfo{author}{Killoran, N.}
\newblock \bibinfo{title}{{Evaluating analytic gradients on quantum hardware}}.
\newblock \emph{\bibinfo{journal}{Physical Review A}}
  \textbf{\bibinfo{volume}{99}}, \bibinfo{pages}{032331}
  (\bibinfo{year}{2019}).
\newblock \urlprefix\url{https://link.aps.org/doi/10.1103/PhysRevA.99.032331}.

\bibitem{Qiskit}
\bibinfo{author}{ANIS, M. D.~S.} \emph{et~al.}
\newblock \bibinfo{title}{{Qiskit: An Open-source Framework for Quantum
  Computing}} (\bibinfo{year}{2021}).

\bibitem{spall1997one}
\bibinfo{author}{Spall, J.~C.}
\newblock \bibinfo{title}{{A one-measurement form of simultaneous perturbation
  stochastic approximation}}.
\newblock \emph{\bibinfo{journal}{Automatica}} \textbf{\bibinfo{volume}{33}},
  \bibinfo{pages}{109--112} (\bibinfo{year}{1997}).

\bibitem{spall2000adaptive}
\bibinfo{author}{Spall, J.}
\newblock \bibinfo{title}{{Adaptive stochastic approximation by the
  simultaneous perturbation method}}.
\newblock \emph{\bibinfo{journal}{IEEE Transactions on Automatic Control}}
  \textbf{\bibinfo{volume}{45}}, \bibinfo{pages}{1839--1853}
  (\bibinfo{year}{2000}).
\newblock \urlprefix\url{http://ieeexplore.ieee.org/document/880982/}.

\bibitem{fisher1936iris}
\bibinfo{author}{Fisher, R.~A.} \& \bibinfo{author}{Marshall, M.}
\newblock \bibinfo{title}{{Iris data set}} (\bibinfo{year}{1936}).

\bibitem{deng2012mnist}
\bibinfo{author}{Deng, L.}
\newblock \bibinfo{title}{The mnist database of handwritten digit images for
  machine learning research}.
\newblock \emph{\bibinfo{journal}{IEEE Signal Processing Magazine}}
  \textbf{\bibinfo{volume}{29}}, \bibinfo{pages}{141--142}
  (\bibinfo{year}{2012}).

\bibitem{holmes2021connecting}
\bibinfo{author}{Holmes, Z.}, \bibinfo{author}{Sharma, K.},
  \bibinfo{author}{Cerezo, M.} \& \bibinfo{author}{Coles, P.~J.}
\newblock \bibinfo{title}{{Connecting Ansatz Expressibility to Gradient
  Magnitudes and Barren Plateaus}}.
\newblock \emph{\bibinfo{journal}{PRX Quantum}} \textbf{\bibinfo{volume}{3}},
  \bibinfo{pages}{010313} (\bibinfo{year}{2022}).
\newblock \urlprefix\url{https://link.aps.org/doi/10.1103/PRXQuantum.3.010313}.

\bibitem{mcclean2018barren}
\bibinfo{author}{McClean, J.~R.}, \bibinfo{author}{Boixo, S.},
  \bibinfo{author}{Smelyanskiy, V.~N.}, \bibinfo{author}{Babbush, R.} \&
  \bibinfo{author}{Neven, H.}
\newblock \bibinfo{title}{{Barren plateaus in quantum neural network training
  landscapes}}.
\newblock \emph{\bibinfo{journal}{Nature Communications}}
  \textbf{\bibinfo{volume}{9}}, \bibinfo{pages}{4812} (\bibinfo{year}{2018}).
\newblock \urlprefix\url{http://www.nature.com/articles/s41467-018-07090-4}.

\bibitem{ls-svm}
\bibinfo{author}{Suykens, J. A.~K.} \& \bibinfo{author}{Vandewalle, J.}
\newblock \bibinfo{title}{{Least Squares Support Vector Machine Classifiers}}.
\newblock \emph{\bibinfo{journal}{Neural Processing Letters}}
  \textbf{\bibinfo{volume}{9}}, \bibinfo{pages}{293--300}
  (\bibinfo{year}{1999}).
\newblock \urlprefix\url{https://doi.org/10.1023/A:1018628609742}.

\bibitem{caro2022generalization}
\bibinfo{author}{Caro, M.~C.} \emph{et~al.}
\newblock \bibinfo{title}{Generalization in quantum machine learning from few
  training data}.
\newblock \emph{\bibinfo{journal}{Nature communications}}
  \textbf{\bibinfo{volume}{13}}, \bibinfo{pages}{1--11} (\bibinfo{year}{2022}).

\bibitem{qka2021}
\bibinfo{author}{Glick, J.~R.} \emph{et~al.}
\newblock \bibinfo{title}{{Covariant quantum kernels for data with group
  structure}}.
\newblock \emph{\bibinfo{journal}{arXiv preprint}}  (\bibinfo{year}{2021}).
\newblock \urlprefix\url{https://arxiv.org/abs/2105.03406}.

\bibitem{qiskit_uc}
\bibinfo{author}{Bergholm, V.}, \bibinfo{author}{Vartiainen, J.~J.},
  \bibinfo{author}{M{\"{o}}tt{\"{o}}nen, M.} \& \bibinfo{author}{Salomaa,
  M.~M.}
\newblock \bibinfo{title}{{Quantum circuits with uniformly controlled one-qubit
  gates}}.
\newblock \emph{\bibinfo{journal}{Physical Review A}}
  \textbf{\bibinfo{volume}{71}}, \bibinfo{pages}{052330}
  (\bibinfo{year}{2005}).
\newblock \urlprefix\url{https://link.aps.org/doi/10.1103/PhysRevA.71.052330}.

\bibitem{steinwart2008support}
\bibinfo{author}{Steinwart, I.} \& \bibinfo{author}{Christmann, A.}
\newblock \emph{\bibinfo{title}{{Support vector machines}}}
  (\bibinfo{publisher}{Springer Science {\&} Business Media},
  \bibinfo{year}{2008}).

\end{thebibliography}

\section*{Acknowledgment}
    This work was supported by the Samsung Research Funding \& Incubation Center of Samsung Electronics under Project Number SRFC-TF2003-01. 
    We acknowledge the use of IBM Quantum services for this work. The views expressed are those of the authors and do not reflect the official policy or position of IBM or the IBM Quantum team.

\section*{Author contributions}
    S.P. contributed to the development and experimental verification of the theoretical and circuit models; D.K.P. contributed to developing the initial concept and the experimental verification; J.K.R. contributed to the development and validation of the main concept and organization of this work. All co-authors contributed to the writing of the manuscript.
    
\section*{Data Availability}
    The numerical data generated in this work are available from the corresponding author upon reasonable request. \url{https://github.com/Siheon-Park/QUIC-Projects}

\section*{Additional information}
    The authors declare no competing interests. 

\clearpage

\setcounter{page}{1}
\renewcommand{\thepage}{S\arabic{page}}
\setcounter{section}{0}
\renewcommand{\thesection}{S\arabic{section}}
\setcounter{table}{0}
\renewcommand{\thetable}{S\arabic{table}}  
\setcounter{figure}{0}
\renewcommand{\thefigure}{S\arabic{figure}}
\setcounter{equation}{0}
\renewcommand{\theequation}{S\arabic{equation}}






%
%

\appendix 

\section*{Supplementary Information}
\section{Review of quantum support vector machine}
    Classification is a fundamental problem in machine learning. The goal of $L$-class classification is to infer the most likely class label of an unseen data point $\hat{\mathbf{x}} \in \mathbb{C}^N$ can be described as, given a labelled data set $\mathcal{S} = \left\{ \left(\mathbf{x}_i, y_i\right) \right\}_{i=0}^{M-1}$ $\subset$ $\mathbb{C}^N\times\{0,1,\ldots,L-1\}$ Although the data is real-valued in usual machine learning tasks, we allow complex-valued data without loss of generality. Support Vector Machine (SVM) is a supervised machine learning algorithm that classifies data to several classes by optimizing separating hyperplanes. \cite{boser1992training, svm} SVM is one of the most robust classifier for having global minimum guaranteed by convex optimization\cite{steinwart2008support}, and has been showing excellent performance not only in classification, but also in regression and other numerous fields. Also, it is known to be efficient for non-linear classification and regression for high-dimensional data with kernel trick. In this paper, we will focus on binary classification where only two class exists for the data (i.e. $L=2$) since a multi-class classification can be achieved with multiple binary SVM by one-vs-all or one-vs-one scheme. Also, for notation simplicity, we assume $N=2^n$, $M=2^m$ $\left(n,m \in \mathbbm{N}_0\right)$.
    \subsection{Hard-Margin SVM}
        For given dataset $\mathcal{S}$, $y\in \left\{-1, 1\right\}^N$, suppose the dataset is linearly separable, that is there exists a hyperplane $\langle{\mathbf{w}},{\mathbf{x}}\rangle+b=0$ such that
        \begin{equation}
            \begin{cases}
                \langle{\mathbf{w}},{\mathbf{x}_i}\rangle+b\ge1 \ &\forall{y_i=1}\\
                \langle{\mathbf{w}},{\mathbf{x}_i}\rangle+b\le-1 \ &\forall{y_i=-1}
            \end{cases}
        \end{equation}
        The margin, distance between two parallel supporting hyperplanes, that we want to maximize is $2/\norm{\mathbf{w}}$. Since $\arg\max{2/\norm{\mathbf{w}}} = \arg\min{\norm{\mathbf{w}}^2/2}$, selecting optimal hyperplane $\langle{\mathbf{w}^\star},{\mathbf{x}}\rangle+b^\star=0$ and its parameters $\left(\mathbf{w}^\star, b^\star\right)$ is equivalent to solving a quadratic optimization problem \eqref{eq:hard_primal}.
        \begin{equation}\label{eq:hard_primal}
                p^\star = \min_{\mathbf{w}, b}{\frac{1}{2}\norm{\mathbf{w}}^2} :y_i(\langle{\mathbf{w}},{\mathbf{x}_i}\rangle+b)\ge1 \ \forall{i\in\{0,\dots,N-1\}}
        \end{equation}
        Instead of solving this primal problem, solving dual problem is much easier.
        The dual problem of \emph{\eqref{eq:hard_primal}} is
        \begin{equation}\label{eq:hard_dual}
            d^\star =\max_{\boldsymbol{\beta}\ge0}\left\{\sum_{i=1}^{M-1}{\beta_i}-\frac{1}{2}{\beta_i\beta_j y_i y_j\langle{\mathbf{x}_i, \mathbf{x}_j}\rangle}\right\} :\sum_{i=0}^{M-1}{\beta_i y_i}=0
        \end{equation}
        where the optimal solution of primal and dual problems satisfies Karush–Kuhn–Tucker conditions.
        \begin{equation}\label{eq:KKT_condtion_stationary}
            \begin{aligned}
                \mathbf{w}^\star=\sum_{i=1}^{M-1}{\beta^\star_i y_i \mathbf{x}_i}, \sum_{i=1}^{M-1}{\beta^\star_i y_i}=0
            \end{aligned}
        \end{equation}
        \begin{equation}\label{eq:KKT_condition_complementary_slackness}
            \beta^\star_i \left[y_i\left(\langle{\mathbf{w}^\star},{\mathbf{x}_i}\rangle+b^\star\right)-1\right]=0
        \end{equation}
        \begin{equation}\label{eq:KKT_condition_feasibility}
            y_i\left(\langle{\mathbf{w}^\star},{\mathbf{x}_i}\rangle+b^\star\right)\ge1, \beta^\star_i\ge0
        \end{equation}
        By investigating complementary slackness condition \eqref{eq:KKT_condition_complementary_slackness}, it is clear that only the data on supporting hyperplanes can have non-zero $\beta^\star_i$ since $y_i \left(\langle{\mathbf{w}^\star},{\mathbf{x}_i}\rangle+b^\star\right)=1$ for those data. They are referred to as support vectors. Note that from this relationship one can find optimal bias that disappeared in dual problem.
        \begin{equation}
            b^\star = y_j-\sum_{i=1}^{M-1}{\beta_i y_i \langle \mathbf{x}_i, \mathbf{x}_j \rangle}\ \forall{j: \alpha^\star_j>0}
        \end{equation}
        Since only support vectors are on supporting hyperplane, it can be considered as a set of linear combinations of support vectors. Thus, they are named as \emph{support} vectors.
        The estimated label $\hat{y}$ of test data $\hat{\mathbf{x}}$ is determined by the relative location of test data with respect to the separating hyperplane $\left\{\mathbf{x}:\langle{\mathbf{w}^\star},{\mathbf{x}}\rangle+b^\star=0\right\}$.
        \begin{equation}\label{eq:hard_class}
            \hat{y} = \mathrm{sgn}\left\{{\sum_{i=0}^{M-1}{\langle{\mathbf{x}_i},{\hat{\mathbf{x}}}} \rangle+b^\star}\right\}
        \end{equation}
        From \eqref{eq:hard_class}, we can treat Lagrange multiplier $\boldsymbol{\beta}$ as normalized weights of the labelled data. 
        \subsection{Soft-Margin SVM}\label{sec:c-svm}
        C-SVM introduce slack variables $\xi_i$'s as violation between supporting hyperplane and outlier data in between two supporting hyperplanes. The goal is to minimize this violation while maximizing distance between supporting hyperplanes.
        \begin{equation}\label{eq:soft_primal}
            p^\star=\min_{\mathbf{w}, b, \boldsymbol{\xi}}{\frac{1}{2}\norm{\mathbf{w}}^2+\frac{C}{2}\sum_{i=1}^{M-1}{\xi_i^2}}:
            y_i\left(\langle{\mathbf{w}},{\mathbf{x}_i}\rangle+b\right)\ge1-\xi_i \forall{i\in\left\{0,\dots,N-1\right\}}
        \end{equation}
        Here, hyperparameter $C$ is user-defined value that controls over-fitting and under-fitting. We have adapted the formulation in Ref. \cite{liu2021rigorous} instead of original formulation in Ref. \cite{svm}. We can interpret the term $\frac{1}{2}\norm{\mathbf{w}}^2$ as regularizing term, and $1/C$ as its hyperparameter. With the same logic in case of hard-margin SVM, solving dual problem is much easier.
        The dual problem of \emph{\eqref{eq:soft_primal}} is
        \begin{equation}\label{eq:soft_dual}
            d^\star=\max_{\boldsymbol{\beta}\ge0}{\sum_{i=1}^{M-1}{\beta_i}-\frac{1}{2}\sum_{i,j=0}^{M-1}{\beta_i\beta_jy_i y_j \langle{\mathbf{x}_i},\mathbf{x}_j\rangle}}-\frac{1}{2C}\sum_{i=0}^{M-1}\beta_i^2
            :\sum_{i=0}^{M-1}\beta_i y_i=0
        \end{equation}
        where the optimal solution of primal and dual problems satisfies Karush–Kuhn–Tucker conditions.
        \begin{equation}\label{eq:soft_KKT_stationary}
            \mathbf{w}^\star=\sum_{i=1}^{M-1}{\beta^\star_i y_i \mathbf{x}_i},
            \sum_{i=1}^{M-1}{\beta^\star_i y_i}=0,
            \xi^\star_i=\frac{1}{C}\beta_i^\star
        \end{equation}
        \begin{equation}\label{eq:soft_KKT_complementary_slackenes}
            \alpha^\star_i \left[y_i\left(\langle{\mathbf{w}^\star},{\mathbf{x}_i}\rangle+b^\star\right)-1+\xi^\star_i\right]=0
        \end{equation}
        \begin{equation}\label{eq:soft_KKT_feasibilty}
            y_i\left(\langle{\mathbf{w}^\star},{\mathbf{x}_i}\rangle+b^\star\right)\ge1-\xi^\star_i,\ \alpha^\star_i\ge0
        \end{equation}
        By investigating complementary slackness condition in \eqref{eq:soft_KKT_complementary_slackenes}, it is clear that only the data on and in between supporting hyperplanes can have non-zero $\beta^\star_i$ since $y_i\left(\langle{\mathbf{w}^\star},{\mathbf{x}_i}\rangle+b^\star\right)=1-\xi_i$ for those data. Therefore, with the same logic as before, optimal bias is obtained manually.
        \begin{equation}\label{eq:optimal_b}
            b^{\star}=y_q(1-C^{-1}\beta^\star_q)-\sum_{i=0}^{M-1}{\beta^{\star}_i y_i k(\mathbf{x}_q, \mathbf{x}_i)}
        \end{equation}
        However, there may be no examples exactly lying on the supporting hyperplane. In this case, we can approximate $b^\star$ as median value of absolute difference $\abs{y_i-\langle{\mathbf{w}^\star}{\mathbf{x}_i}\rangle}$ among all training data.\cite{mml-book} Estimating test label would be the same as that of the Hard-margin SVM \eqref{eq:hard_class}.
        
        \subsection{Kernel Trick}\label{sec:kenel_trick}
        Even with the soft margin assumption, dataset may not be linearly separated efficiently. In this case, we can define feature map $\phi:\mathcal{X}\mapsto\mathcal{H}$ to map data to high dimensional Hilbert space $\mathcal{H}$. With the sophisticated feature map, we expect mapped dataset $\mathbf{x}\rightarrow\phi\left(\mathbf{x}\right)$ is linearly separable in $\mathcal{H}$.
        \begin{equation}\label{eq:dual_phi_obj}
            d^\star=\max_{\boldsymbol{\beta}\ge0}{\sum_{i=1}^{M-1}{\beta_i}-\frac{1}{2}\sum_{i,j=0}^{M-1}{\beta_i\beta_jy_i y_j \langle\phi({\mathbf{x}_i}),\phi(\mathbf{x}_j)\rangle}}-\frac{1}{2C}\sum_{i=0}^{M-1}\beta_i^2
            :\sum_{i=0}^{M-1}\beta_i y_i=0
        \end{equation}
        \begin{equation}\label{eq:class_phi}
            \hat{y} = \mathrm{sgn}\left\{{\sum_{i=0}^{M-1}{\langle\phi({\mathbf{x}_i}),\phi({\hat{\mathbf{x}}})} \rangle+b^\star}\right\}
        \end{equation}
        However, defining feature map may not be practical nor possible. For example, the dimension of $\mathcal{H}$ should be infinite in order to linearly separate any mapped dataset. In addition, the \emph{closeness} between two mapped data is not clear. Note that both Equation \eqref{eq:dual_phi_obj} and \eqref{eq:class_phi} require calculation of inner product between $\phi$ instead of $\phi$ itself. Thus, rather than constructing ill-defined vector map, we define kernel function $k:\mathcal{X}\cross\mathcal{X}\mapsto\mathbb{C}$.
        \begin{equation}\label{eq:kernel_def}
            k({\mathbf{x}},{\mathbf{y}})=\langle \phi({\mathbf{x}}),\phi({\mathbf{y}})\rangle_{\mathcal{H}}
        \end{equation}
        We can consider the kernel as the measure of \emph{similarity}, for it is defined as inner product between mapped vectors. It has been proven that there exists unique feature map $\phi$ for positive semi-definite kernel $k({\cdot},{\cdot})=\langle\phi(\cdot),\phi(\cdot)\rangle_{\mathcal{H}}$.
        Consider the formulations \eqref{eq:soft_dual} and \eqref{eq:hard_class}. They imply that the inner product between examples only matter for given dataset. Therefore, if we define some features $\phi(\mathbf{x}_i)$ to represent $\mathbf{x}_i$, then the dual SVM formulation \eqref{eq:soft_dual} and classifying equation \eqref{eq:hard_class} are almost the same except for inner product part. 

        Since $\phi({\cdot})$ can be highly non-linear mapping, we can construct non-linear classifier on given dataset using SVM. Note that SVM only solves linear classification problem originally. Users can now have more degree of freedom from the arbitrary selection on the feature mapping to classify examples. However, $\phi({\cdot})$ may map original data to very high-dimensional Hilbert space so that calculating not only features themselves but also inner product of features explicitly may cost severe computation resources. It has been proven that there exist a unique feature map for any positive semi-definite (PSD) kernel. Therefore users can define a kernel to solve dual SVM problem. This is known as kernel trick\cite{svm, boyd2004convex, mml-book}.
        \begin{equation}\label{eq:dual_SVM_kernel}
            d^\star=\max_{\boldsymbol{\beta}\ge0}{\sum_{i=1}^{M-1}{\beta_i}-\frac{1}{2}\sum_{i,j=0}^{M-1}{\beta_i\beta_jy_i y_j k({\mathbf{x}_i},\mathbf{x}_j)}}-\frac{1}{2C}\sum_{i=0}^{M-1}\beta_i^2
            :\sum_{i=0}^{M-1}\beta_i y_i=0
        \end{equation}
        \begin{equation}\label{eq:class_kernel}
            \hat{y} = \mathrm{sgn}\left\{{\sum_{i=0}^{M-1}{k({\mathbf{x}_i},{\hat{\mathbf{x}}})}+b^\star}\right\}
        \end{equation}

        There are several popular choice for kernel; polynomial kernel and Gaussian radial basis function kernel, for examples,
        \begin{equation}\label{eq:poly_kernel}
            k({\mathbf{x}},{\mathbf{y}}) = \left(\langle{\mathbf{x}},{\mathbf{y}\rangle}+c\right)^d,
        \end{equation}
        \begin{equation}\label{eq:rbf_kernel}
            k({\mathbf{x}},{\mathbf{y}}) = e^{-\frac{\gamma}{2}\norm{\mathbf{x}-\mathbf{y}}^2}.
        \end{equation}
        Polynomial kernel is defined as \eqref{eq:poly_kernel} with kernel hyperparameter bias $c$, and order $d$. Gaussian radial basis function kernel(RBF kernel) is defined as \eqref{eq:rbf_kernel} with kernel hyperparameter inverse of standard deviation $\gamma$.
        
\section{Proof of Eqs. (15) and (16)}\label{sec:proof}
    The proof of Eqs. (15) is as follows: 
    \begin{equation}\label{eq:proof_primal_QASVM}
        \begin{aligned}
            \ket{\psi} =& \sum_{i=0}^{M-1}{\sqrt{\alpha_i}\ket{0,i,\phi(\mathbf{x}_i),y_i, \phi(\hat{\mathbf{x}})}}\\
            \Longrightarrow & \frac{1}{\sqrt{2}}\sum_{i=0}^{M-1}{\sqrt{\alpha_i}\ket{0,i,\phi(\mathbf{x}_i),y_i, \phi(\hat{\mathbf{x}})}}+
            \frac{1}{\sqrt{2}}\sum_{i=0}^{M-1}{\sqrt{\alpha_i}\ket{1,i,\phi(\mathbf{x}_i),y_i,\phi(\hat{\mathbf{x}})}}\\
            \Longrightarrow & \frac{1}{\sqrt{2}}\sum_{i=0}^{M-1}{\sqrt{\alpha_i}\ket{0,i,\phi(\mathbf{x}_i),y_i,\phi(\hat{\mathbf{x}})}}+
            \frac{1}{\sqrt{2}}\sum_{i=0}^{M-1}{\sqrt{\alpha_i}\ket{1,i,\phi(\hat{\mathbf{x}}),y_i,\phi(\mathbf{x}_i)}}\\
            \Longrightarrow & \frac{1}{2}\sum_{i=0}^{M-1}{\sqrt{\alpha_i}\left(\ket{0,i,\phi(\mathbf{x}_i),y_i,\phi(\hat{\mathbf{x}})}+\ket{0,i,\phi(\hat{\mathbf{x}}),y_i,\phi(\mathbf{x}_i)}\right)}\\
            +& \frac{1}{2}\sum_{i=0}^{M-1}{\sqrt{\alpha_i}\left(\ket{1,i,\phi(\mathbf{x}_i),y_i,\phi(\hat{\mathbf{x}})}-\ket{1,i,\phi(\hat{\mathbf{x}}),y_i,\phi(\mathbf{x}_i)}\right)}\\
            &\xrightarrow{\expval{Z_aZ_y}}
            \sum_{i=0}^{M-1}\alpha_iy_i{\abs{\braket{\phi(\mathbf{x}_i)}{\phi(\hat{\mathbf{x}})}}^2},\ 
            \xrightarrow{\expval{I_aZ_y}} \sum_{i=0}^{M-1}\alpha_iy_i{}
        \end{aligned}
    \end{equation}
    The proof of Eqs. (16) is as follows: 
    \begin{equation}\label{eq:proof_dual_QASVM}
        \begin{aligned}
            \ket{\psi} =& \sum_{i,j=0}^{M-1}{\sqrt{\alpha_i}\sqrt{\alpha_j}\ket{0,i,\phi(\mathbf{x}_i),y_i, j,\phi(\mathbf{x}_j),y_j}}\\
            \Longrightarrow & \frac{1}{\sqrt{2}}\sum_{i,j=0}^{M-1}{\sqrt{\alpha_i}\sqrt{\alpha_j}\ket{0,i,\phi(\mathbf{x}_i),y_i,j,\phi(\mathbf{x}_j),y_j}}+
            \frac{1}{\sqrt{2}}\sum_{i,j=0}^{M-1}{\sqrt{\alpha_i}\sqrt{\alpha_j}\ket{1,i,\phi(\mathbf{x}_i),y_i,j,\phi(\mathbf{x}_j),y_j}}\\
            \Longrightarrow & \frac{1}{\sqrt{2}}\sum_{i,j=0}^{M-1}{\sqrt{\alpha_i}\sqrt{\alpha_j}\ket{0,i,\phi(\mathbf{x}_i),y_i,j,\phi(\mathbf{x}_j),y_j}}+
            \frac{1}{\sqrt{2}}\sum_{i,j=0}^{M-1}{\sqrt{\alpha_i}\sqrt{\alpha_j}\ket{1,i,\phi(\mathbf{x}_j),y_i,j,\phi(\mathbf{x}_i),y_j}}\\
            \Longrightarrow & \frac{1}{2}\sum_{i,j=0}^{M-1}{\sqrt{\alpha_i}\sqrt{\alpha_j}\left(\ket{0,i,\phi(\mathbf{x}_i),y_i,j,\phi(\mathbf{x}_j),y_j}+\ket{0,i,\phi(\mathbf{x}_j),y_i,j,\phi(\mathbf{x}_i),y_j}\right)}\\
            +& \frac{1}{2}\sum_{i,j=0}^{M-1}{\sqrt{\alpha_i}\sqrt{\alpha_j}\left(\ket{0,i,\phi(\mathbf{x}_i),y_i,j,\phi(\mathbf{x}_j),y_j}-\ket{0,i,\phi(\mathbf{x}_j),y_i,j,\phi(\mathbf{x}_i),y_j}\right)}\\
            &\xrightarrow{\expval{Z_aZ_{y_0}Z_{y_1}}}
            \sum_{i,j=0}^{M-1}\alpha_i\alpha_jy_iy_j{\abs{\braket{\phi(\mathbf{x}_i)}{\phi(\mathbf{x}_j)}}^2}, \ 
            \xrightarrow{\expval{I_aZ_{y_0}Z_{y_1}}}
            \sum_{i,j=0}^{M-1}\alpha_i\alpha_jy_iy_j{}=\left(\sum_{i=0}^{M-1}\alpha_iy_i{}\right)^2
        \end{aligned}
    \end{equation}
    For more detail, see \cite{blank2020quantum, park2020theory}.
    \clearpage
    
\section{Supplementary Figures}
    In this section, we provide supplementary figures mentioned in the main text.
    \begin{figure}[ht]
        \centering
        \includegraphics[width=0.5\textwidth]{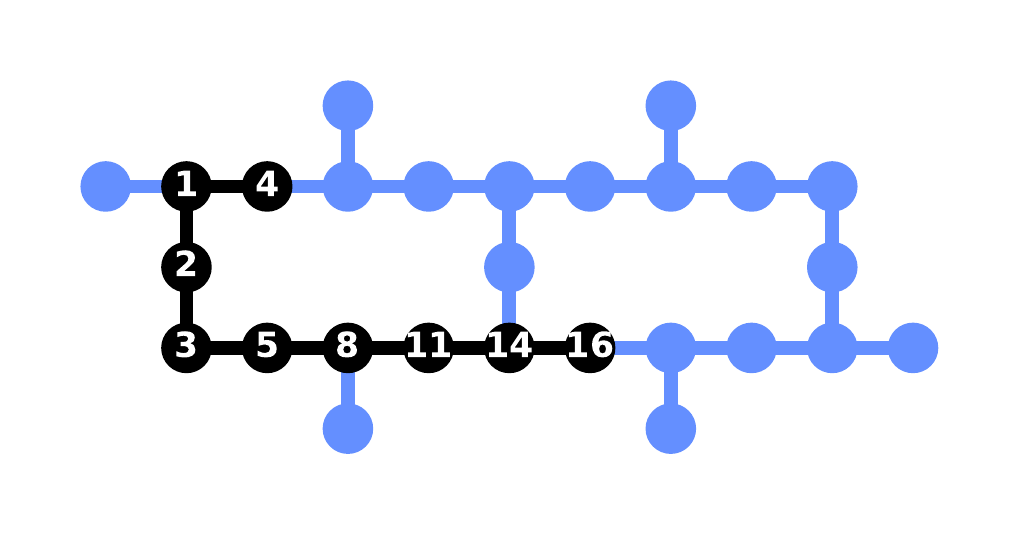}
        \caption{\textit{ibm\_montreal} qubits configurations. We used qubits colored in black for QASVM with toy data set }
    \end{figure}
    \begin{figure}[ht]
        \centering
        \begin{subfigure}{0.3\textwidth}
            \centering
            \includegraphics[width=\textwidth]{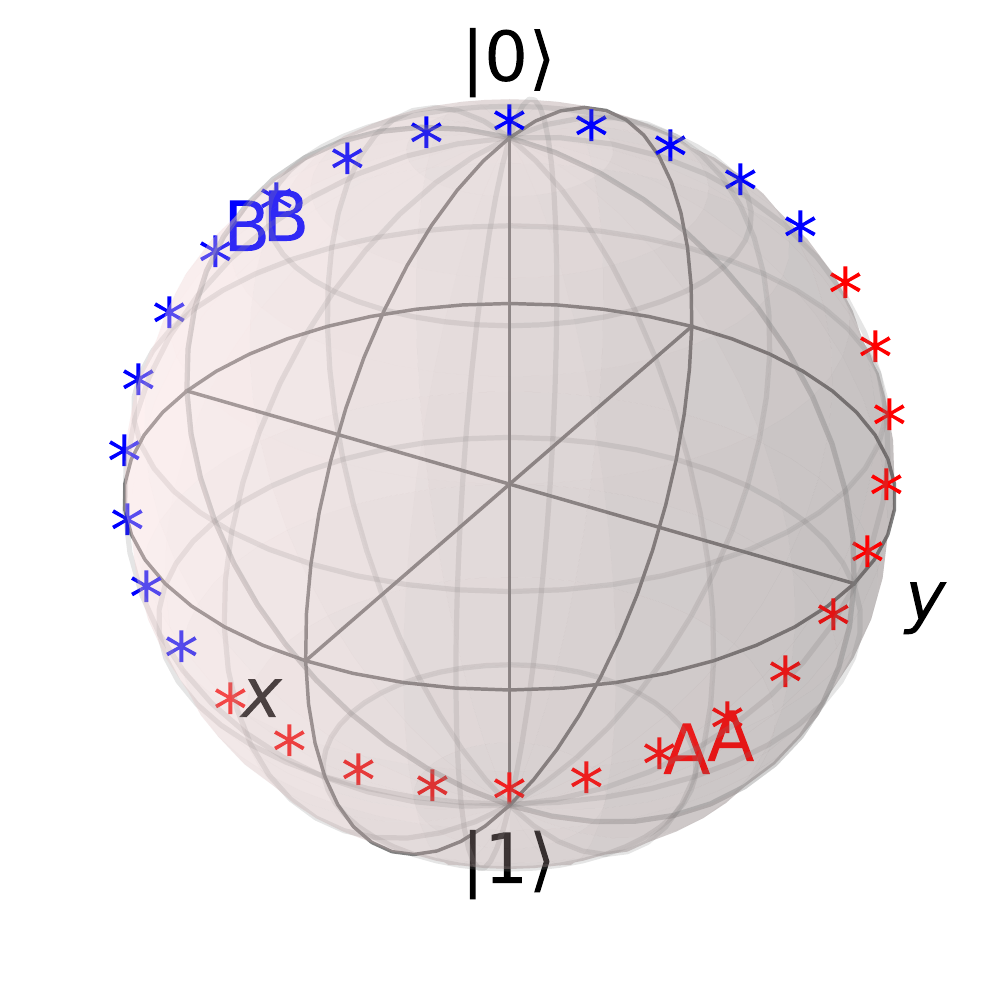}
            \caption{Balanced toy dataset}
        \end{subfigure}
        \begin{subfigure}{0.65\textwidth}
            \centering
            \includegraphics[width=\textwidth]{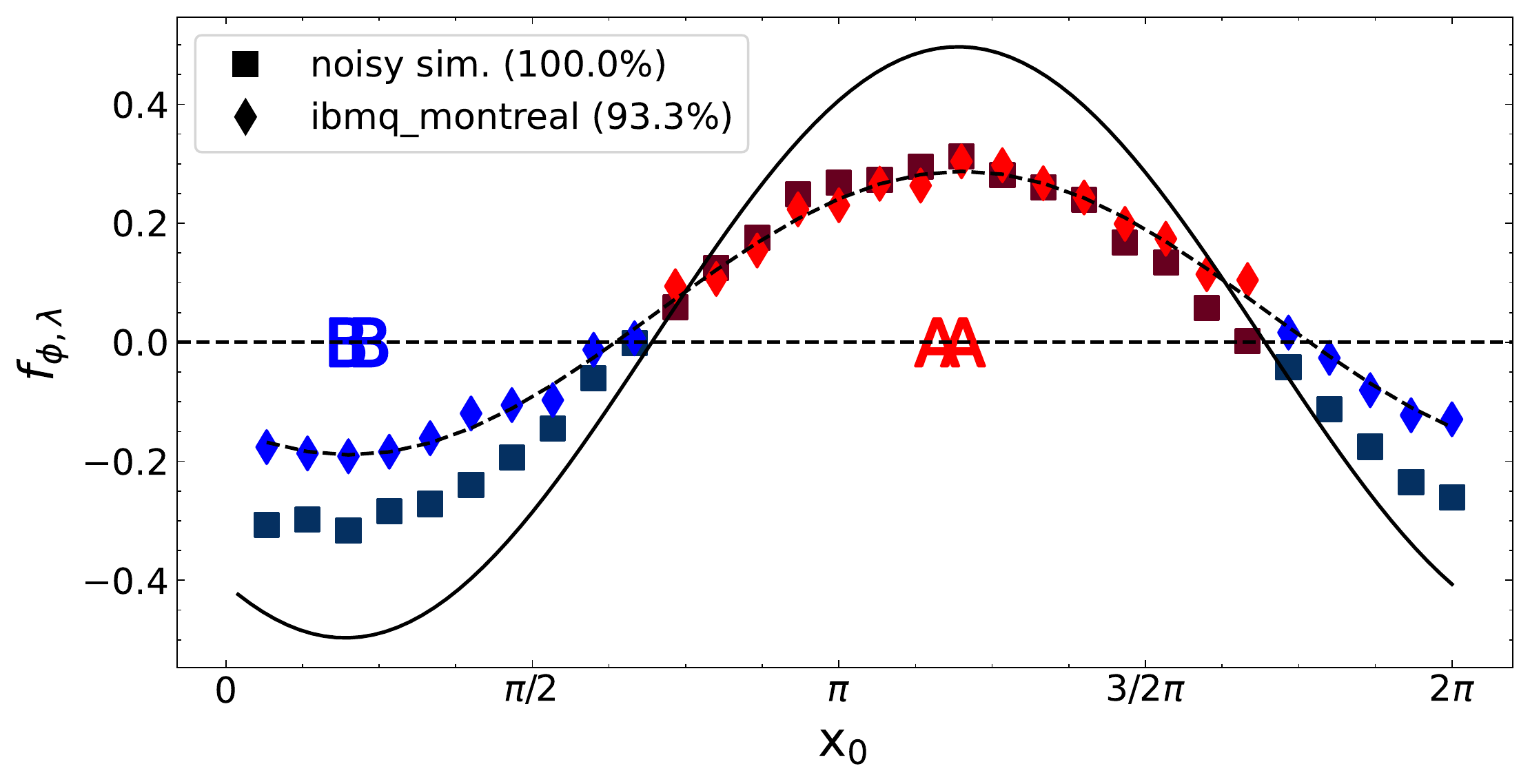}
            \caption{Classification result}
        \end{subfigure}
        \caption{Experiments on \textit{ibmq\_montreal} cloud NISQ processor with different example dataset. Here, the training data set is balanced.}
    \end{figure}
    \begin{figure}[ht]
        \centering
        \includegraphics[width=0.5\textwidth]{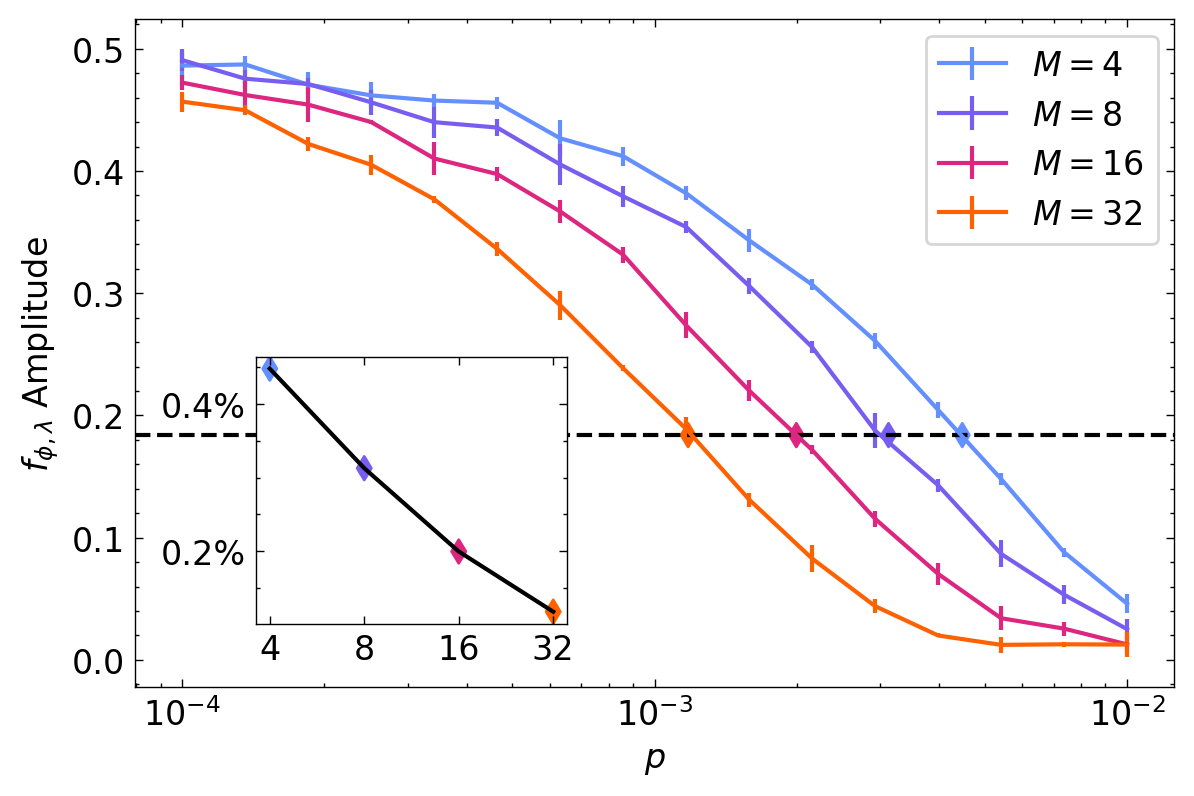}
        \caption{Noise robustness simulation result on depolarizing error. $f_{\phi, \lambda}$ amplitude represents amplitude of sine-fitted function of measured decision values. $p$ is depolarizing error rate of single-qubit gate. We set the error rate of two-qubit gates 10 times higher than that of single-qubit gates.}
    \end{figure}
    \begin{figure}[ht]
        \centering
        \begin{subfigure}{\textwidth}
            \centering
            \includegraphics[width=\textwidth]{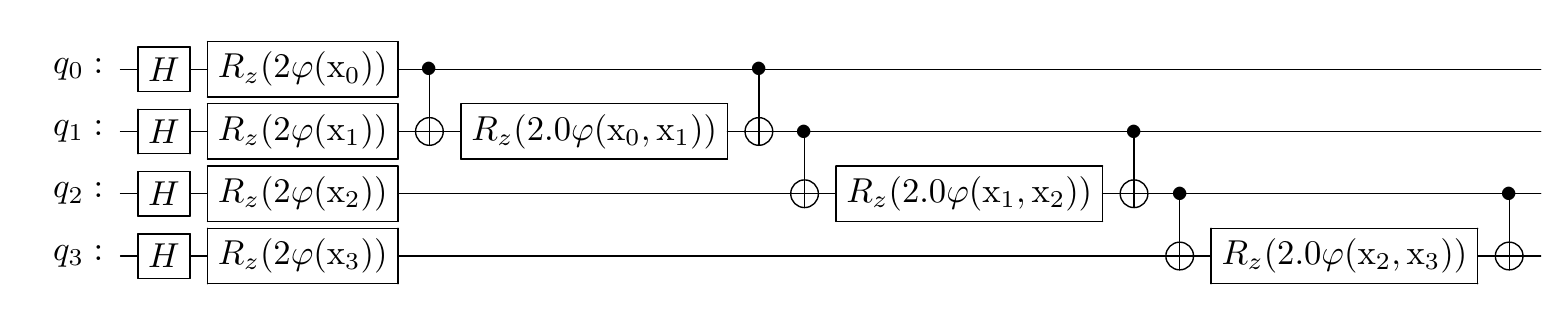}
            \caption{\textit{ZZFeatureMap} with linear entanglement}
            \label{fig:linear}
        \end{subfigure}
        \begin{subfigure}{\textwidth}
            \centering
            \includegraphics[width=\textwidth]{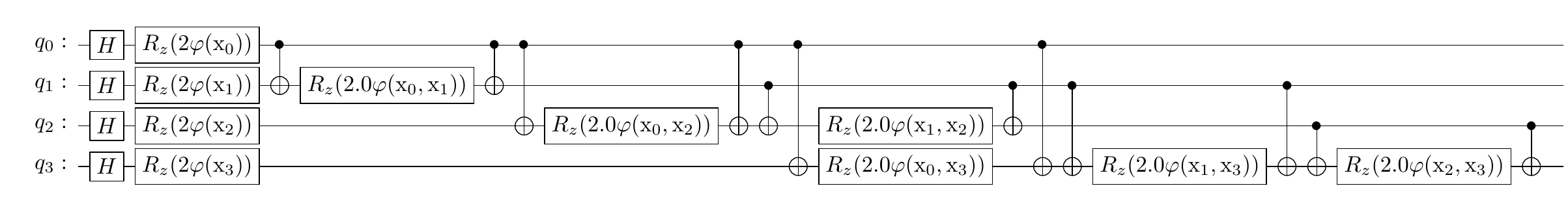}
            \caption{\textit{ZZFeatureMap} with all-to-all entanglement}
            \label{fig:alltoall}
        \end{subfigure}
        \caption{\textit{ZZfeatureMap} quantum feature map from Ref.\cite{ibm-qsvm}}
    \end{figure}
    \begin{figure}[ht]
        \centering
        \begin{subfigure}[b]{0.45\textwidth}
            \centering
            \includegraphics[width=\textwidth]{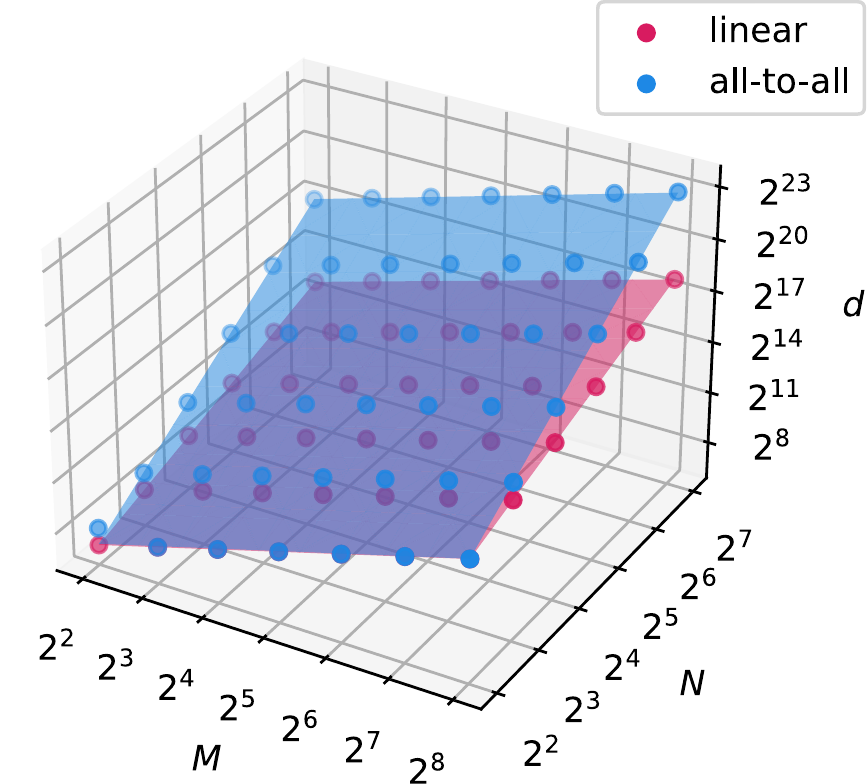}
            \caption*{{\bf a} Loss circuit depth}
            \label{fig:numanal_depth}
        \end{subfigure}
        \hfill
        \begin{subfigure}[b]{0.45\textwidth}
            \centering
            \sffamily
            \textsf{\scriptsize Classification Accuracy}
            \includegraphics[width=\textwidth]{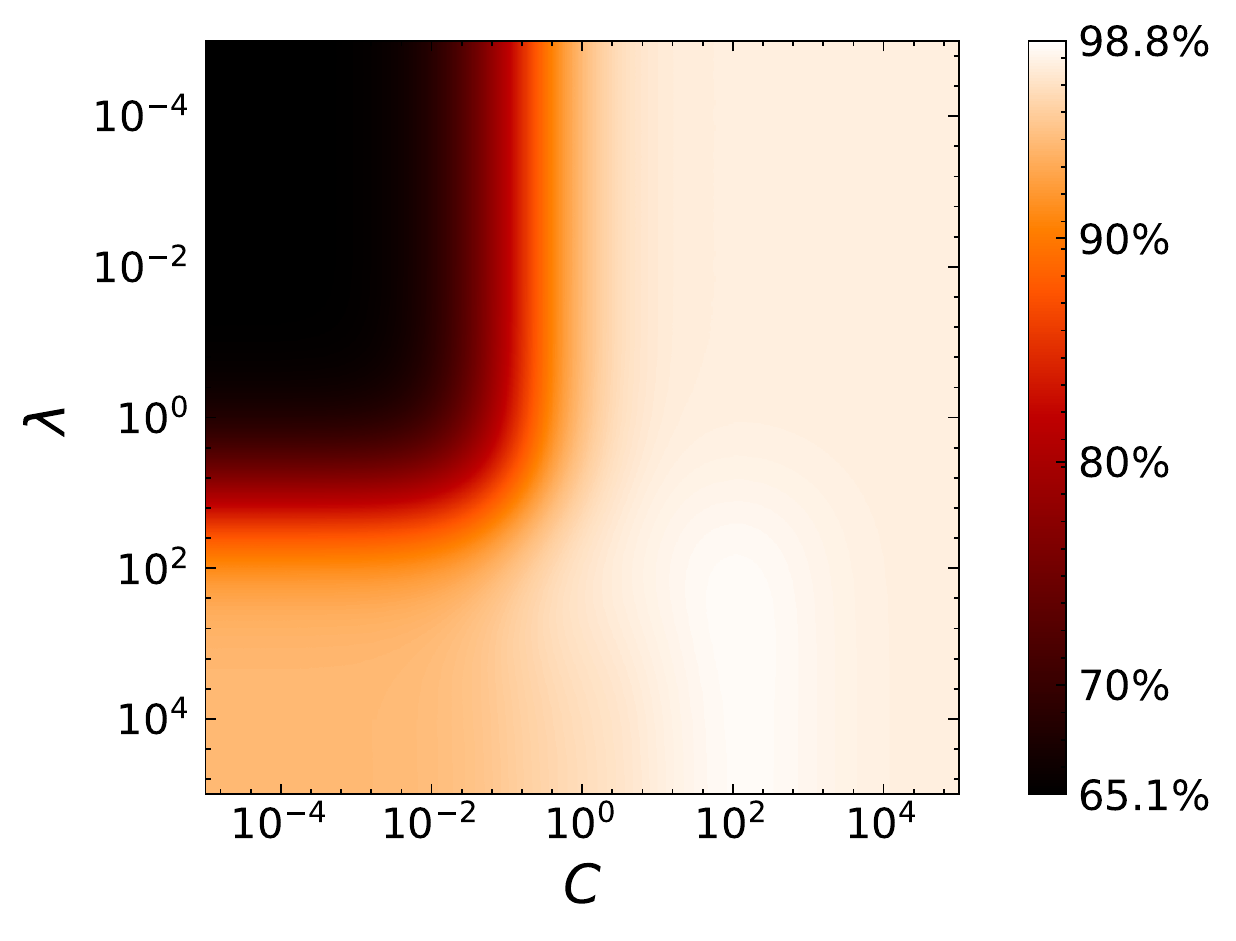}
            \caption*{{\bf b} Hyperparameter selection}
            \label{fig:hyperparameters}
        \end{subfigure}
        \caption{Numerical order analysis. 
            {\bf a}, depth of quantum circuit $d$. The closed circles indicate simulated samples, where transparent planes are the linear fitting. In the legend, `linear` indicates the use of the feature map in Fig. \ref{fig:linear}, and `all-to-all' indicates the use of the feature map in Fig. \ref{fig:alltoall}. The results show linear dependence on $M$. The circuits are decomposed with a universal gate set $\{R_x, R_y, R_z, CNOT\}$. 
            {\bf b}, classification accuracy of iris data set with a $M=64$ training data set displayed as a function of $\lambda$ and $C$.}
        \label{fig:numanal_setup} 
    \end{figure}  
    \begin{figure}[ht]
        \centering
        \begin{subfigure}{0.3\textwidth}
            \centering
            \includegraphics[width=\textwidth]{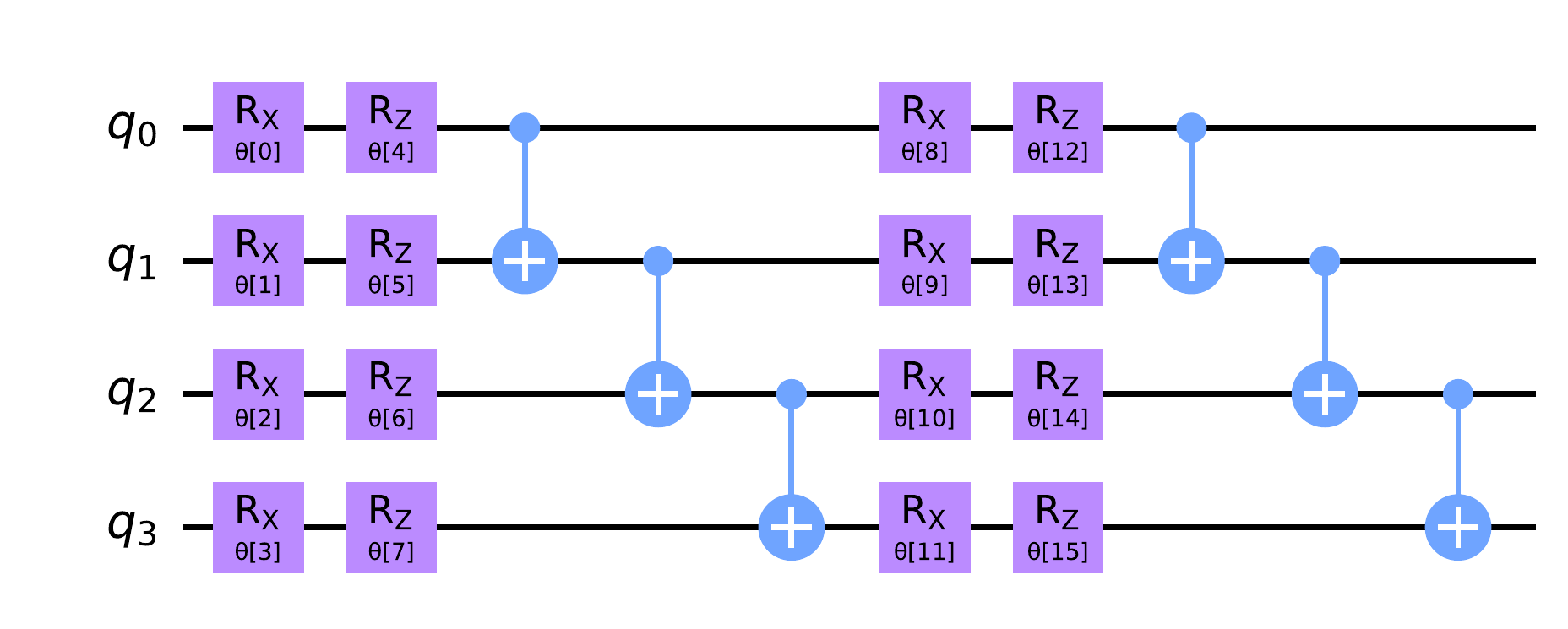}
            \caption{Circuit \#2}
        \end{subfigure}
        \hfill
        \begin{subfigure}{0.3\textwidth}
            \centering
            \includegraphics[width=\textwidth]{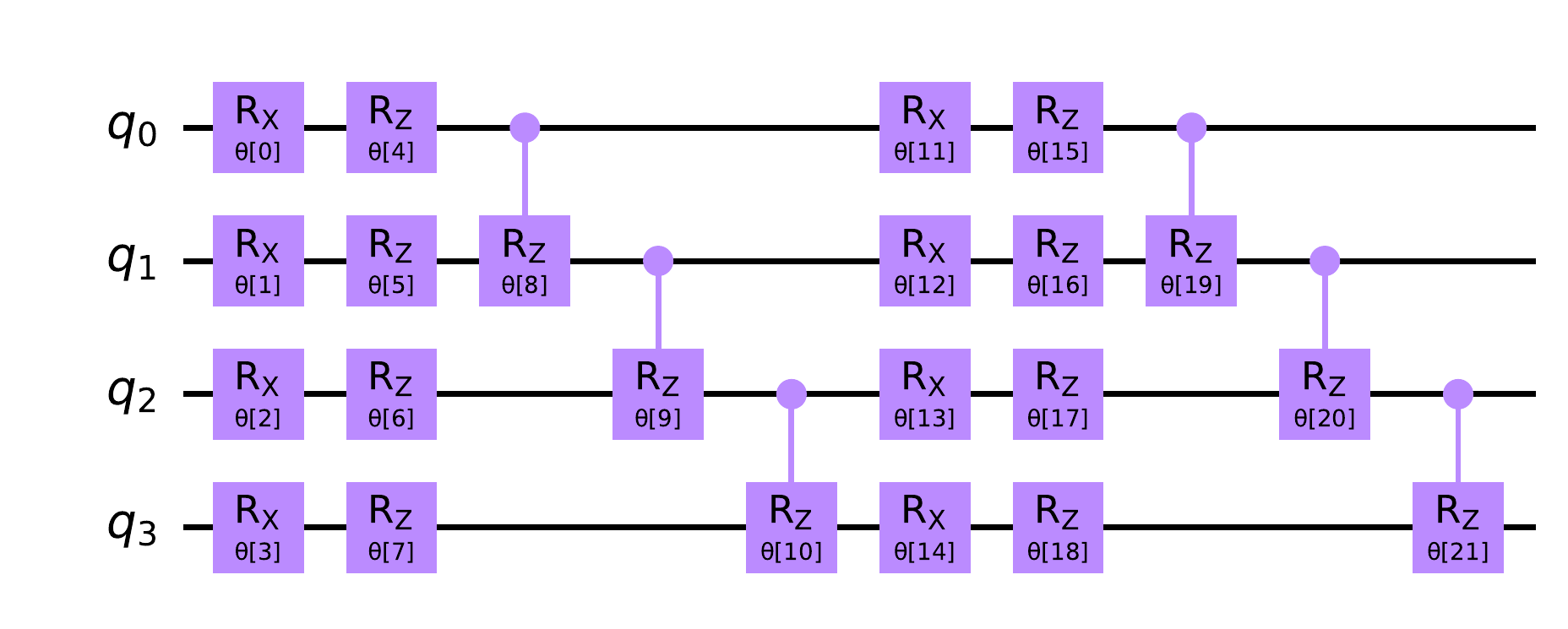}
            \caption{Circuit \#3}
        \end{subfigure}
        \hfill
        \begin{subfigure}{0.3\textwidth}
            \centering
            \includegraphics[width=\textwidth]{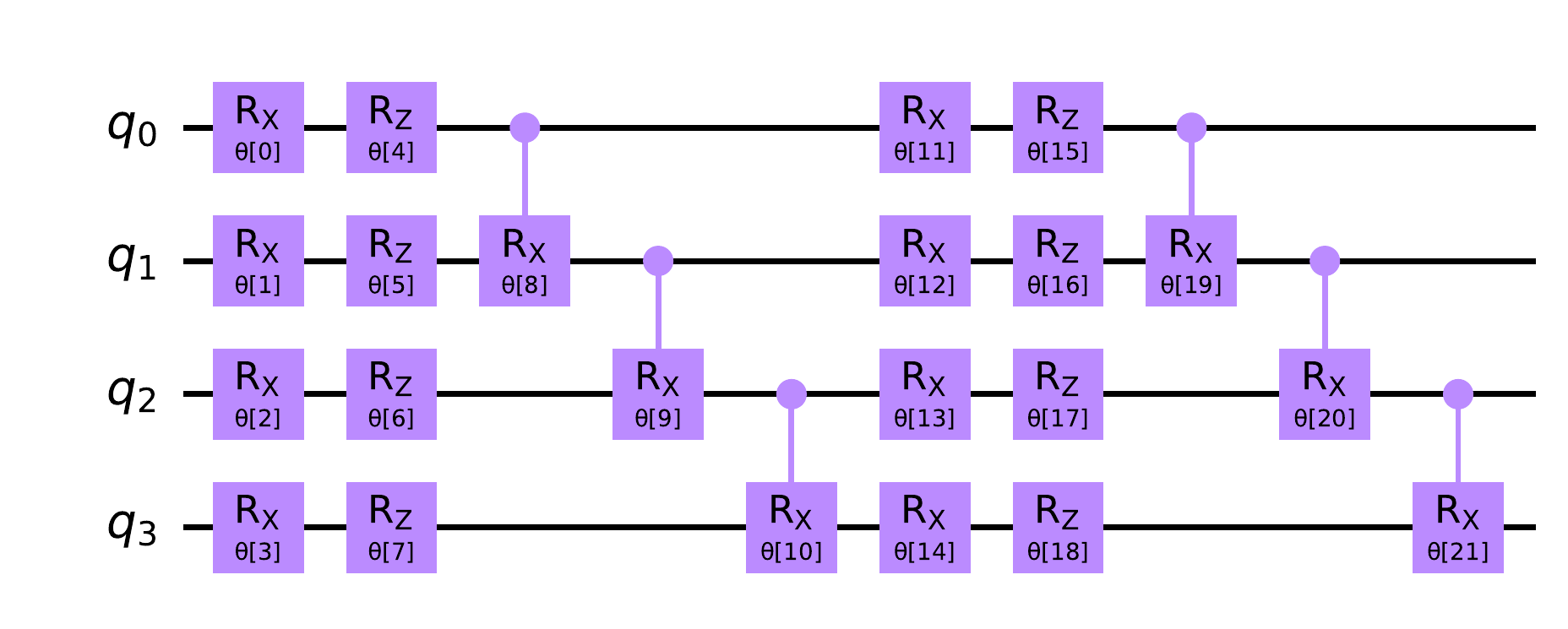}
            \caption{Circuit \#4}
        \end{subfigure}
        \hfill
        \begin{subfigure}{\textwidth}
            \centering
            \includegraphics[width=\textwidth]{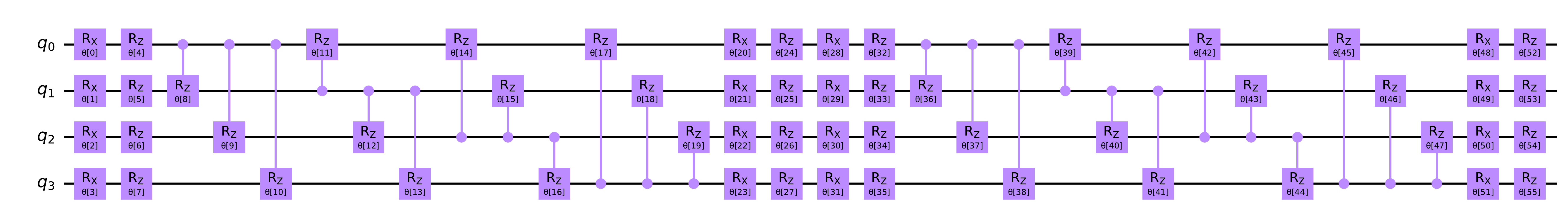}
            \caption{Circuit \#5}
        \end{subfigure}
        \hfill
        \begin{subfigure}{\textwidth}
            \centering
            \includegraphics[width=\textwidth]{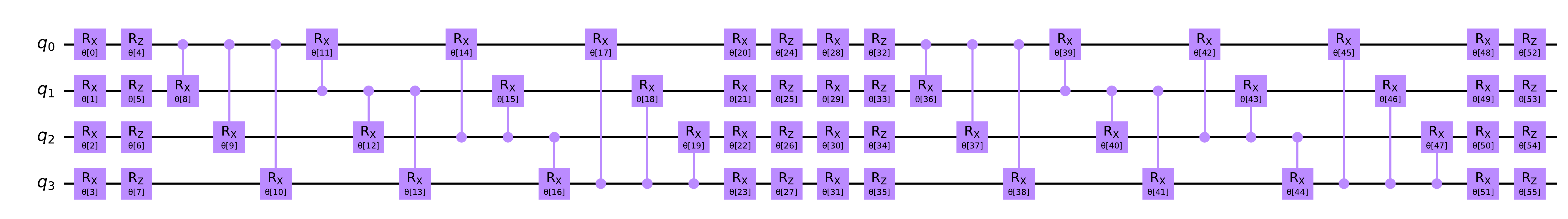}
            \caption{Circuit \#6}
        \end{subfigure}
        \hfill
        \begin{subfigure}{0.18\textwidth}
            \centering
            \includegraphics[width=\textwidth]{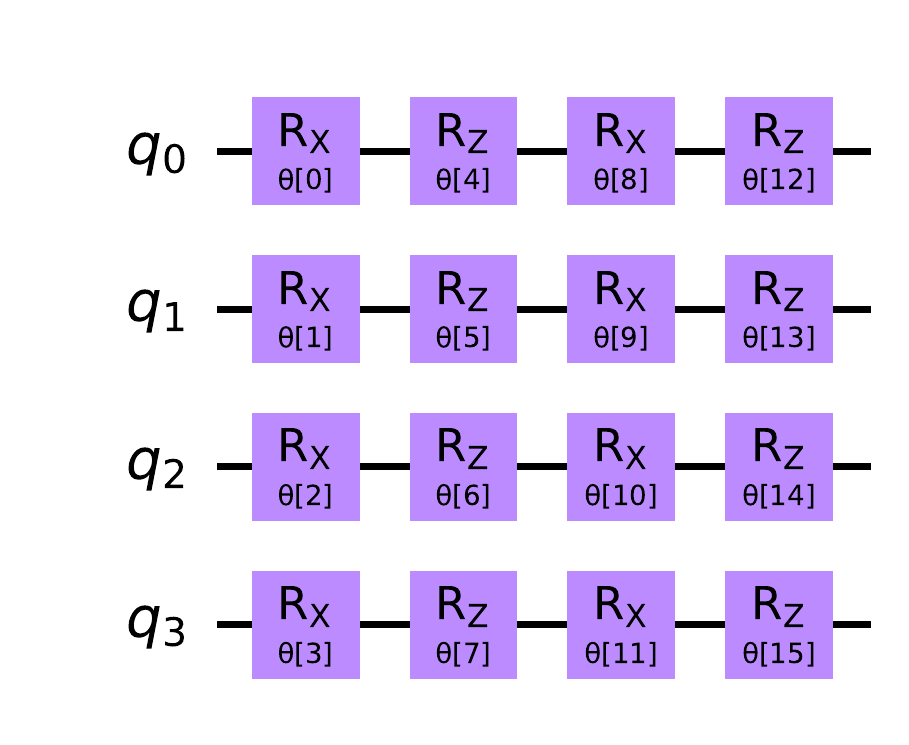}
            \caption{Circuit \#1}
        \end{subfigure}
        \hfill
        \begin{subfigure}{0.4\textwidth}
            \centering
            \includegraphics[width=\textwidth]{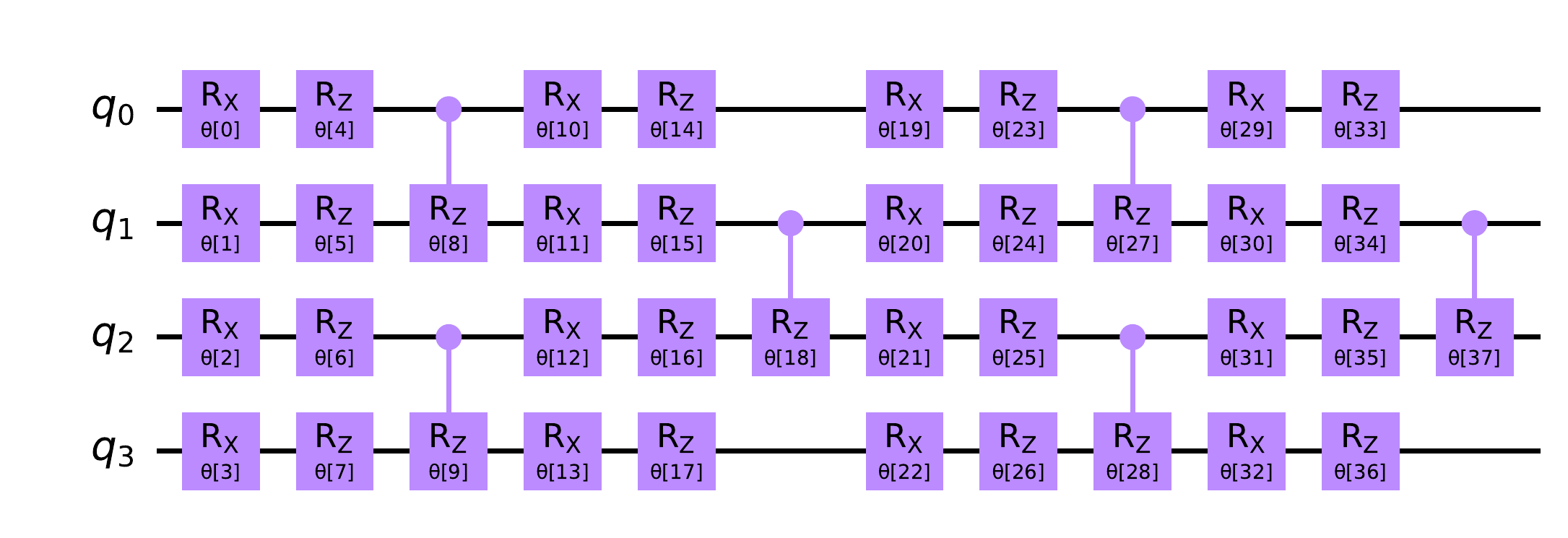}
            \caption{Circuit \#7}
        \end{subfigure}
        \hfill
        \begin{subfigure}{0.4\textwidth}
            \centering
            \includegraphics[width=\textwidth]{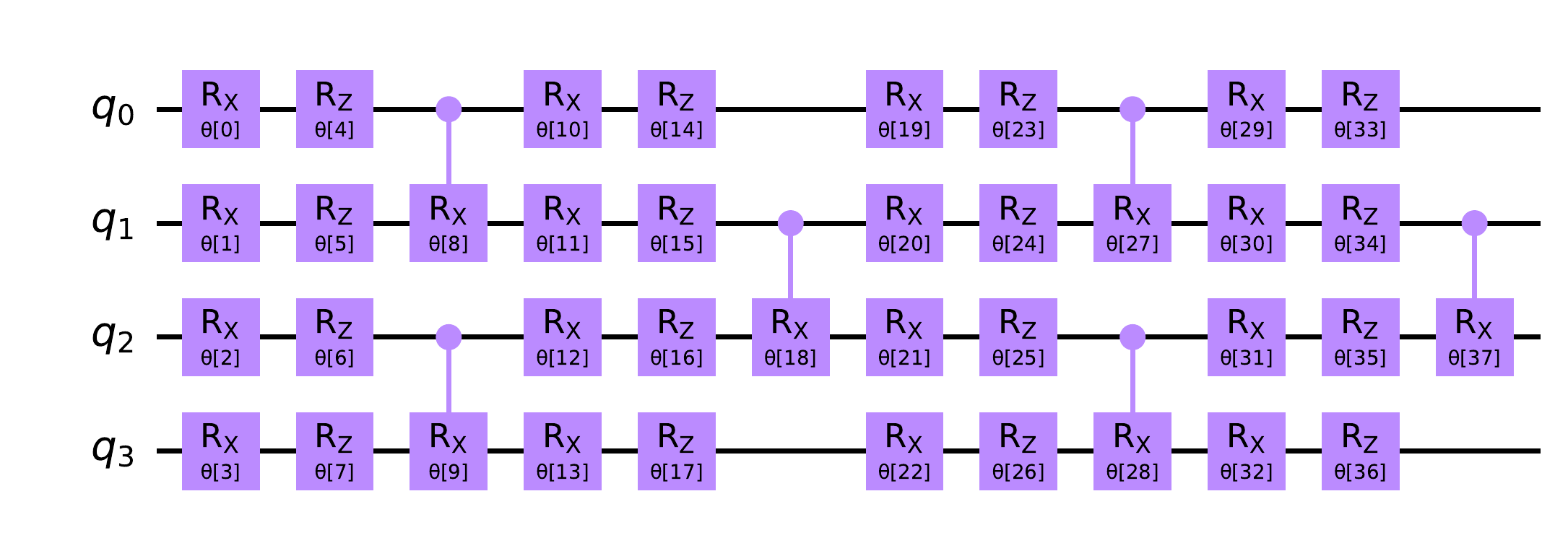}
            \caption{Circuit \#8}
        \end{subfigure}
        \hfill
        \begin{subfigure}{0.49\textwidth}
            \centering
            \includegraphics[width=\textwidth]{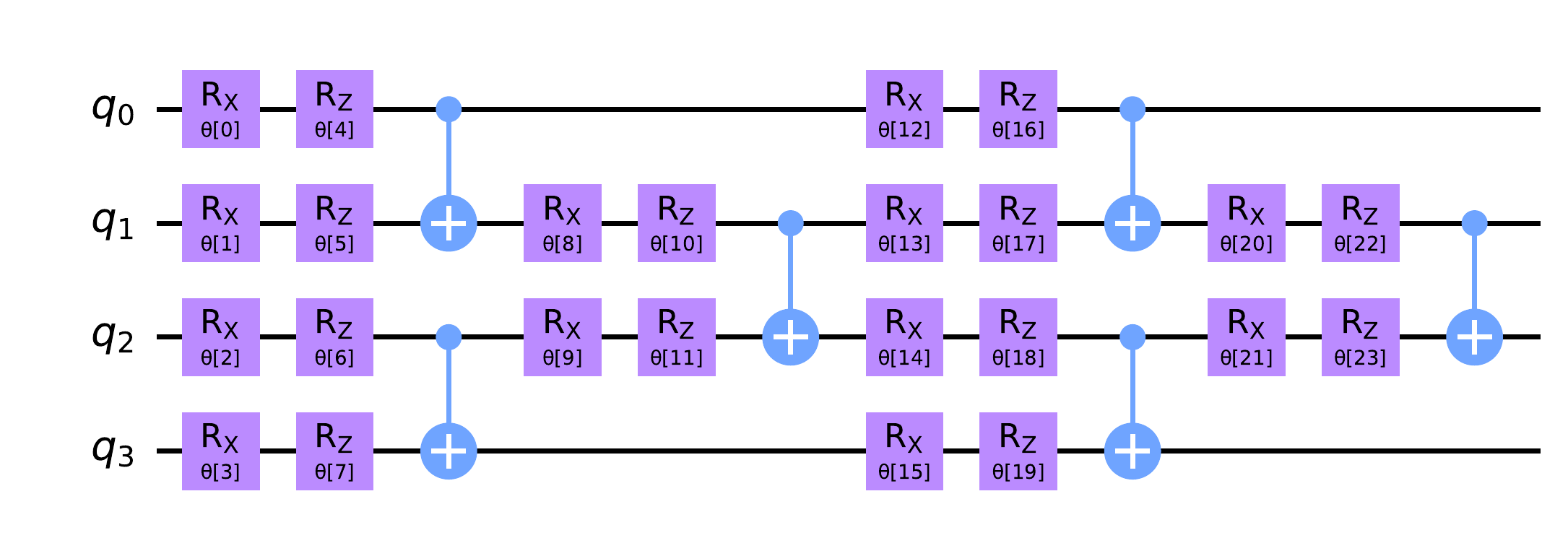}
            \caption{Circuit \#11}
        \end{subfigure}
        \hfill
        \begin{subfigure}{0.49\textwidth}
            \centering
            \includegraphics[width=\textwidth]{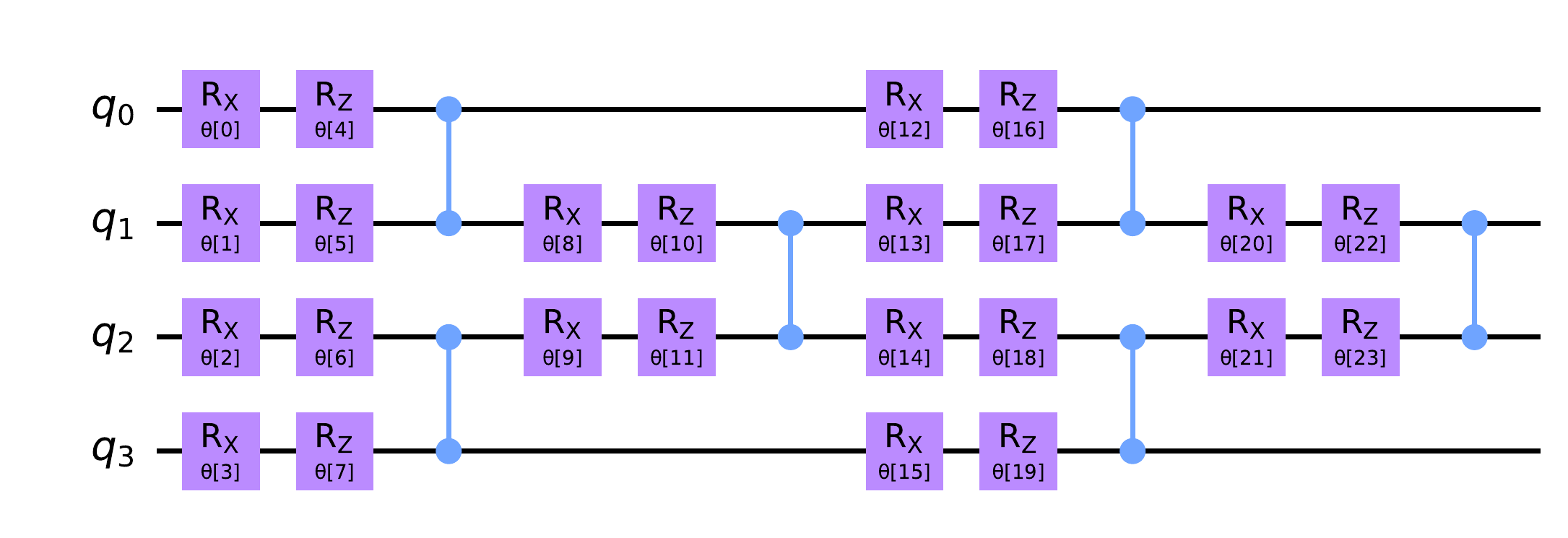}
            \caption{Circuit \#12}
        \end{subfigure}
        \hfill
        \begin{subfigure}{0.3\textwidth}
            \centering
            \includegraphics[width=\textwidth]{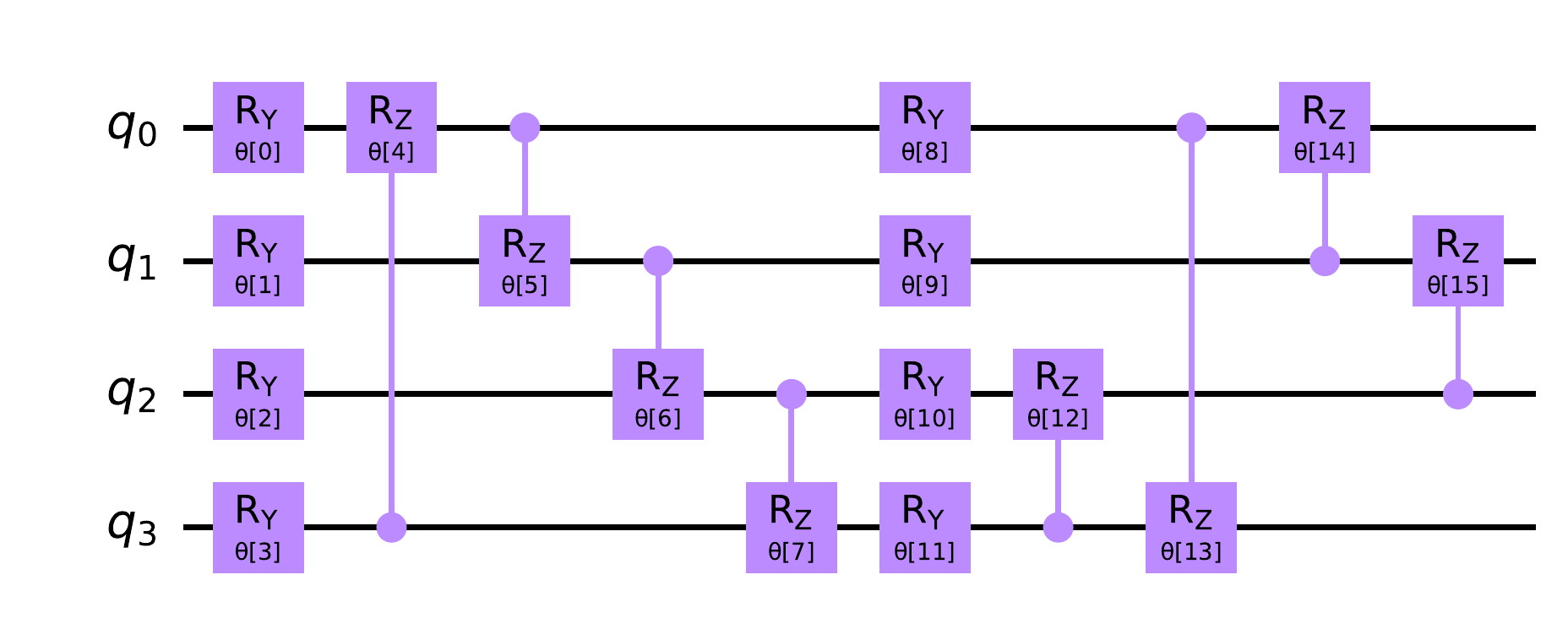}
            \caption{Circuit \#13}
        \end{subfigure}
        \hfill
        \begin{subfigure}{0.3\textwidth}
            \centering
            \includegraphics[width=\textwidth]{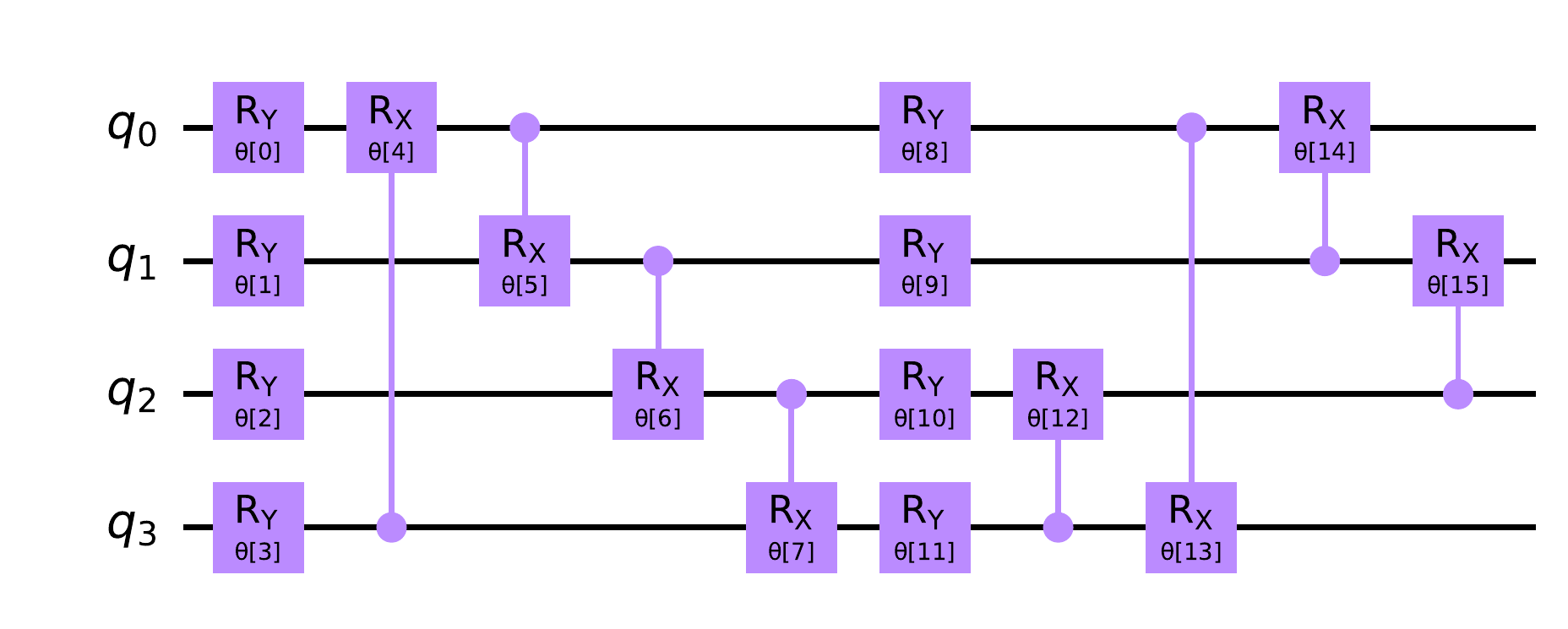}
            \caption{Circuit \#14}
        \end{subfigure}
        \hfill
        \begin{subfigure}{0.3\textwidth}
            \centering
            \includegraphics[width=\textwidth]{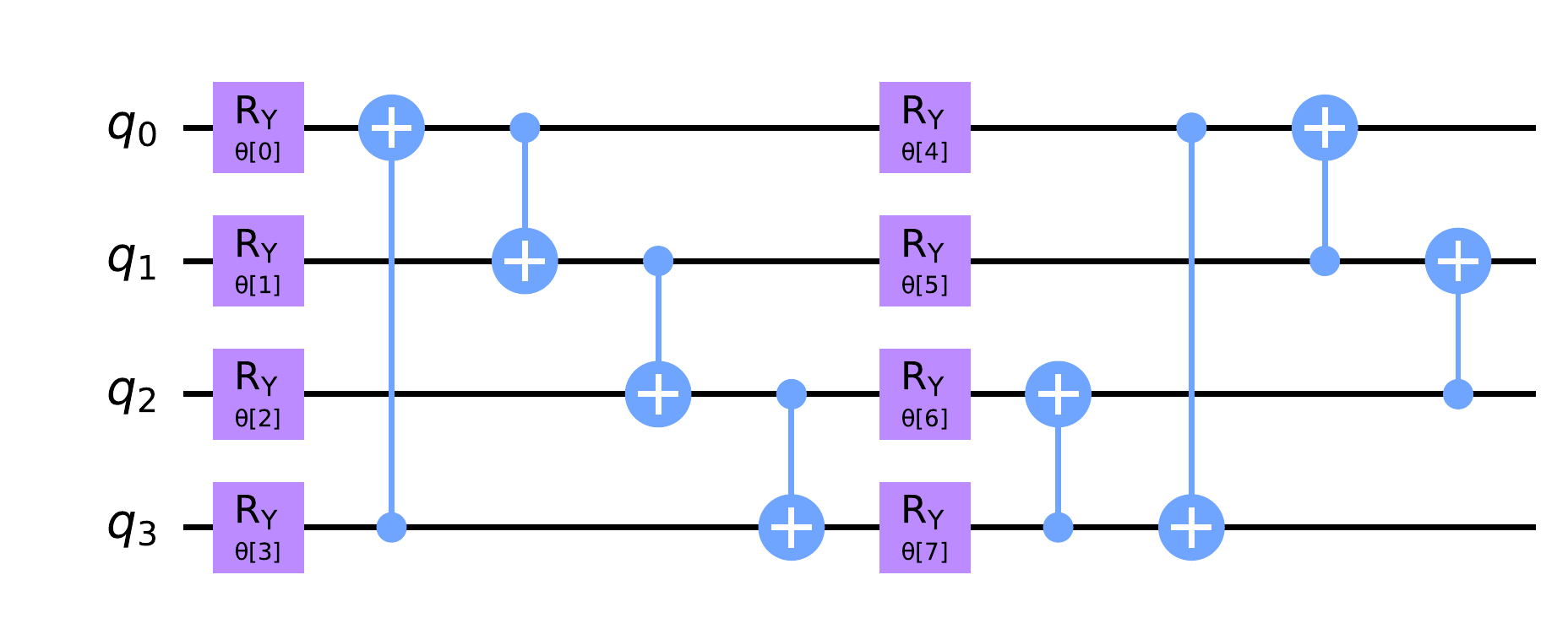}
            \caption{Circuit \#15}
        \end{subfigure}
        \hfill
        \begin{subfigure}{0.35\textwidth}
            \centering
            \includegraphics[width=\textwidth]{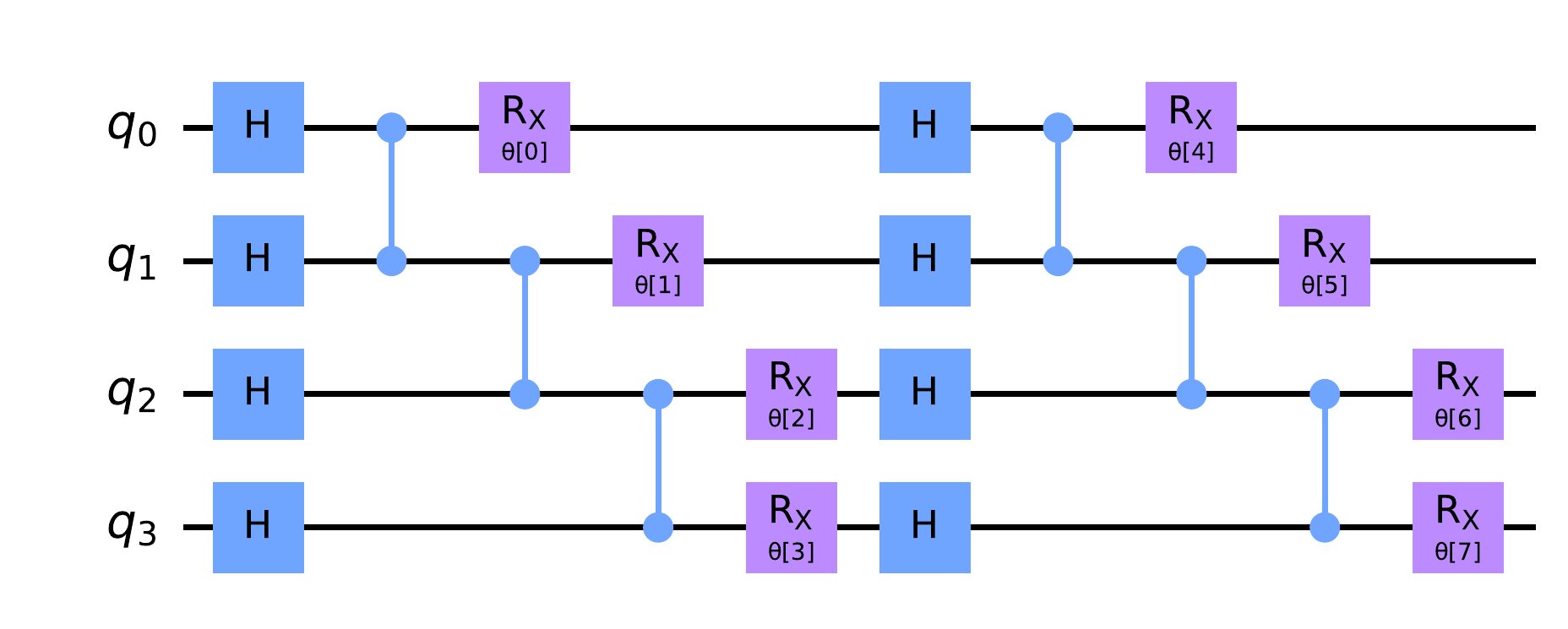}
            \caption{Circuit \#9}
        \end{subfigure}
        \hfill
        \begin{subfigure}{0.25\textwidth}
            \centering
            \includegraphics[width=\textwidth]{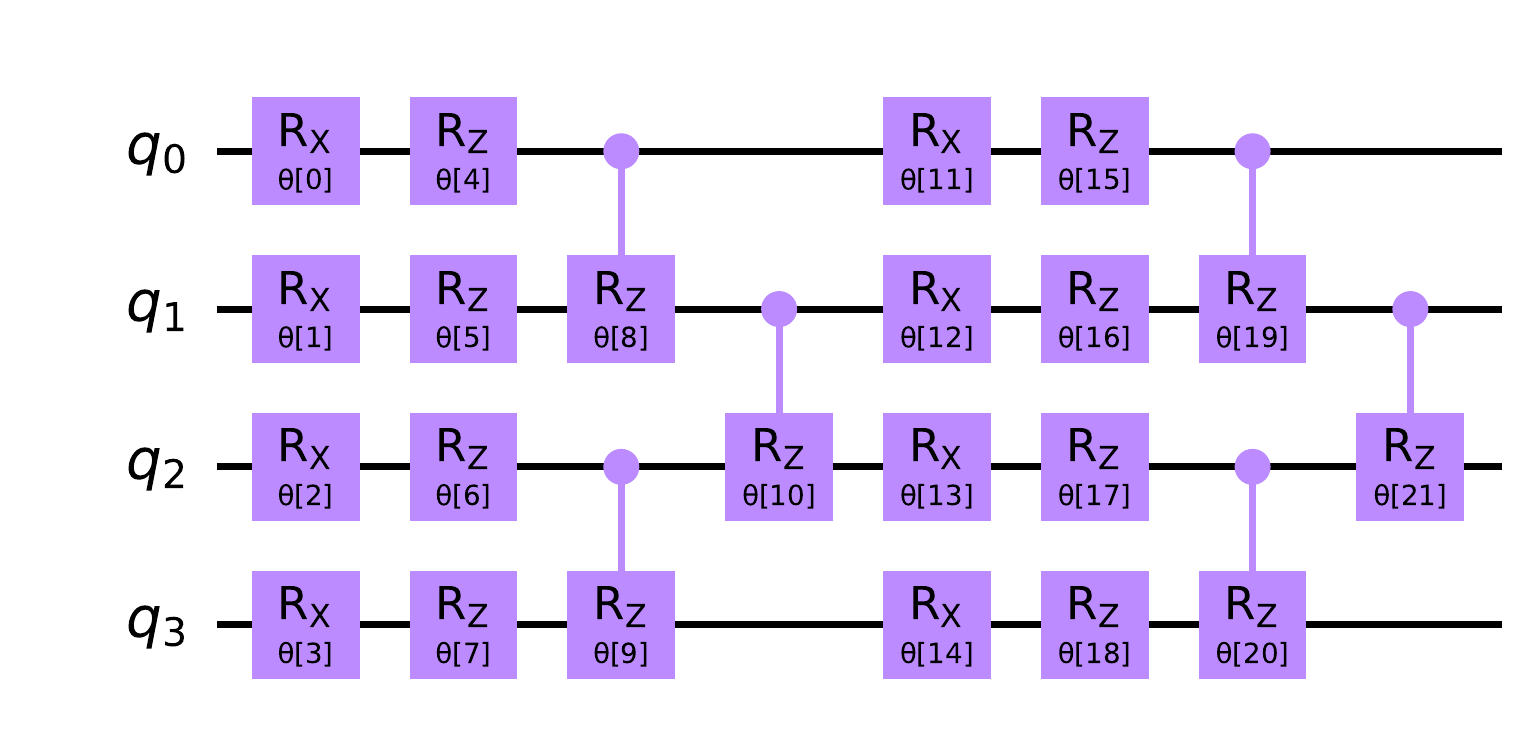}
            \caption{Circuit \#16}
        \end{subfigure}
        \hfill
        \begin{subfigure}{0.37\textwidth}
            \centering
            \includegraphics[width=\textwidth]{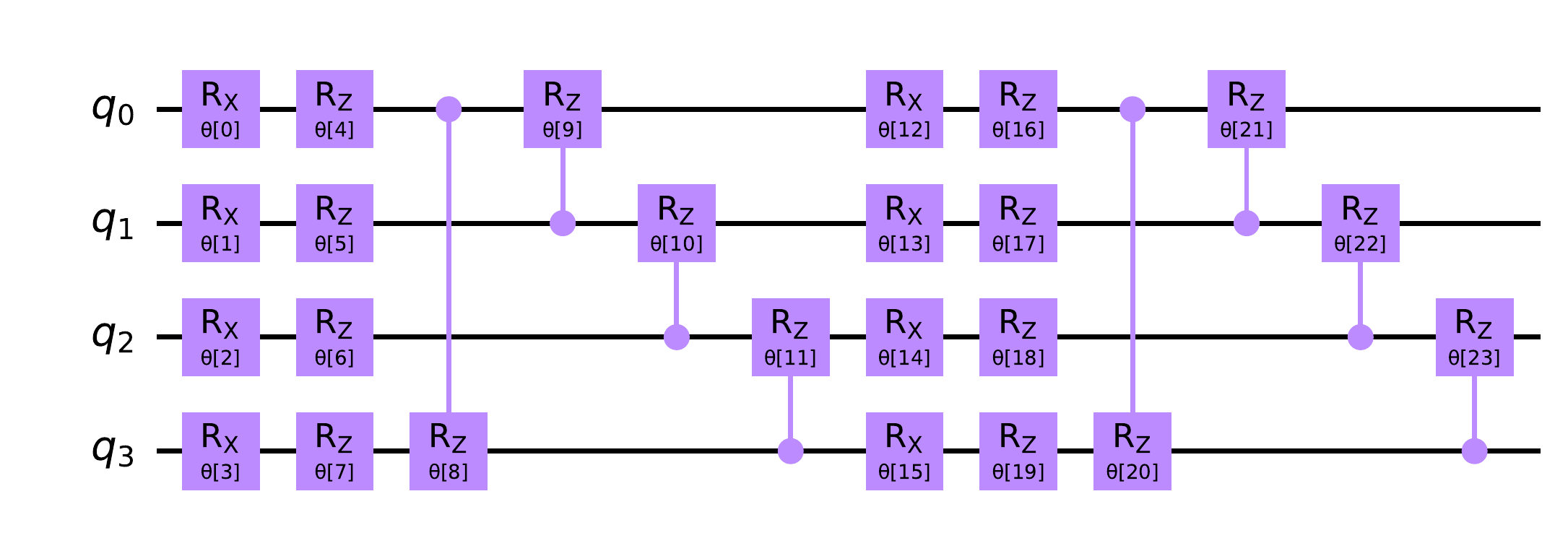}
            \caption{Circuit \#18}
        \end{subfigure}
        \hfill
        \begin{subfigure}{0.35\textwidth}
            \centering
            \includegraphics[width=\textwidth]{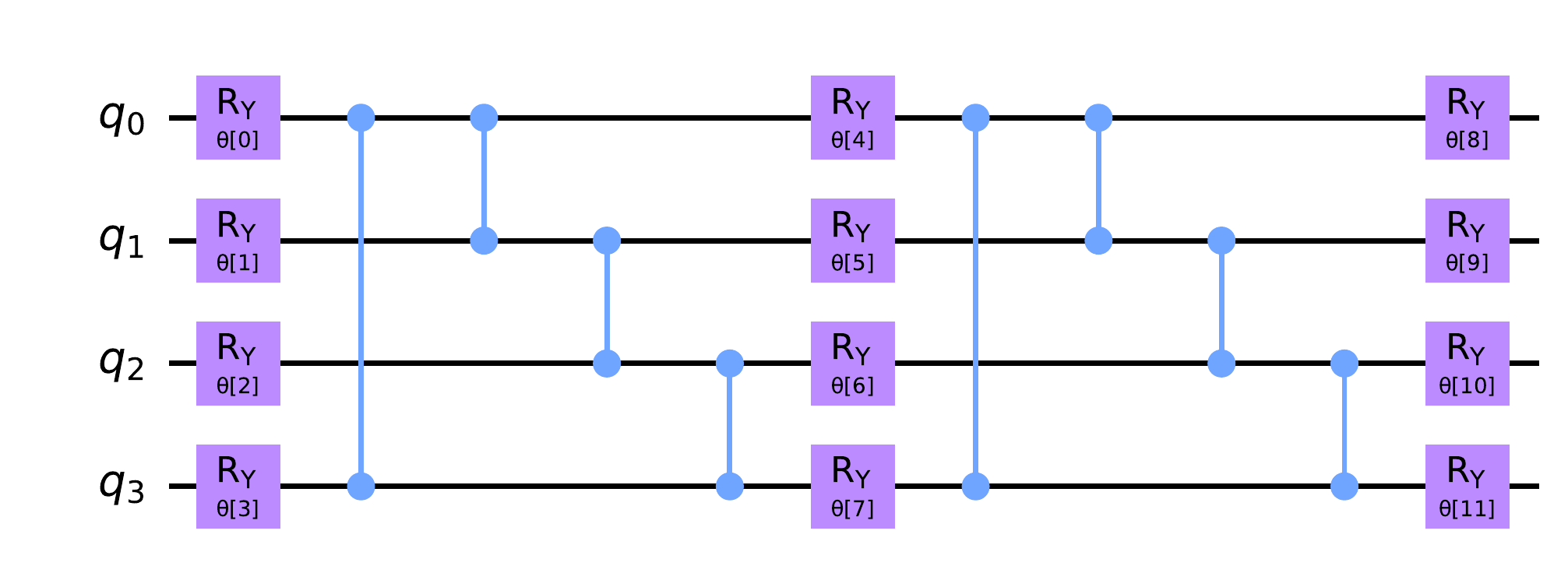}
            \caption{Circuit \#10}
        \end{subfigure}
        \hfill
        \begin{subfigure}{0.25\textwidth}
            \centering
            \includegraphics[width=\textwidth]{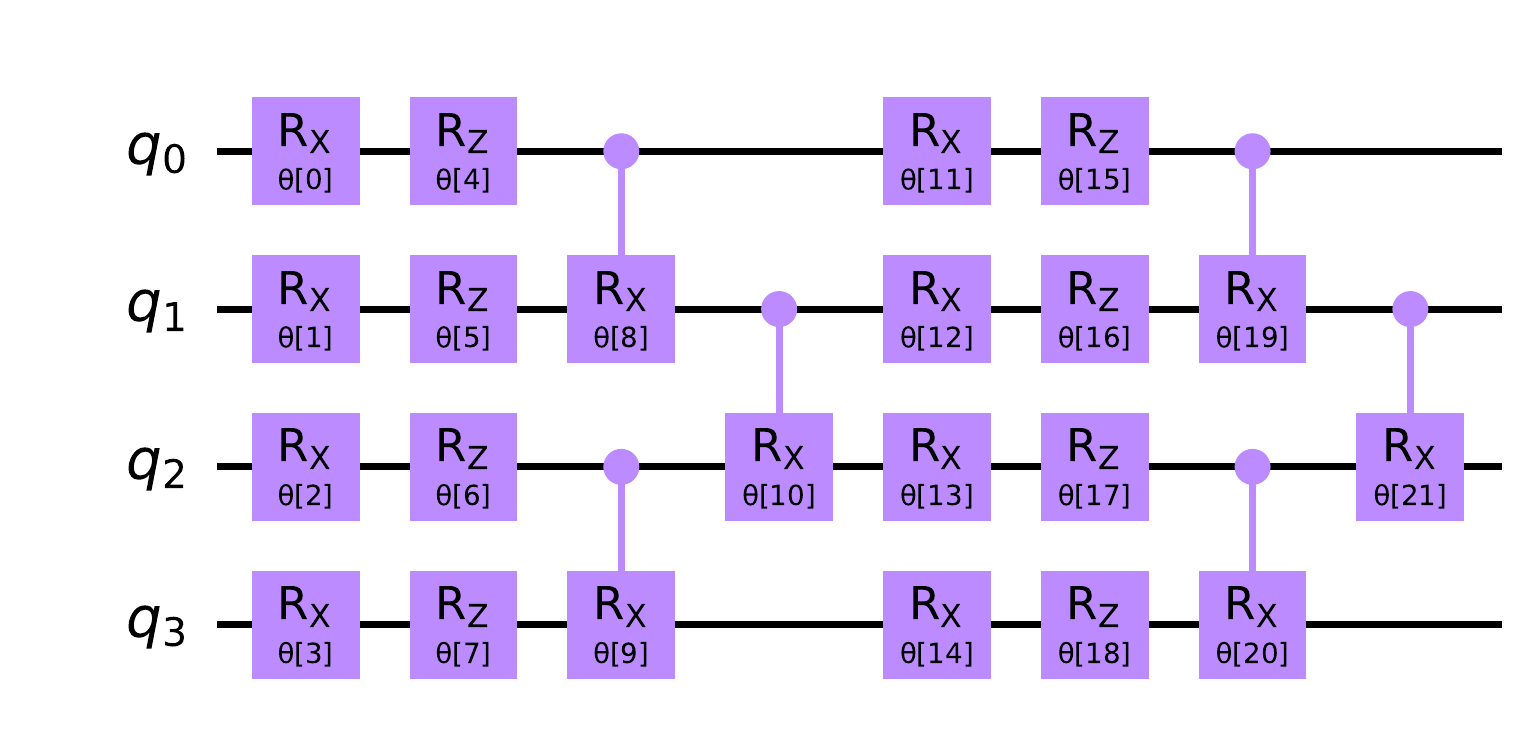}
            \caption{Circuit \#17}
        \end{subfigure}
        \hfill
        \begin{subfigure}{0.37\textwidth}
            \centering
            \includegraphics[width=\textwidth]{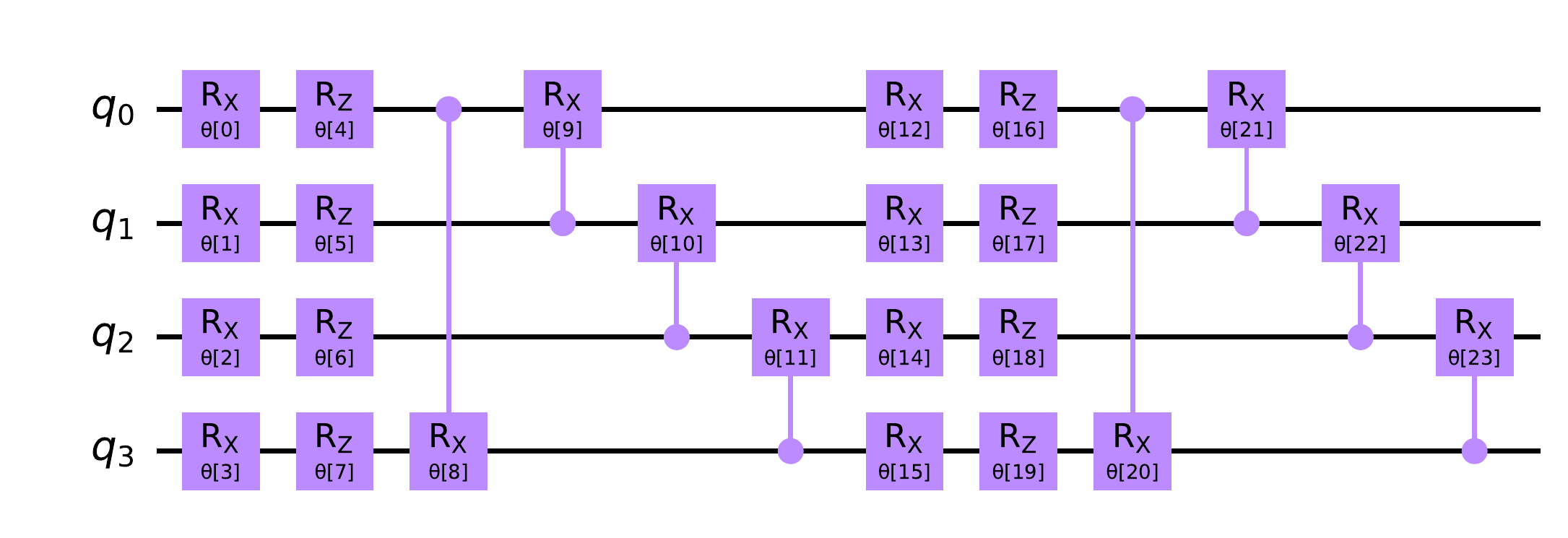}
            \caption{Circuit \#19}
        \end{subfigure}
        \hfill
        \caption{19 sample PQC architecture provided in Ref.~\citenum{sim2019expressibility}. Each circuit has 4 qubits and 2 layers.}
    \end{figure}

    \begin{figure}[ht]
        \centering
        \includegraphics[width=\textwidth]{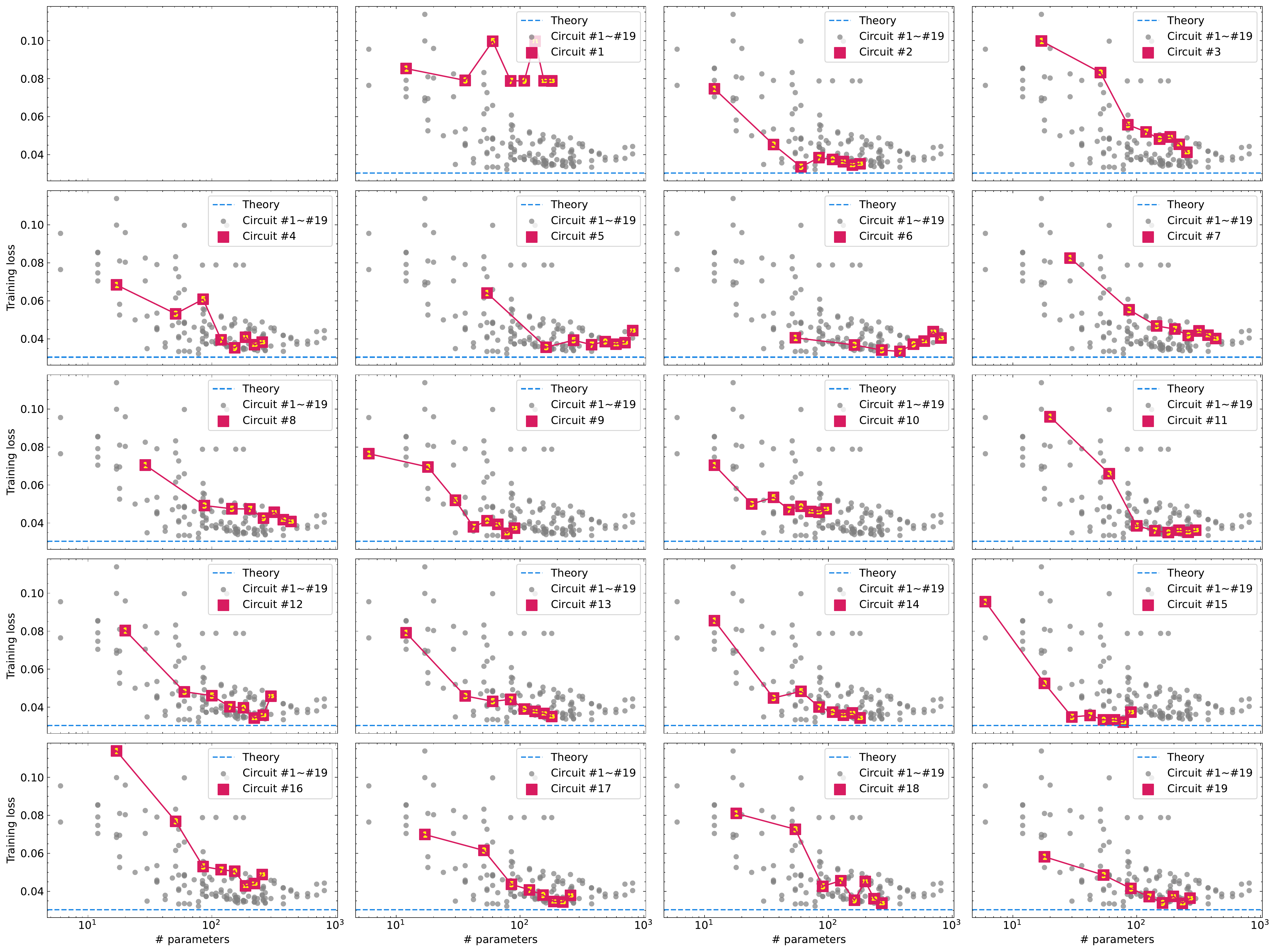}
        \caption{Training results on iris data set ($M=64$)}
    \end{figure}
    \begin{figure}[ht]
        \centering
        \includegraphics[width=\textwidth]{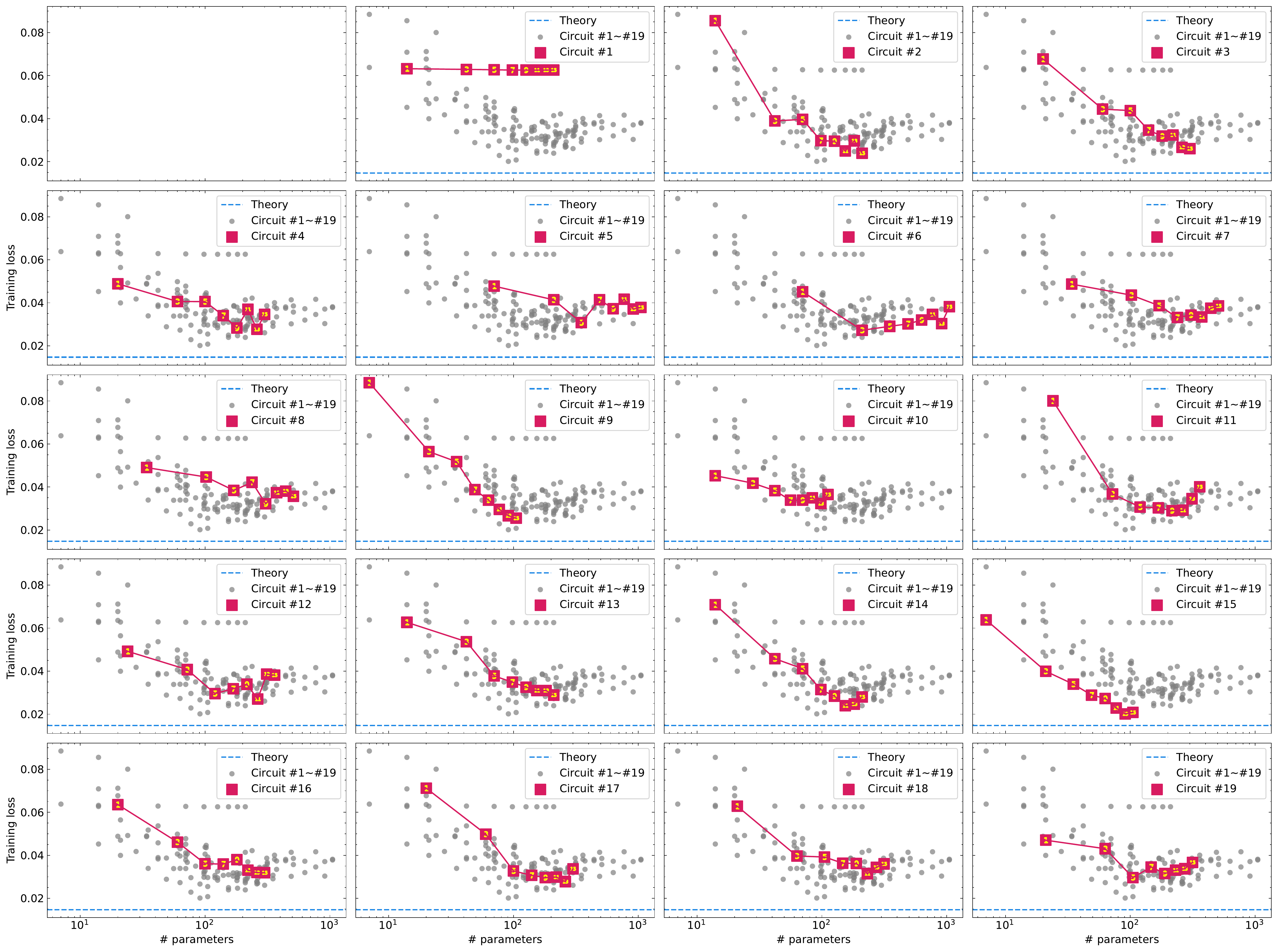}
        \caption{Training results on iris data set ($M=128$)}
    \end{figure}
    \begin{figure}[ht]
        \centering
        \begin{subfigure}{0.45\textwidth}
            \centering
            \includegraphics[width=\textwidth]{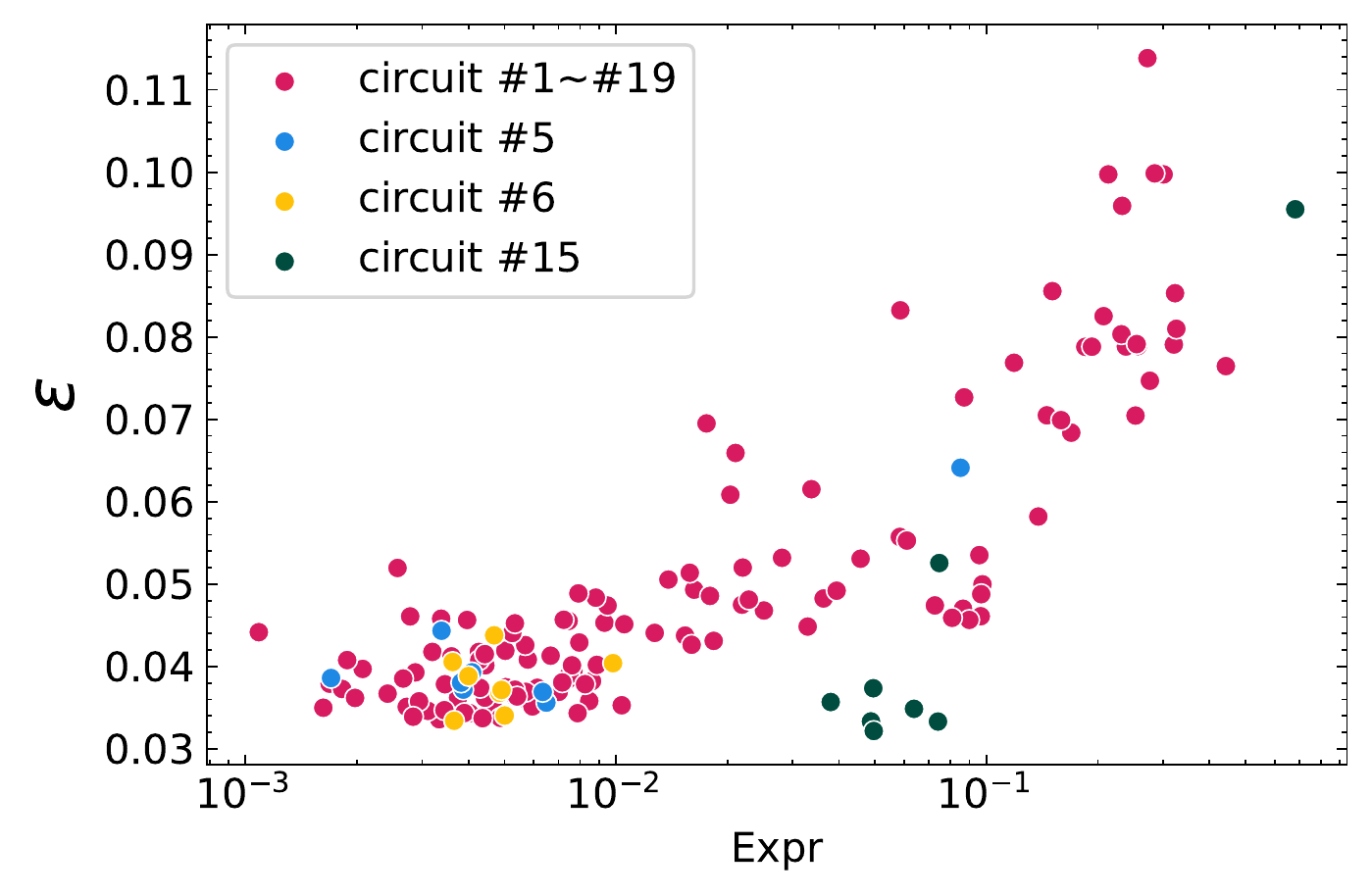}
            \caption{$M=64$}
        \end{subfigure}
        \begin{subfigure}{0.45\textwidth}
            \centering
            \includegraphics[width=\textwidth]{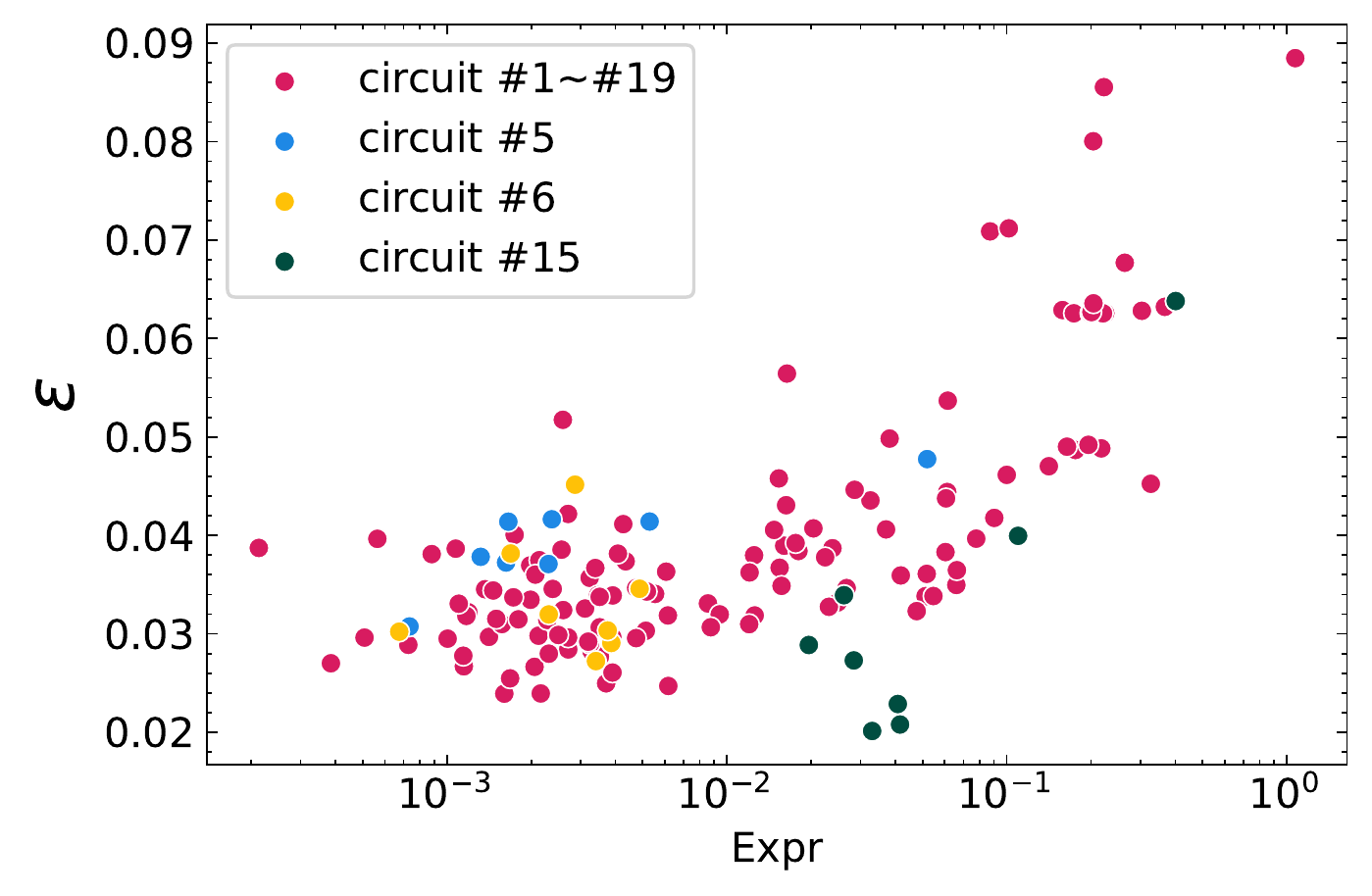}
            \caption{$M=128$}
        \end{subfigure}
        \caption{Decision value error on iris data set. \textsf{Expr} refers to expressibility of PQCs.\cite{sim2019expressibility} Circuit \#5 and \#6 has $L\times\log(M)^2$ parameters.}
    \end{figure}
    \begin{figure}[t]
        \centering
        \begin{subfigure}{0.49\textwidth}
            \centering
            \includegraphics[width=\textwidth]{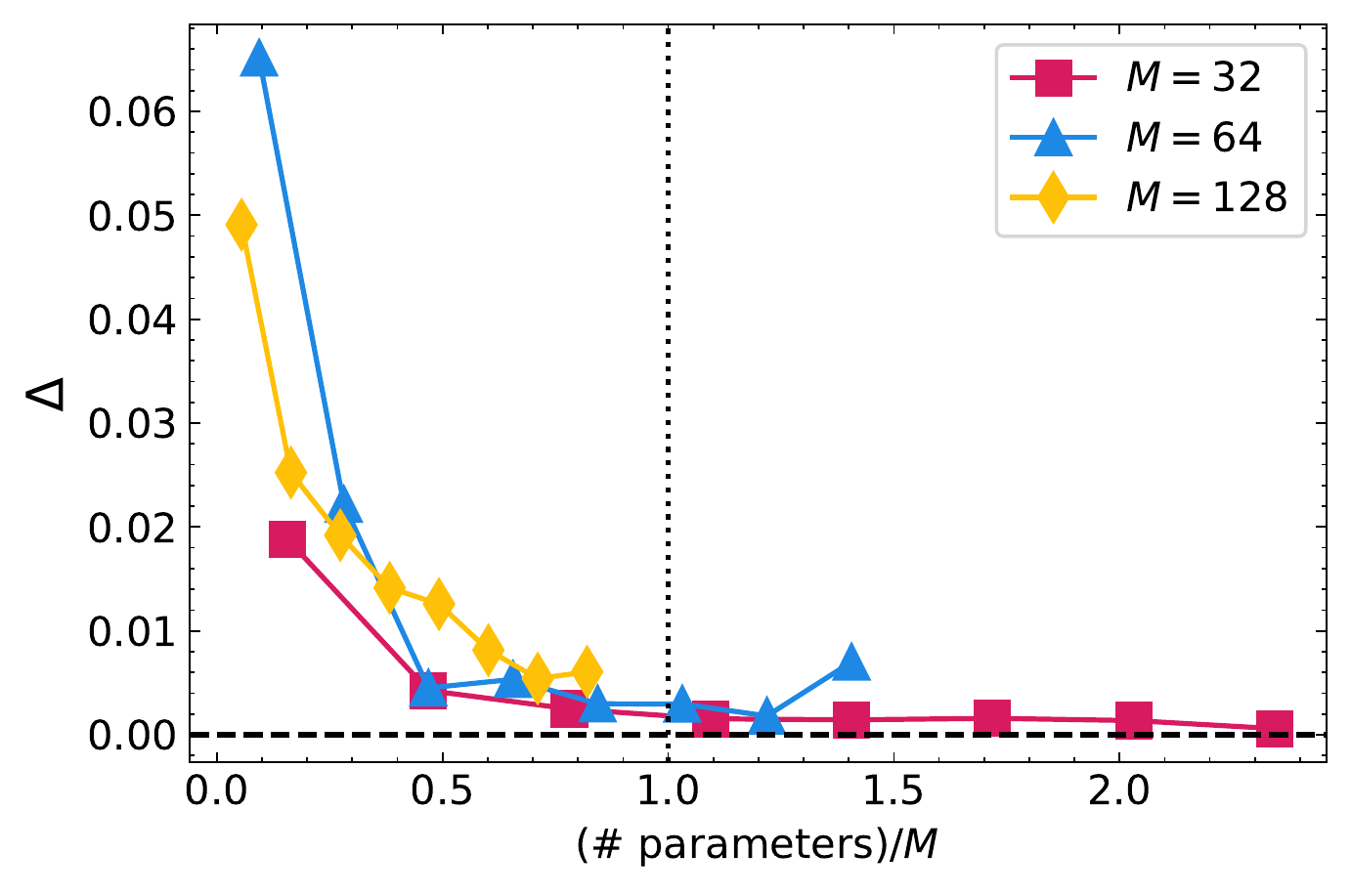}
            \caption{Loss on training data \label{fig:last_cost_avg}}
        \end{subfigure}
        \hfill
        \begin{subfigure}{0.49\textwidth}
            \centering
            \includegraphics[width=\textwidth]{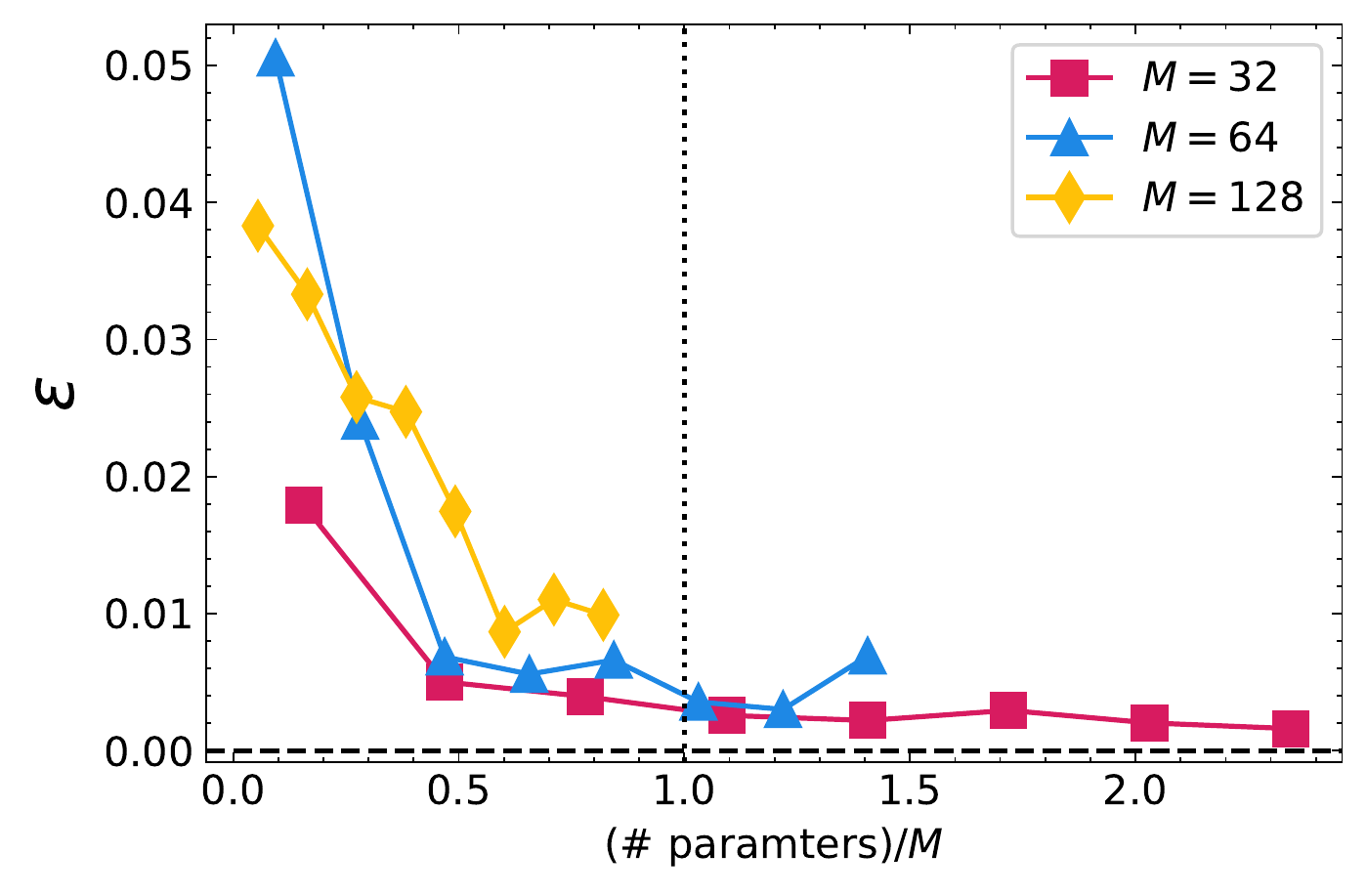}
            \caption{Accuracy on test data\label{fig:ferror}}
        \end{subfigure}
        \hfill
        \begin{subfigure}{0.49\textwidth}
            \centering
            \includegraphics[width=\textwidth]{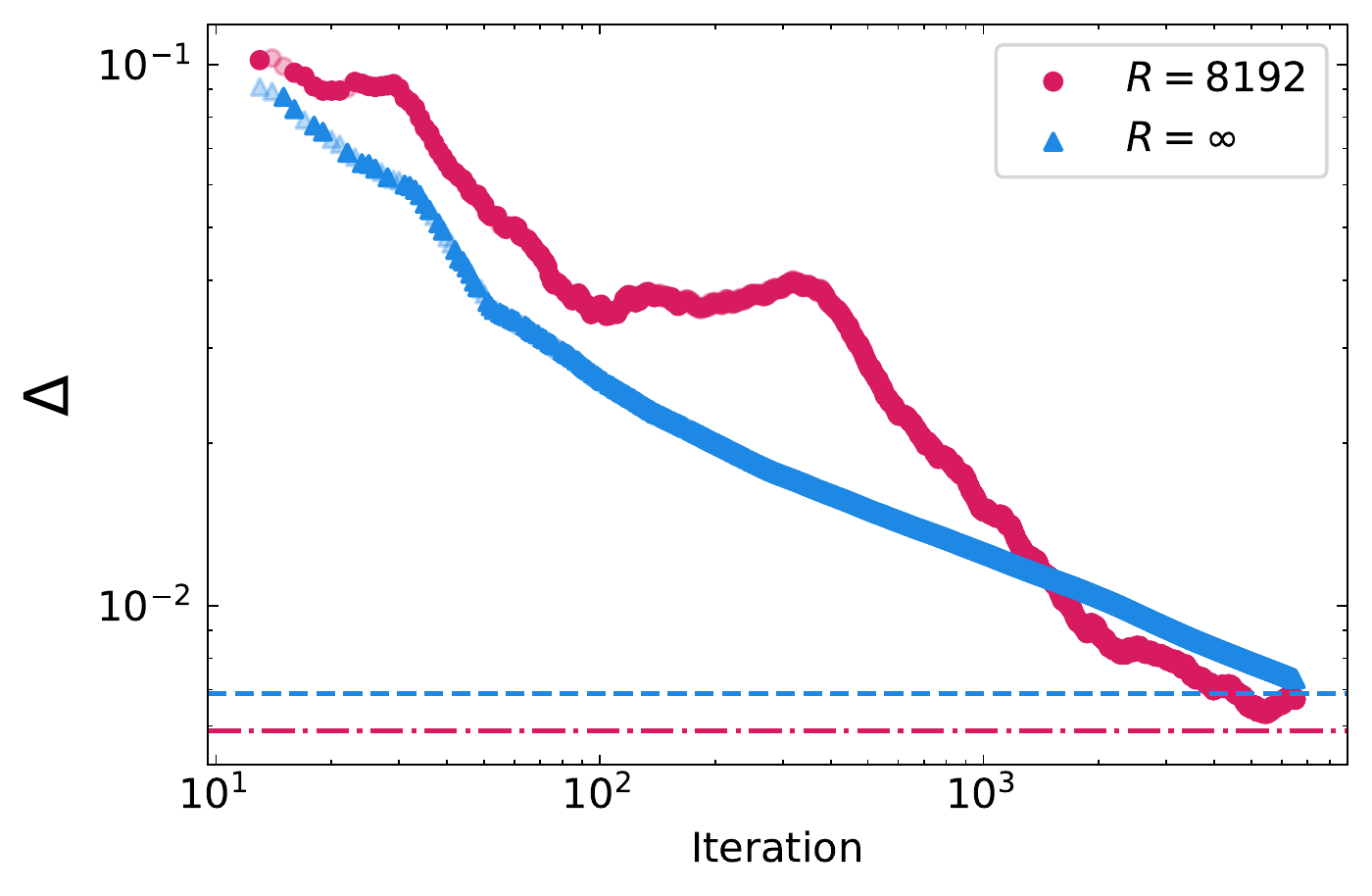}
            \caption{Training loss convergence\label{fig:stat_no_error}}
        \end{subfigure}
        \hfill
        \begin{subfigure}{0.49\textwidth}
            \centering
            \includegraphics[width=\textwidth]{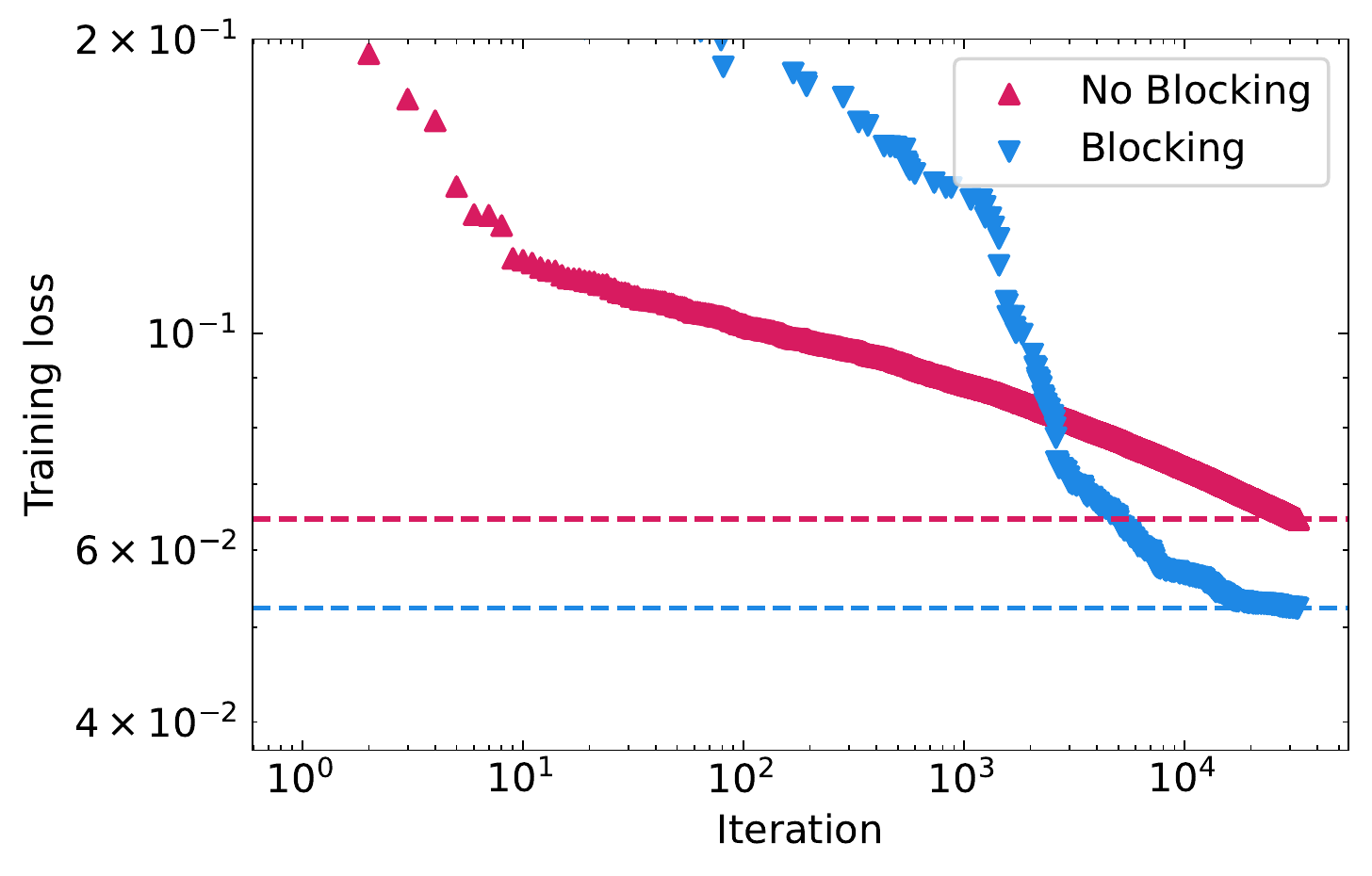}
            \caption{Blocking technique for SPSA}
        \end{subfigure}
        \caption{Numerical Analysis on iris data set ($\lambda=C=10^4$). In {\bf a} and {\bf b}, after $2^{10}$ iterations, the residual loss of training $\Delta$ and average decision value error $\mathcal{E}$ are plotted as a function of normalized number of parameters with respect to $M$, respectively. In {\bf c}, coarse-grained $\Delta$ averaged over $10^{-0.1}t$ to $10^{0.1}t$ at iteration $t$ is shown in a log scale plot. In {\bf d}, Training loss converges faster when blocking technique for SPSA is used.}
    \end{figure}
    \begin{figure}[ht]
        \centering
        \begin{subfigure}{0.45\textwidth}
            \centering
            \includegraphics[width=\textwidth]{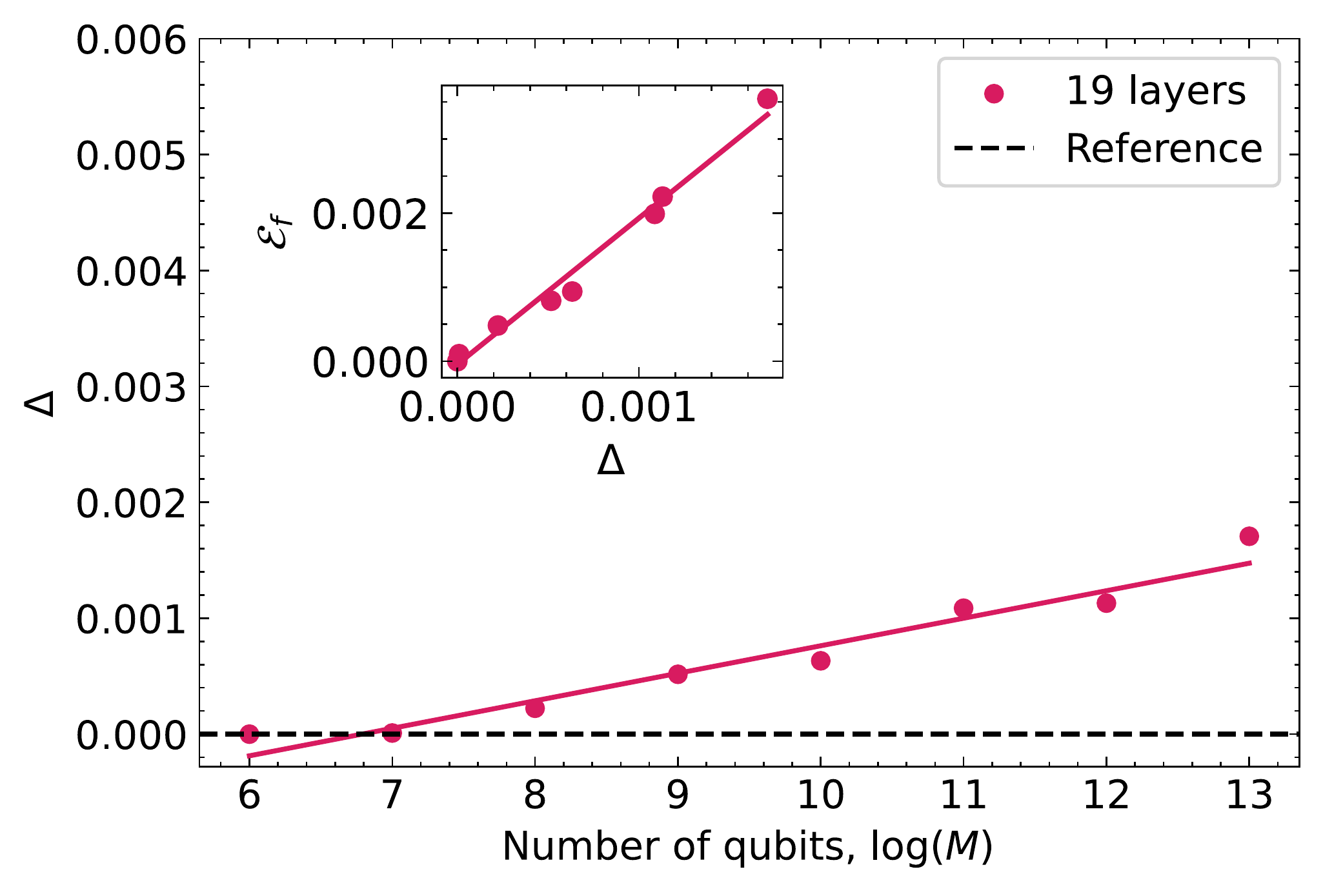}
            \caption{Trainability}
        \end{subfigure}
        \hfill
        \begin{subfigure}{0.45\textwidth}
            \centering
            \includegraphics[width=\textwidth]{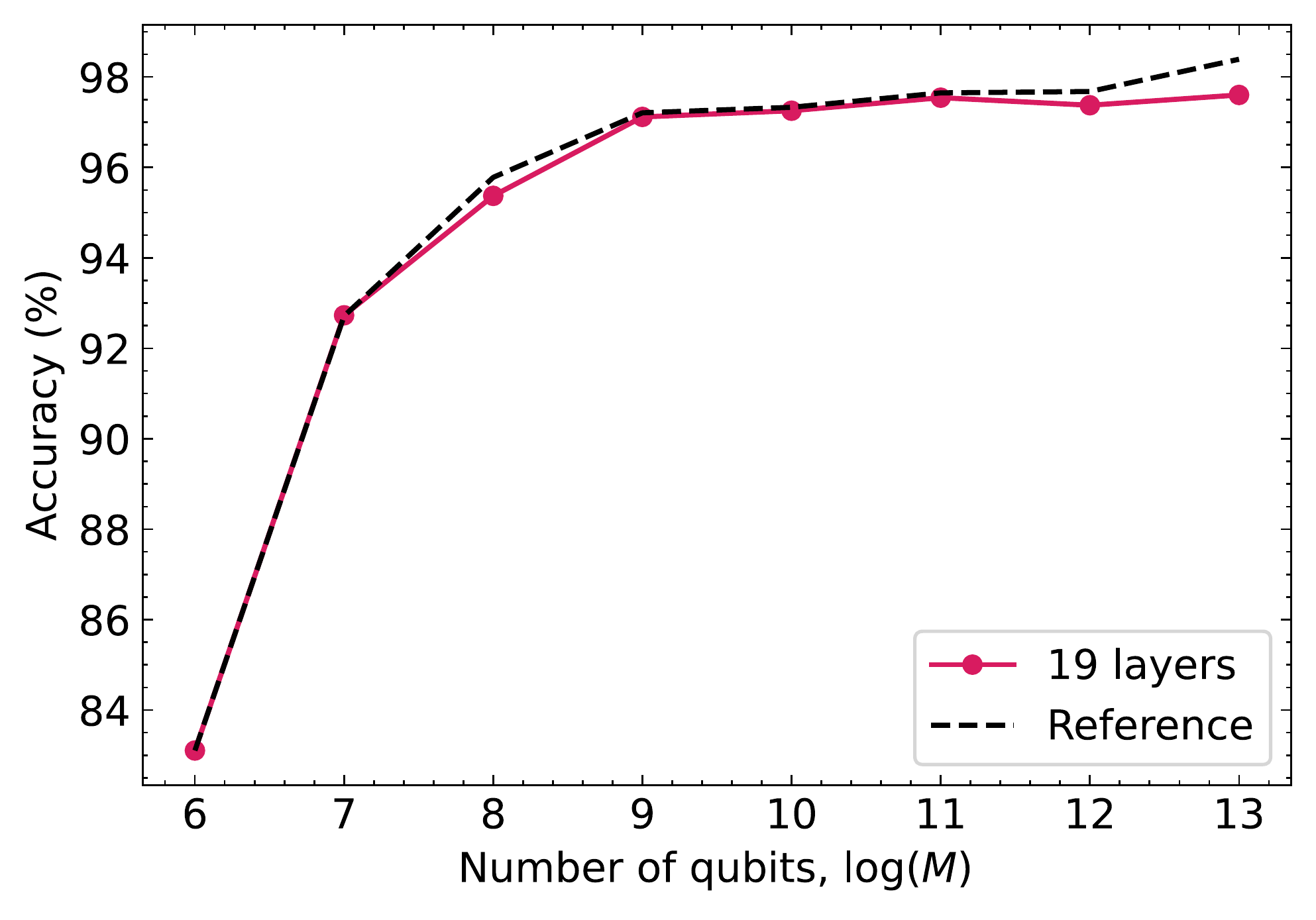}
            \caption{Classification Accuracy}
        \end{subfigure}
        \caption{Numerical Analysis on the MNIST dataset ($\lambda=C=10^4$). In the legend, `19 layers' refers to the results of VQASVM with the PQC having 19 layers, whereas `Reference' denotes the results obtained by convex optimization. \textbf{\textsf{(a)}} $\mathcal{E}_f$ is the error on decision function; $\mathcal{E}_f=\sum_{\mathbf{\hat{x}}\in\mathcal{T}}|{f_{\lambda,\phi}(\mathbf{\hat{x}};\boldsymbol{\theta}^\star,\mathcal{S})-h(\mathbf{\hat{x}};\mathcal{S})}|/\abs{\mathcal{T}}$ where $\mathcal{T}$ is the set of test data and $h$ is the theoretical decision function obtained by convex optimization. }
    \end{figure}
    
\end{document}